\documentclass{aa}

\usepackage[T1]{fontenc}
\usepackage{amsmath}
\usepackage{amsfonts}
\usepackage{url}
\usepackage{bm}
\usepackage{graphicx}
\usepackage{amssymb}
\usepackage{soul}
\usepackage{booktabs}
\usepackage{array}
\usepackage[utf8]{inputenc}
\inputencoding{latin1}
\inputencoding{utf8}
\usepackage{appendix}
\usepackage{diagbox}
\usepackage[table]{xcolor}
\usepackage{caption}
\usepackage{adjustbox} 

\usepackage[normalem]{ulem}
\bibpunct{(}{)}{;}{a}{}{,} 

\def\lsim{ \lower .75ex\hbox{$\sim$} \llap{\raise .27ex \hbox{$<$}} }
\def\gsim{ \lower .75ex \hbox{$\sim$} \llap{\raise .27ex \hbox{$>$}} }

\newcommand{\bi}{\begin{itemize}}
\newcommand{\ei}{\end{itemize}}

\DeclareMathOperator{\sech}{sech}
\newcommand{\citeg}[1]{(e.g., \citealt{#1})}

\definecolor{cleangray}{gray}{0.85} 

\usepackage[colorlinks=true,linkcolor=blue,citecolor=blue]{hyperref}

\title{How do recollimation-induced instabilities shape the propagation of hydrodynamic relativistic jets?} 

\author{
A. Costa\inst{1,2}
\and G. Bodo\inst{2}
\and F. Tavecchio\inst{3}
\and P. Rossi\inst{2}
\and P. Coppi\inst{4}
\and A. Sciaccaluga\inst{3}
\and S. Boula\inst{3}
}

\institute{
DiSAT, Università dell’Insubria, Via Valleggio 11, I-22100 Como, Italy\\
\email{agnese.costa@inaf.it}
\and
INAF -- Osservatorio Astrofisico di Torino, Strada Osservatorio 20, 10025 Pino Torinese, Italy
\and
INAF -- Osservatorio Astronomico di Brera, via E. Bianchi 46, 23807 Merate, Italy
\and
Department of Astronomy, Yale University, PO Box 208101, New Haven, CT 06520-8101, USA
}
\date{}

\begin{document}



\abstract
{Recollimation is a phenomenon of particular importance in the dynamical evolution of jets and in the emission of high-energy radiation. Additionally, the full comprehension of this phenomenon provides insights into fundamental properties of jets in the vicinity of the Active Galactic Nucleus (AGN). Three-dimensional (magneto-)hydrodynamic simulations revealed that the jet conditions at recollimation favor the growth of strong instabilities, challenging the traditional view—supported from two-dimensional simulations—of confined jets undergoing a series of recollimation and reflection shocks.}{In order to investigate the stability of relativistic jets in Active Galactic Nuclei at recollimation sites, we perform a set of long duration three-dimensional relativistic hydrodynamic simulation, to focus on the development of hydrodynamical instabilities. We explore the non-linear growth of the instabilities and their effects on the physical jet properties as a function of the initial jet parameters.}{We perform two-dimensional and three-dimensional relativistic hydrodynamic simulations using the state-of-the-art PLUTO code. We assume that an initially free-expanding jet is collimated from an external medium, and we explore the role of the jet Lorentz factor, temperature, opening angle and the jet-environment density-contrast on the jet deceleration and entrainment. The parameter space is designed to describe low-power, weakly magnetized jets at small distances from the core (around the parsec scale).}{All collimating jets we simulated develop instabilities. Recollimation instabilities decelerate the jet, heat it, entrain external material, and move the recollimation point to shorter distances from the core. This is true for both conical and cylindrical jets. The instabilities, that are first triggered by the centrifugal instability, appear to be less disruptive in the case of narrower, denser, more relativistic, and warmer jets. These results provide valuable insights into the complex processes governing AGN jets and could be used to model the properties of low-power, weakly magnetized jetted AGNs.}{}{}


\titlerunning{Recollimation shock instabilities}

\maketitle

\section{Introduction}

Relativistic jets released from Active Galactic Nuclei (AGN) are the most powerful persistent emitters of the Universe, with powers even up to $10^{46}$ erg/s \citeg{blandford19}. They are extremely fascinating objects where the large-scale physical properties of the bulk determine the conditions for particle acceleration and the non-thermal radiation observed across the whole electromagnetic spectrum \citeg{romero17}, from the radio band up to TeV energies, and in the form of multi-messenger emission, with at least one neutrino detected from a blazar \citep{txs18_neutrino,ansoldi18}. 

AGN jets have been observed on a variety of length scales spanning various decades, proving excellent stability, up to distances from the core of the order of the Mpc \citep{willis74} in the case of the most powerful sources. Jetted AGNs are canonically divided in the radio in edge-darkened Fanaroff Riley type I (FRI) and edge-brightened Fanaroff Riley type II (FRII) \citep{FR74}, respectively corresponding less and more powerful sources. In addition, it was recently realized that the largest part of jetted AGNs are in the form of even less powerful, more edge-darkened sources \citep{bestheck12}, named Fanaroff Riley Type 0 (FR0) \citep{ghisellini11,sadler14,baldi15}, where the jets have bulk speeds of the order of $0.5c$ or less at parsec scales \citep{cheng18,baldi21b,cheng21,giovannini23}. Differences in their morphologies have been partially attributed to FRI and FR0 jets being more subject to instabilities \citeg{rossi08,laing14,perucho14,rossi20,costa23}. In fact, it has been extensively observed that a broad range of magneto-hydrodynamic instabilities can affect the jet propagation, including the Kelvin-Helmholtz, current driven, pressure driven, centrifugal, Rayleigh-Taylor \citeg{birkinshaw96,bodo13,kim18,gourgkom18b,begelman19,musso24}. 

In this context, recent research has emphasized the role of recollimation shocks in promoting the development of instabilities that can even disrupt the jet relativistic flow \citeg{matsumasa13,gourgkom18}. Recollimation shocks are an extremely interesting area of focus interconnected to a web of unsolved problems, both dynamical and radiative. Generally, observations cannot resolve well the areas in the vicinity of the black hole, with the important exceptions of a few sources, such as BL Lac and M87 among low-power radio-galaxies \citeg{cohen14,walker18,hada}. These sources exhibit jets undergoing an early phase of acceleration, associated with a fast expansion, then reaching relativistic velocities, and becoming more collimated at greater distance; nevertheless, the jet properties are not well constrained at such a small distance. Most of the high energy radiation is produced in those inner regions, as required from the fast variability seen in many blazars \citep{tavecchio11,aleksic14}, through various possible acceleration and emission mechanisms, highly dependent on the local jet conditions. Evidence mounted that in low power AGNs there has to be some critical event at short distances from the nucleus, that makes the jet change its configuration, and that is associated with strong non-thermal emission, possibly consisting of a recollimation shock \citep{bromlev07,nalewajko09,bodotav18}.

Contrary to the well established view of recollimated jets undergoing a series of recollimation and reflection shocks, based on theoretical works \citep{komfalle97} and supported  2D axisymmetric simulations \citeg{mizuno15}, 3D simulations \citeg{matsumasa13,gourgkom18} showed that the region downstream of recollimation shock is strongly subject to instabilities. The first works focused on the Rayleigh-Taylor instability (RTI) developing at a heavy jet-light environment interface, as well as the Richtmyer–Meshkov instability (RMI) growing at reflection shock from the RTI-induced corrugations of the contact discontinuity  \citep{matsumasa13}. Later, it was understood that, regardless of the jet relative density with respect to the external medium, the centrifugal instability (CFI) grows due to the curvature of the streamlines downstream of the recollimation shock \citep{gourgkom18}. These first studies on the CFI showed some differences in the instability impact between lighter and heavier jets, but did not fully explore the parameter space, nor the instability evolution in depth. There have been following works that further studied the jet stability at recollimation. Some papers analyzed the role of magnetic fields \citeg{matsumoto21,GottliebMHD,hu24}. \cite{costa23} modeled the peculiar features of FR0s through the lens of recollimation-induced instabilities, \cite{GottliebHD} focused on GRB jets, \cite{perucho18} focused on a case-study for X-ray binaries, \cite{ayan24} discussed the recollimation dynamics at large distances from the core, \cite{abolbrom23} explored more the parameter space, but focusing on RTI-induced instabilities without discussing the CFI, and a comprehensive parameter study designed to describe recollimation instabilities in AGN jets is lacking.  

This work aims at shedding light on the recollimation machine in the framework of jets near their inner core, where blazars are expected to emit and radio-galaxies become collimated. We pursued a program of 3D relativistic hydrodynamic, high resolution simulations with the PLUTO code \citep{M07}, dedicated to study the onset and effects of instabilities in AGN jets. Given that low-power jets are typically more susceptible to instabilities, the setup and the parameters were chosen to reflect the observed properties of canonical FR0s and weak FRIs. In Section \ref{setup} we introduce the physical problem of a relativistic hydrodynamic conical jet, collimated  the external medium, and we define the relevant parameters that we vary in our simulations. In Section \ref{numerical} we describe the numerical approach and the simulation setup. In Section \ref{2D_results} we describe the results from axisymmetric simulations and compare them with the general view of axisymmetric collimated jets. In Section \ref{details}, we study the early phases of the instability growth for an exemplary setup. In Section \ref{bubble} we analyze the later evolution of instabilities and the feedback they have on the stable part of the jet, focusing on two different cases.
In Section \ref{decentr} we analyze the structure of the quasi-stationary configuration that the jets reach at the end of the simulations and we discuss the differences in deceleration and entrainment across different setups. 
In Section \ref{remarks} we present a cylindrical setup in order to make connections with previous works.
Finally, in Section \ref{discussion} we conclude with a summary of our results, a discussion of the implications, and an overview to future implementations. 

\section{Problem description \label{setup}}

There is evidence that AGN jets feature quasi-stationary structures, likely corresponding to recollimation shocks, around parsec scale distances from the central engine \citeg{bromlev09,cohen14,casadio21}. Since evolutionary times are extremely long ($\gtrsim10^7$ years) \citep{begelman89}, we assume that the cocoon resulting from the jet head bow shock has had the time to disperse at parsec scales, and that it it does not influence the jet dynamics. For this reason, we simulate a jet that is \textit{naked}, meaning that it is directly in contact with the external material, without a cocoon. We then expect that, after a first region of free propagation during which the jet has a conical shape, the jet enters a recollimation region, developing a series of recollimation shocks. Our purpose is to study the stability of this recollimation shock chain through numerical simulations. We have to specify that in our approach we do not simulate the jet propagation, since we should run the simulation until the jet head leaves the computational box and the cocoon disperses, as has been done, for example, in \cite{abolbrom23}. Instead, we first carry out 2D axisymmetric simulations to find steady solutions; and then we study the stability of this steady configuration by performing 3D simulations, thus relaxing the axisymmetry constraint. In this way we can better disentangle the evolutionary processes associated with jet propagation and instability growth and compare our results with those obtained in \cite{gourgkom18,matsumoto21} through an analogous procedure.

\subsection{Fluid equations}
The system is described by the set of relativistic hydrodynamic equations (RHD), expressing conservation of mass, momentum and energy, that can be written as 
\begin{equation}
    \frac{\partial }{\partial t} \left( \begin{matrix}
        d\\
        \mathbf{m}\\
        e_t
    \end{matrix} \right) + \boldsymbol{\nabla}\cdot \left( \begin{matrix}
        d\mathbf{v}\\
        \mathbf{mv}+p\mathbf{I}\\
        \mathbf{m}
    \end{matrix}\right) = \left( \begin{matrix}
        0\\
        \mathbf{f}_g\\
        \mathbf{f}_g\cdot \mathbf{v}
    \end{matrix} \right),
    \label{eq:pluto_rhd}
\end{equation}
where $d = \rho \Gamma$, $\mathbf{m}=\rho h \Gamma^2 \mathbf{v}$ is the momentum per unit volume, $e_t= \rho h \Gamma^2 - p$ is the energy density, $\rho$ is the proper density, $p$ is the pressure, $\mathbf{v}$ is the three-velocity, $\mathbf{f}_g$  is a force density, $\Gamma$ and $h$ are respectively the Lorentz factor and the specific enthalpy and $\mathbf{I}$ is the unit $3 \times 3$ tensor. We close the set of equations with the Taub-Matthews equation of state \citep{mignoneTM}: 
\begin{equation}
    h = \frac{5}{2}\mathcal{T}+\sqrt{\frac{9}{4}\mathcal{T}^2+1},\quad\text{with}\,\mathcal{T}=p/(\rho c^2),
\end{equation}
that approximates the Synge EoS of a single-specie relativistic perfect fluid \citep{synge}. 

\subsection{The environment \label{environment}}
The jet is injected in a static, gravitationally stratified external medium,
characterized  by density and pressure that decay with distance from the central engine as power laws of index $\eta$:
\begin{align}
  \rho_{\text{ext}}(z)=\rho_{\text{ext},0}\left(\frac{z}{z_0}\right)^{-\eta}, \quad
p_{\text{ext}}(z) = p_{\text{ext},0} \left(\frac{z}{z_0}\right)^{-\eta},
\end{align}
where the subscript `ext' refers to external quantities, and we adopted the notation $q_{\text{ext},0} = q_{\text{ext}}(z_0)$ for the values of quantities at $z_0$, that represents the $z$ coordinate of the jet injection (see Fig. \ref{fig:sketch}).  
The external medium is maintained in hydrostatic equilibrium through an external force per unit volume given by
\begin{equation}
    \mathbf{f}_g = \frac{\partial p_{\text{ext}}(z)}{\partial z} \mathbf{\hat{e}}_z = - \frac{1}{2}\frac{p_{\text{ext},0}}{z_0} \left(\frac{z}{z_0}\right)^{-\eta-1} \mathbf{\hat{e}}_z .
\end{equation}
This environment should represent a wind or the warm interstellar medium still influenced the central engine attractive gravitational force, able to confine the jet. For this reason, the index $\eta$ should be smaller than $2$ \citep{komfalle97}, and and we set $\eta = 0.5$, determining a quasi-cylindrical propagation, to simulate a region where the jet transitions from a conical to a more collimated profile \citeg{porkom15}.

\subsection{Jet parameters and units}
Our equations are nondimensionalized by using $z_0$ as the unit of length, $c$ as the unit of velocity and $\rho_{\text{ext},0}$ as the unit of density. If we choose specific values for $z_0$ and $\rho_{\text{ext},0}$ we can expess all quantities in physical units. We chose the constants $z_0$ and $\rho_{ext}(z_0)$  to be in agreement with typical properties of low-power radio galaxies at recollimation/blazar scales \citeg{heckman14, russell15, boccardi21, casadio21}. In all simulations, we adopted $z_0=1$ pc, $\rho_{\text{ext},0}=\rho_0=1 \, m_p\mbox{cm}^{-3}$. It is however important to note that the results of hydrodynamic simulations are scale invariant and therefore do not depend on these specific choices  and it is possible to re-scale them depending on the appropriate astrophysical conditions one wants to consider.

The axisymmetric steady state is characterized by a few fundamental parameters \citep{mukh20}, that we define at $z_0$, that is the length scale of the simulations and represents the jet distance from its cone vertex (see Fig. \ref{fig:sketch}), assumed to be comparable with the distance from the central engine. In table \ref{tab:prameters} we list the values of each parameter in the different simulations we performed. To distinguish between different cases, we use acronyms that indicate the values of the parameters that characterize the simulated jet (see the first row of table \ref{tab:prameters}).

We list below the parameters, denoting with the subscript `j' the jet quantities:

\textit{a) Lorentz factor} $\Gamma_j$. The value of the Lorentz factor in AGN jets can vary between  $3$ to $50$ in the most relativistic sources \citeg{ghisellini93,lister13,ghita15}. We adopt $\Gamma_j =5$ in most of the setups, denoted with the letter `S' (slow), but we increase it to $\Gamma_j = 15$ in one case, called `Ff' (fast). 

\begin{table*}[t]
\centering
\scalebox{0.9}{ 
\begin{tabular}{|c|c|c|c|c|c|c|c|c|c|}
\hline
\diagbox[width=2cm,height=1.1cm]{Param.}{Case} & \textbf{SLC2} & \textbf{FfLC2} & \textbf{SLW2} & \textbf{SLWw2} & \textbf{SHC2} & \textbf{SHC05} & \textbf{SHCcyl}\\
\hline
\hline
\rowcolor{cleangray} 
$\Gamma_j$ & 5 & 15 & 5 & 5 & 5 &5  & 5 \\
\hline
$\theta_{j}$ & 0.2 & 0.2 & 0.2 & 0.2 & 0.2 & 0.05 & 0 \\
\hline
\rowcolor{cleangray} 
$\nu$ & $7.6 \times 10^{-6}$ & $7.6 \times 10^{-6}$ & $7.6 \times 10^{-6}$ & $7.6 \times 10^{-6}$ & $7.6 \times 10^{-5}$  &  $7.6 \times 10^{-5}$ &  $7.6 \times 10^{-5}$ \\
\hline
$p_r$ & $10^{-3}$ & $10^{-3}$ & $1$ & $10$ & $10^{-2}$ & $10^{-3}$ & $ 10^{-2}$ \\
\hline
\hline
\rowcolor{cleangray} 
$r_{j,0}$ (pc) & 0.2 & 0.2 & 0.2 & 0.2 & 0.2 & 0.05 & 0.05 \\
\hline
$L_j $ (erg/s)&  $1.0 \times10^{40}$ & $9.5 \times 10^{40}$ & $2.2 \times 10^{40}$ & $ 1.6\times 10^{41}$ & $ 1.0\times 10^{41}$ & $6.5 \times 10^{39}$ & $ 6.5\times10^{39}$ \\
\hline\rowcolor{cleangray} 
$\mathcal{T}_j$ & $3.9 \times 10^{-4}$ & $3.9 \times 10^{-4}$ & $3.9 \times 10^{-1}$ & $3.9 $ & $3.9 \times 10^{-4}$ & $3.9 \times 10^{-5}$ &  $3.9 \times 10^{-4}$ \\
\hline
\end{tabular}
}
\vspace{0.5cm}
\caption{List of parameters in the different cases we simulated. The names of the cases indicate the jet properties: L/H= light/heavy, C/W/Ww = cold/warm/very warm, S/Ff = slow/fast, 2/05/cyl: wide/narrow/cylindrical.}
\label{tab:prameters}
\end{table*}

\textit{b) Jet radius / jet opening angle} $r_0$, $\theta_j$. The relation between the two quantities in conical jets is $r_0 = z_0 \theta_j$. The value of the opening angle  is usually thought to be very small, but at small distances from the central engine it can be quite large, depending on the position where the transition from a parabolic to conical profile occurs \citeg{walker18}. We compare results for $\theta_j = 0.2$ and $\theta_j = 0.05$, named with $2$ or $05$. In one case we simulate a cylindrical jet, denoted with `cyl', and in that case we use $r_0 = 0.05\, z_0$.

\textit{c) Density contrast} $\nu = \rho_{j,0}/\rho_{\text{ext},0}$. This is the value that, in cold jets, once the Lorentz factor and the jet opening angle have been fixed, determines the power of the source. The jet power, using the values of $z_0$  and $\rho_{\text{ext},0}$ discussed above, is
\begin{align}
    L_j =& (\pi z_0^2\theta_j^2)\rho_j h_j c^2 \Gamma_j^2 v_j  =\\
    \simeq & 1.2\times 10^{40}\left(\frac{\theta_j}{0.2}\right)^2\left(\frac{\nu}{10^{-5}}\right)\left(\frac{h_j}{1}\right)\left(\frac{\Gamma_j}{5}\right)^2\,\mbox{erg s$^{-1}$},
\end{align}
that would correspond to a weak relativistic jet compatible with a FR0 source \citeg{baldi23}. In order to simulate FR0 and FRI jets, we choose as standard value $\nu = 7.6 \times 10^{-6}$ in all simulation named with `L' (light) and we increase it by a factor $10$ in a few simulations, denoted with `H' (heavy). As we shall see, larger values of the density contrast move the recollimation point further away from the origin and the simulations of these cases becomes unfeasible because one would need much larger grid extensions and, in order to maintain an adequate resolution, the computational cost would become unacceptably large.

\textit{d) External temperature} $\mathcal{T}_\text{ext} = p_{\text{ext}}/(\rho_{\text{ext}}c^2)$. In general, for a fully ionized hydrogen gas, the temperature $\mathcal{T}$ is linked to the temperature in Kelvin $T$ through the relation:
\begin{equation}
    T = \frac{m_p c^2}{2 k_B}\mathcal{T} = 5\times 10^{12}\mathcal{T}\,\text{K} .
    \label{eq:temp_real}
\end{equation}
In all the simulations we took $\mathcal{T}_\text{ext} = 3\times 10^{-6}$: with this choice the environment temperature  $T_{\text{ext}} = 1.5\times 10^7$ K. Much higher values would not be supported from the lack of thermal emission at high energies, but we still do not expect very low values at small distances from the central engine.

\textit{e) Jet-environment pressure ratio} at the jet base, $p_r = p_{j,0}/p_{\text{ext},0}$. Since the external environment properties have been chosen, this value implicitly determines the jet temperature
\begin{equation}
    \mathcal{T}_{j,0} = \frac{p_{j,0}}{\rho_{j,0} c^2} = \frac{p_r}{\nu} \mathcal{T}_\text{ext},
\end{equation}
and its specific enthalpy (we remind that $c=1$)
\begin{equation}
    h_{j,0} =  \frac{5}{2}\mathcal{T}_{j,0} + \sqrt{\frac{9}{4}\mathcal{T}_{j,0}^2+1}.
\end{equation}
In our simulations we often consider cold jets, denoted with letter `C' (cold), with $h_{j,0}\simeq 1$ and $p_r =10^{-3} \,- 10^{-2}$. 
We also consider warmer jets, with $p_r = 1$ and $p_r = 10$, called `W' (warm) and `Ww' (very warm) with specific enthalpy respectively $h_{j,0} = 2$ and $h_{j,0} = 16$. 

\section{Numerical approach
}\label{numerical}

The simulations are carried out with the state-of-the-art PLUTO code for astrophysics \citep{M07,mignoneAMR}. In particular, we use the `RHD' module of the code, designated to solve the set of conservative equations of relativistic hydrodynamics (RHD). We adopt a linear reconstruction scheme \citep{M07}, with a second order Runge-Kutta method for time integration \citeg{butcher_rk}, and, in order to being able to capture the instability \citep{matsumasa13}, the HLLC Riemann solver \citep{pluto_hllc}. The details of the numerical grids for the different simulations are discussed in the following subsections.

\begin{figure}[htbp]
    \centering
    \includegraphics[width=3.5cm]{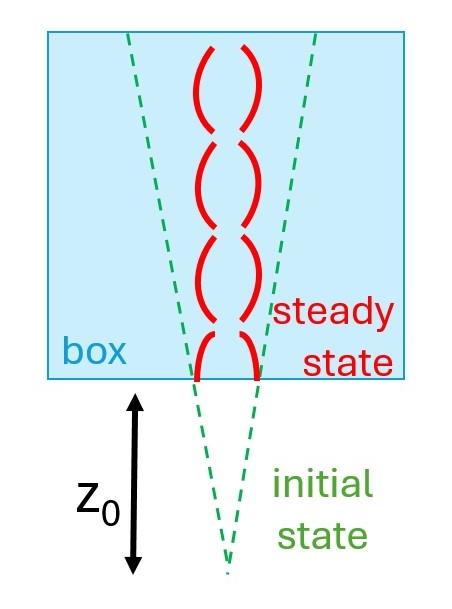}        
    \caption{Sketch of the computational setup (not to scale) for the 2D simulations. We illustrate the free propagation regime (initial configuration, in green) and the recollimated steady state (final state, in red), overlayed on the computational box, at a distance $z_0$ from the cone vertex of the initial conical profile.}
    \label{fig:sketch}
\end{figure}

As discussed above and similarly to what \cite{costa23} did, we first perform 2D axisymmetric simulations to find steady state configurations, that we then use as initial conditions for the 3D runs. 2D simulations are performed with an initial condition of a freely propagating conical jet, with injection at the boundary $z=z_0$ as represented in Fig. \ref{fig:sketch} and following what has been done in \cite{bodotav18}. The jet is then let evolve until a steady state is reached.

\subsection{2D setup \label{2D_setup}}

The 2D axisymmetric simulation are performed in cylindrical coordinates $(r,z)$, assuming axisymmetry. The initial condition at $t=0$ (see Fig. \ref{fig:sketch}) is that of a conical jet of opening angle $\theta_j$ with Lorentz factor $\Gamma_j$ and velocity components
\begin{equation}
  v_z (r,z,t=0)= \sqrt{1- \frac{1}{\Gamma_j^2}} \frac{z}{R}, \label{eq:vz}  
\end{equation}

\begin{equation}
v_r (r,z,t=0) = \sqrt{1- \frac{1}{\Gamma_j^2}} \frac{r}{R},    \label{eq:vr}
\end{equation}
where $R$ is the spherical radius: $R=\sqrt{r^2+z^2}$. As the jet expands, its density decays as $1/R^2$ and its pressure decays adiabatically from their conditions defined at $z_0$ (and $r=0$) as
\begin{gather}
  \rho_j(r,z,t=0) = \nu \left(\frac{R}{R_0}\right)^{-2}, \label{eq:rho}\\ 
 p_j(r,z,t=0) = p_r \mathcal{T}_{\text{ext},0} \left(\frac{R}{R_0}\right)^{-2\gamma},\label{eq:prs}
\end{gather}
where $R_0=z_0$, $\gamma$ is the adiabatic index derived from the EoS. We use $\gamma=5/3$ for cold jets and $\gamma=4/3$ for warm setups (denoted with `W' or `Ww'). In the cylindrical case, SHCcyl, the radial velocity is set to zero, and the quantities do not decay due to the jet expansion, so the jet primitive variables are initialized as

 \begin{align}
     & v_z (r,z,t=0)= \sqrt{1- \frac{1}{\Gamma_j^2}},\\
     & \rho_j(r,z,t=0) = \nu , \\
    & p_j(r,z,t=0) = p_r \mathcal{T}_{\text{ext},0}.
 \end{align}

The external gas is initialized outside the jet as described in sec. \ref{environment}. In order to avoid problems related to numerical noise, we did not treat the interface as a discontinuity, but we smooth all primitive quantities there \citeg{mukh20,abolbrom23}. All the details of the 2D setup can be found in Appendix \ref{2D_details}, including those concerning the grid resolution and domain size, $[0,r_{\max,2D}]\times[z_{\min,2D}, z_{\max,2D}]$ in units of $z_0$, both chosen to ensure a good resolution of the jet in all stages of the simulation. 

The boundary conditions are reflective on the axis ($r = 0$) and outflows at the top and right boundaries. At the bottom boundary we prescribe the same jet-environment profiles, given by Eqs. \ref{eq:smooth} and \ref{eq:smooth_cyl} evaluated at $z = z_0$. In order to minimize spurious reflection effects at the lower boundary just outside the jet injection region in the cold conical cases we use a different setup for the lower boundary as described in the Appendix \ref{2D_details}. This is not necessary in warm jets, since their collimation stage is separated enough from the lower boundary.


We run the 2D simulations until the jet-environment system reaches a steady configuration, with $t_{l,2D}$ that depends on the specific case, and that can be read in Tab. \ref{tab:grid}. The typical value is $3000$ in units of $t_0=z_0/c$. We evaluate that, around $t_{l,2D}$, the positions of the first few recollimation points vary by less than $1-2\%$ over a time interval of $\Delta t = 500$.

\subsection{3D setup \label{3D_setup}}

We use the final steady state obtained in the 2D simulations as the initial conditions for the 3D simulation. The values of all the variables  are projected and interpolated from the 2D cylindrical grid onto a 3D Cartesian grid $[-L_{\max,3D},L_{\max,3D}]\times[-L_{\max,3D},L_{\max,3D}]\times[1,z_{\max,3D}]$. In general, the resolution in the 3D domain has been chosen to ensure that the injection radius is resolved with $30$ points. The domain covers the jet for at least a few recollimation-reflection shocks the grid extension has been increased for the two simulations at highest power, SHC2 and SLWw2, since they recollimate on larger scales. More details on the grid for each case are reported in the Appendix \ref{3D_details}.


The 3D simulations were run until the reaching of a quasi-stationary state. We note that what we refer to as stationary state is not perfectly steady, but it is characterized by an average configuration, with small fluctuations in time. In particular, we determined the reaching of a quasi-stationary configuration when the profiles, along the jet, of averaged quantities show variations smaller than about $5\%$ across $\Delta t\sim50$. The final time $t_{l,3D}$ can vary a lot depending on the case, within a range between $300$ and $640$.

\section{The axisymmetric steady state \label{2D_results}}

\begin{figure}[htbp]
    \centering

        \includegraphics[width=\columnwidth]{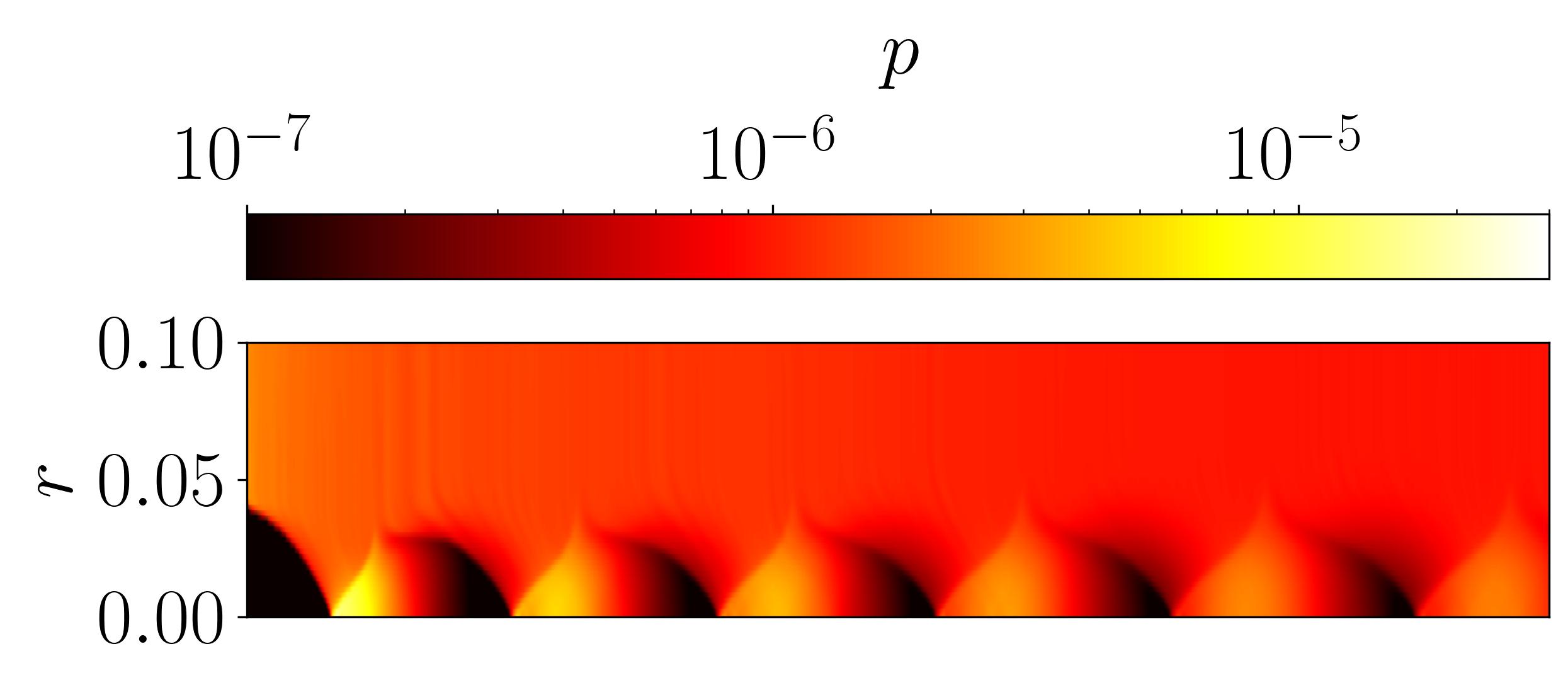} 
        
        \includegraphics[width=\columnwidth]{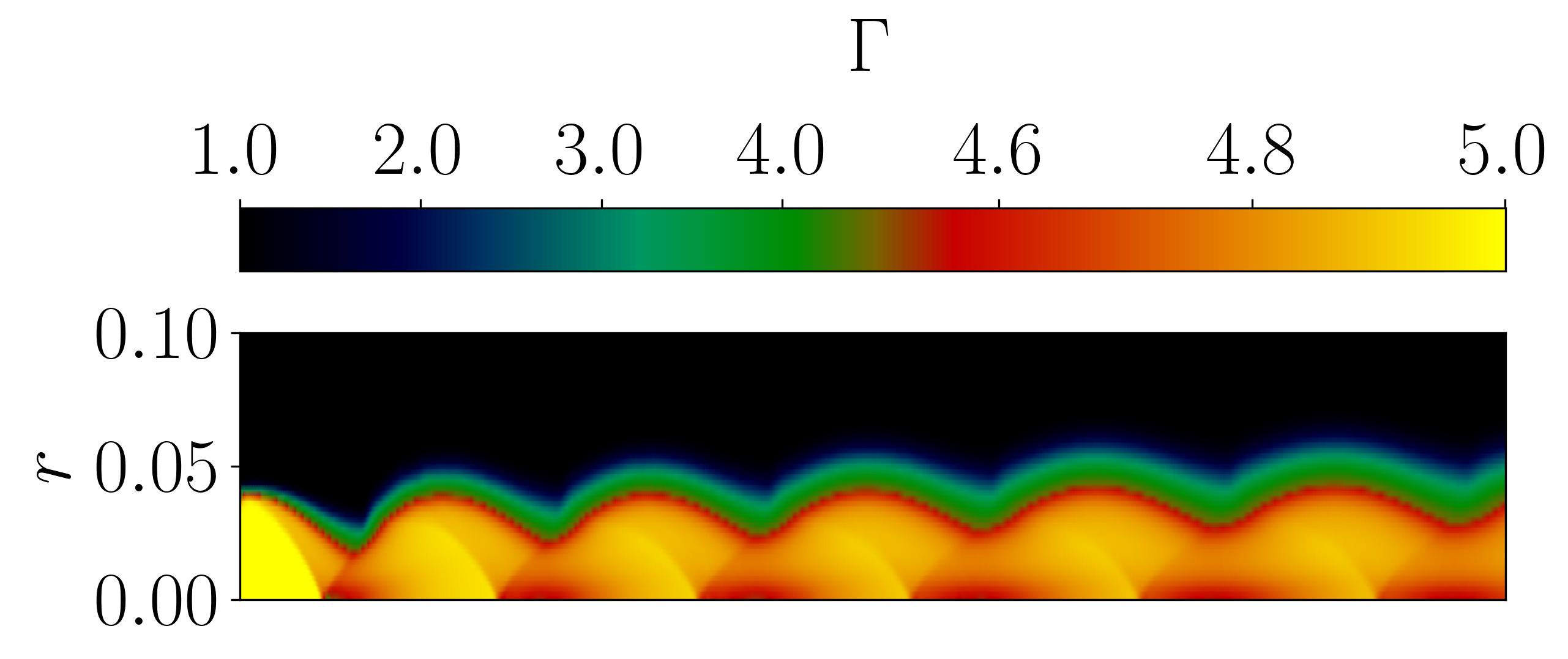}  
        
        \includegraphics[width=\columnwidth]{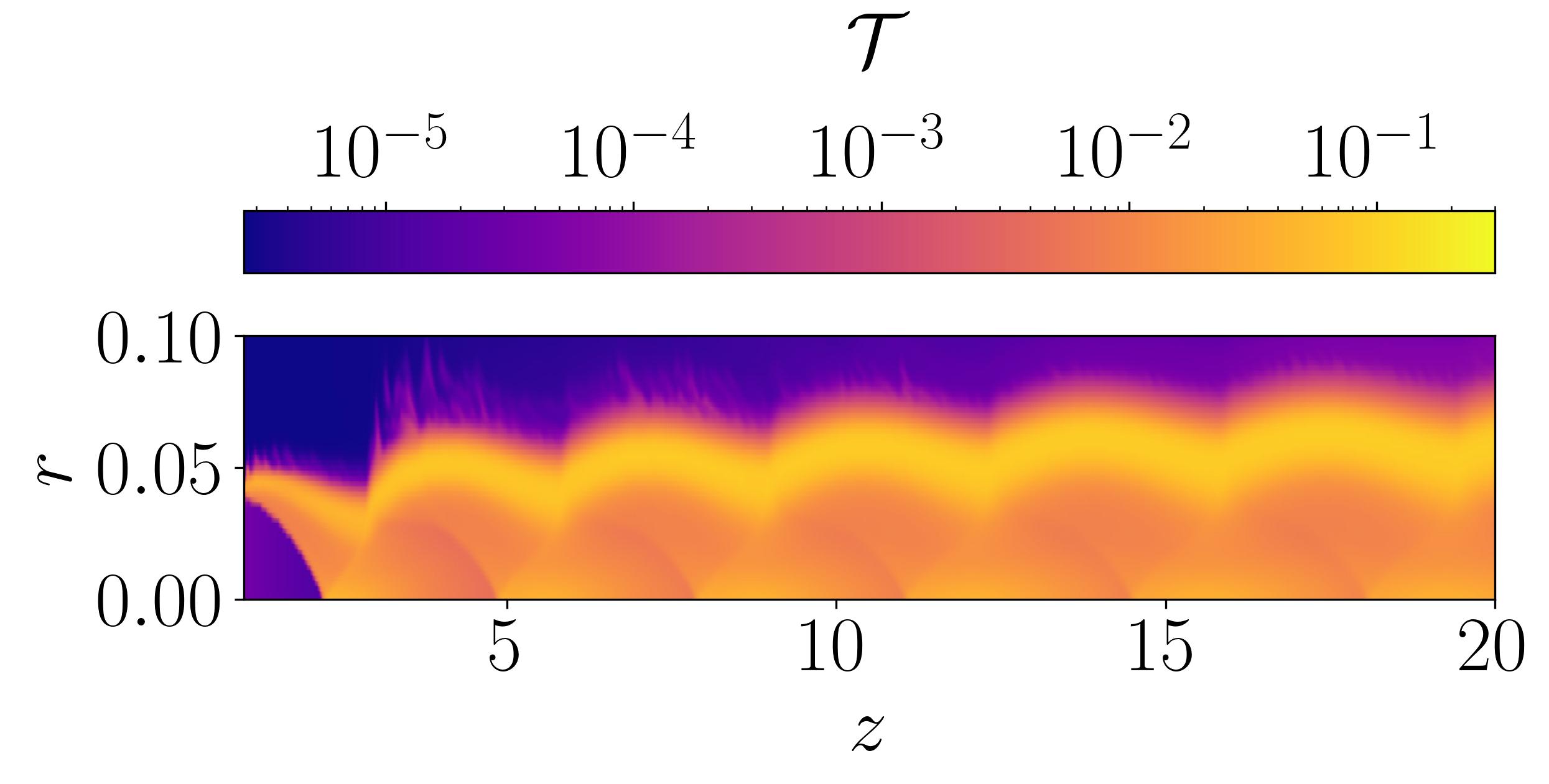}   
     \caption{Pressure (top), Lorentz factor (middle) and temperature (bottom) $(r,z)$ maps at the end of the 2D simulation in case SHC05.}
        \label{fig:2D_maps}
\end{figure}

\begin{figure}[htbp]
    \centering
    \vspace{0.2cm}
    \begin{minipage}{\columnwidth}
        \centering
        \includegraphics[width=\textwidth]{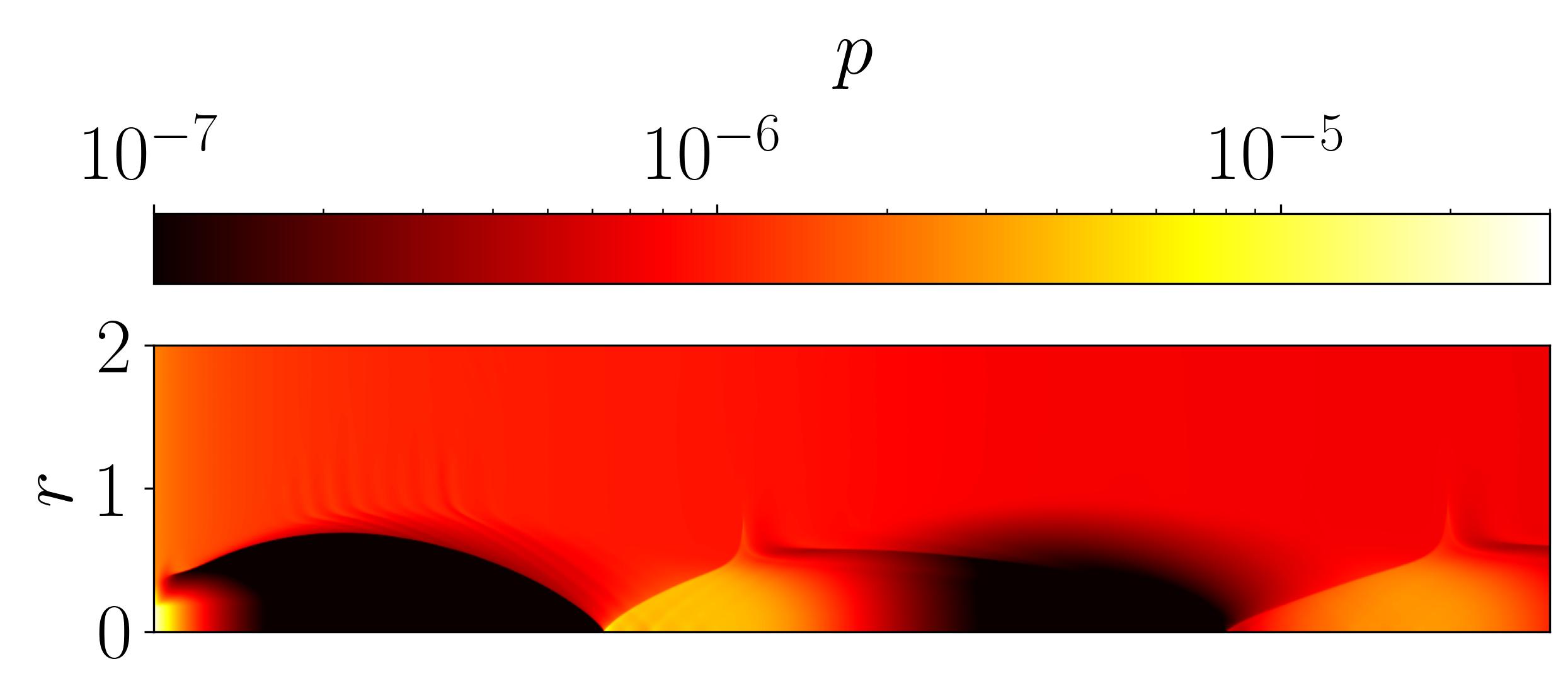}  
    \end{minipage}
    \vspace{0.2cm}
        
    \begin{minipage}{\columnwidth}
        \centering
        \includegraphics[width=\textwidth]{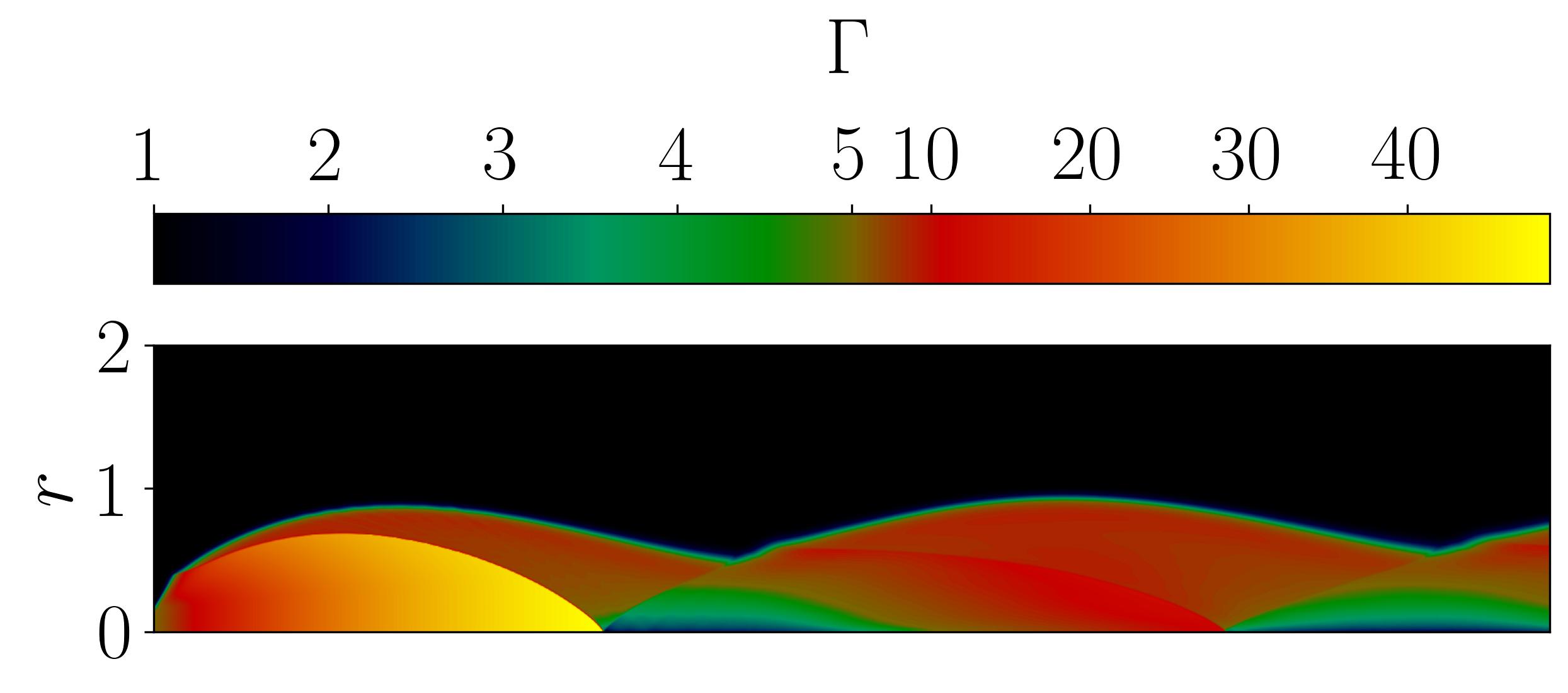}  
    \end{minipage}  
    \vspace{0.2cm}
        
    \begin{minipage}{\columnwidth}
        \centering
        \includegraphics[width=\textwidth]{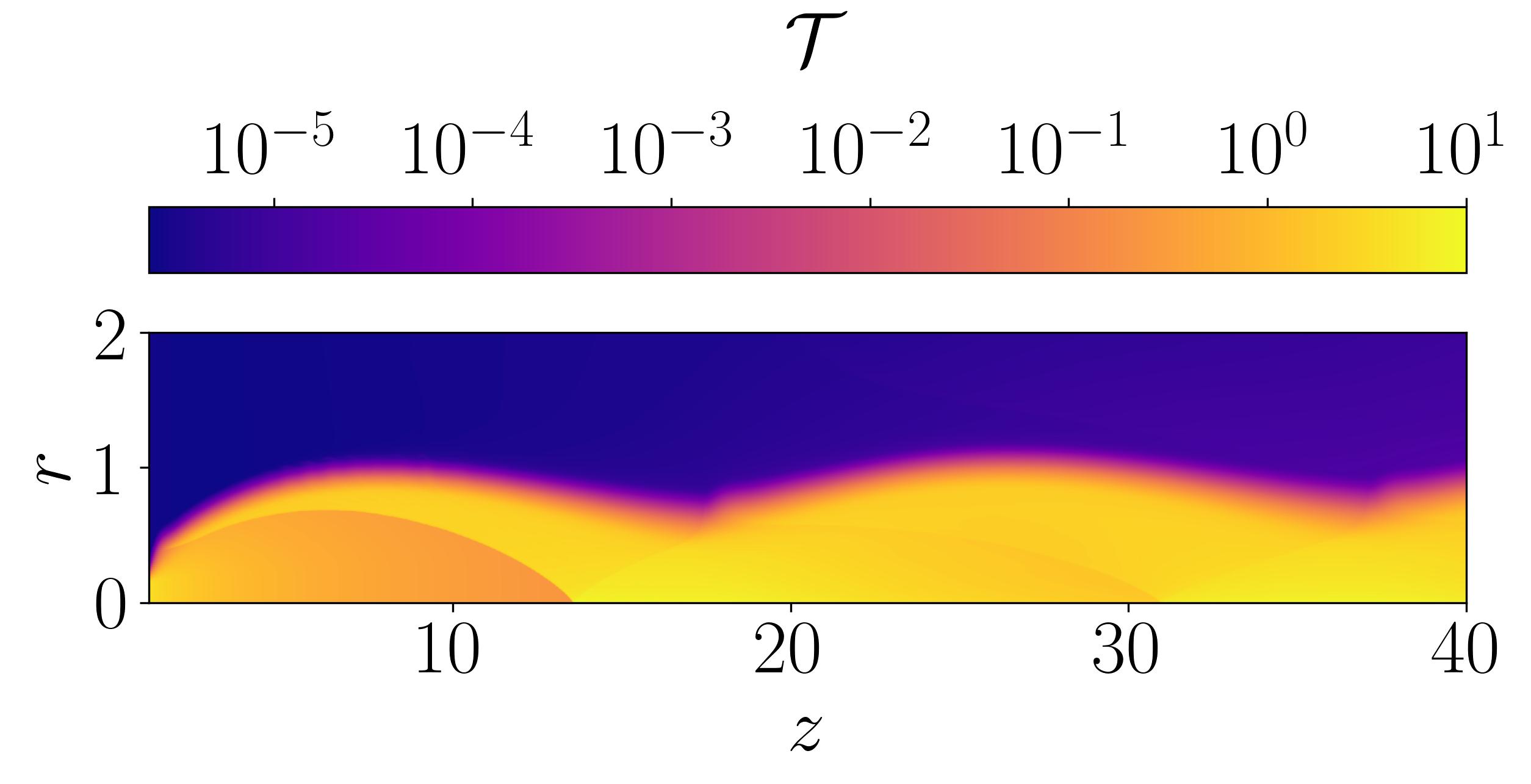}
    \end{minipage}  
    \vspace{0.15cm}
     \caption{$(r,z)$ maps of pressure (top), Lorentz factor(middle), and temperature (bottom) at the end of the 2D simulation in case SLWw2. The scale is different from Fig. \ref{fig:2D_maps}.}
        \label{fig:2D_maps_hot}
\end{figure}
At the end of our 2D simulations we find steady states that are generally in agreement with the predictions of previous axisymmetric simulations and analytical results. We illustrate in Figs. \ref{fig:2D_maps} and \ref{fig:2D_maps_hot} the typical outcome of our 2D conical simulations by showing 2D maps of a few quantities at $t=t_{l,2D}$, and comparing them for two setups that stand at opposite ends in the parameter space we explored, SHC05 and SLWw2: SHC05 is cold, narrow, kinetically dominated and the least powerful one, while SLWw2 is wide, warm, enthalpy dominated, and it is the most powerful jet we simulated.
In the top panels, displaying the pressure, we can observe more easily the shock structure. Downstream of recollimation shocks, the pressure matches that of the external medium. Instead, the pressure reaches its maxima downstream of reflection shocks (on or near the axis), with values that are slightly more than four times those in the external medium at the same distance $z$.
The pressure decreases adiabatically where the jet expands: before the first recollimation shock and after each reflection shock.

The Lorentz factor traces the relativistic jet and a contact discontinuity separates the jet from the external medium. The shocks lead to a decrease of the Lorentz factor that reaches its lowest values where the shocks are strongest: near the axis and, especially in case SHC05, near the contact discontinuity. Indeed, the strength of recollimation and reflection shocks varies with position due to their curved profiles and to the profile of upstream streamlines. The Lorentz factor also increases in general as the jet expands, due to conversion of thermal energy into kinetic energy. This process is important when the fluid temperature is high, like in the region downstream of the reflection shock and before the next recollimation shock. In case SLWw2 this happens also in the first expansion phase after the injection, since the jet in this case is warm. In this first expansion region the jet reaches maximum values of $\Gamma_{\text{SLWw2},\max}\simeq 50$.

The lowest panels of Figs. \ref{fig:2D_maps} and \ref{fig:2D_maps_hot} show that, after the first recollimation shock, the  temperature in the jet becomes much higher than in the environment. Downstream of the following shocks the temperature increases, reaching maximum values of $\mathcal{T}_{\text{SHC05},\max} \simeq 0.03 $ and $\mathcal{T}_{\text{SLWw2},\max} \simeq 11 $ in the two cases, and near the contact discontinuity in case SHC05, where the recollimation shock is strong. In this case, since the jet at the contact discontinuity is in pressure equilibrium with the environment, the high temperature region close to the contact discontinuity corresponds to a region of lower density.
Ripples outside of the SHC05 jet visible in the temperature map (bottom of Fig. \ref{fig:2D_maps}) reflect the presence of small-scale density perturbations, probably due to the KHI.

\subsection{Comparison with the analytical predictions on the recollimation point}

A comparison between Figs. \ref{fig:2D_maps} and \ref{fig:2D_maps_hot} also shows how different is the jet configuration, with more powerful jets generally re-collimating at greater distances and showing a larger expansion. The analytical predictions for the profile of the recollimation shock and for the position of the first recollimation point were derived under the assumption of small opening angles and low jet temperatures, for steady axisymmetric jets \citep{komfalle97}. 
The position of the recollimation point \citep{komfalle97} can be written in terms of the parameters that characterize our setups: the Lorentz factor, the external temperature, the density contrast and the jet opening angle (since it determines the jet nozzle width at fixed $z_0$):
\begin{equation}
    \frac{z_r}{z_0} = \left(\left(\frac{\mu\delta^2\theta^2_j\nu\Gamma_j^2 }{\mathcal{T}_\text{ext}}\right)^{1/2}+1\right)^{1/\delta},
    \label{eq:recpoint_code}
\end{equation}
where $\delta=1-\frac{\eta}{2}$ and $\mu=0.7$.
At higher temperatures, when  $h_j>1$, the assumptions made in the derivation of the above equation is not valid. Nevertheless, we can try to include the effect of jet temperature to a first approximation, by replacing $\nu$ with $\nu h_{j,0}$ (with this substitution we take into account only the effect on the inertia, while neglecting all the other effects). Then we would use 
\begin{equation}
    \frac{z_r}{z_0} = \left(\left(\frac{\mu\delta^2\theta^2_j\nu h_{j,0}\Gamma_j^2 }{\mathcal{T}_\text{ext}}\right)^{1/2}+1\right)^{1/\delta}.
    \label{eq:rec_approx}
\end{equation}


We performed three sets of 2D simulations to compare the outcome of our axisymmetric simulations with the analytical results. All sets use $\nu = 7.6 \times 10^{-6}$, $\eta=0.5$, $\Gamma_j=5$, while we vary the jet opening angle and the jet specific enthalpy. For each set we performed 5 simulations at varying external temperature. 

\begin{figure}[h]
\centering
\includegraphics[width=\columnwidth]{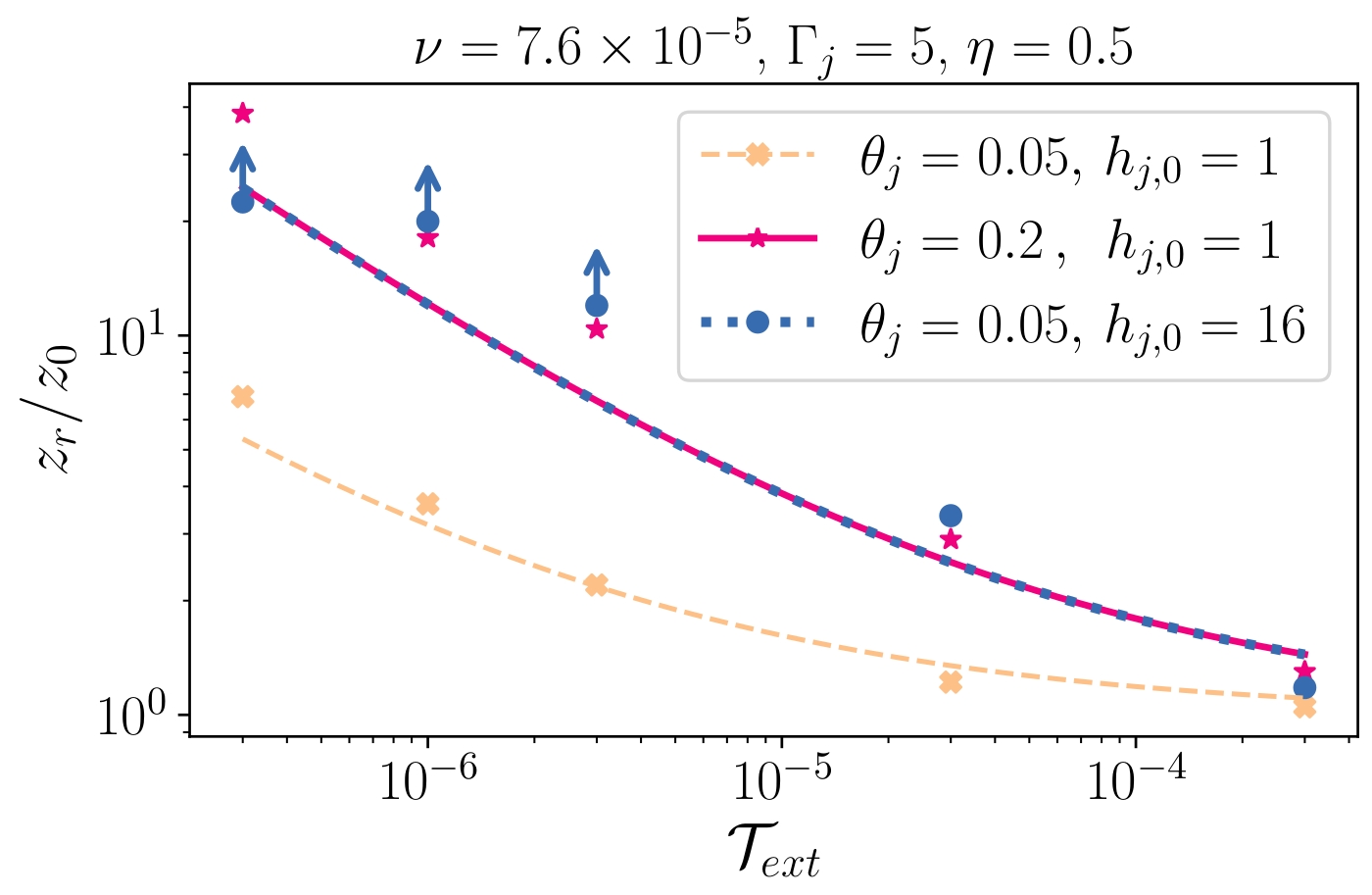}
\caption{Recollimation point as function of the external temperature, comparing analytical (curves) and 2D simulation (points) results. Three sets of cases are shown at varying jet opening angle and specific enthalpy: $\theta_j=0.05$, $h_{j,0}=1$ in yellow, $\theta_j=0.2$, $h_{j,0}=1$ in pink, and $\theta_j=0.05$, $h_{j,0}=16$ in blue.}
\label{fig:recplot}
\end{figure}

Fig. \ref{fig:recplot} plots the position of recollimation point against the external temperature for all sets, as found in 2D simulations (scatter points) and analytically via the modified Eqs. \ref{eq:recpoint_code} and \ref{eq:rec_approx}. The first set is for a cold and narrow jet, with $\theta_j = 0.05$ and $h_{j,0}=1$ (yellow dashed line/crosses). In this case the approximations used for deriving Eqs. \ref{eq:recpoint_code} are better satisfied. The second set differs for the jet opening angle: $\theta_j =0.2$ (pink solid line/stars). The third set simulates a narrow, $\theta_j=0.05$, but hot jet, that has specific enthalpy $h_{j,0} \simeq 16$ (blue dotted line/dots). For this set, three cases with a very high jet-environment pressure ratio do not reach a steady state, but show irregular oscillations of the position of the recollimation point. For these cases we plot the lower position of the recollimation point as a lower limit. We observe that the analytical curves for the second and third set are the same, because $\theta_j^2$ and $h_{j,0}$ contribute to the recollimation point position the same way in Eq. \ref{eq:rec_approx}. 

Eq. \ref{eq:recpoint_code} predicts that, when the external temperature is increased, the recollimation point comes closer to the origin and this behavior is reproduced in our numerical results. The agreement between analytical and numerical results are the best for the cold and narrow jet (case SH05) for which the approximations used in the analytical derivation are better satisfied. In the other two cases the approximations are not satisfied either because the opening angle is wide or the jet is hot and the agreement is worse. For wider jets, the relative difference between the analytical and numerical results are of the order of $10\%$ . 
For warmer jets, as we said, it is difficult to reach a steady state, therefore we can only establish a lower limit of about $10\%$ on the relative difference between numerical results and the analytical expression obtained from Eq. \ref{eq:rec_approx} (dotted blue curve). We observe indeed that our expression, that corrects Eq. \ref{eq:recpoint_code} by including the enthalpy, greatly improves the match between the analytical estimate of the recollimation point and simulations. If we neglected the enthalpy effect, and we used Eq. \ref{eq:recpoint_code} for this case (it would correspond to the dashed yellow curve), the error could be even larger than $80\%$.

\section{The formation of the instability \label{details}}

\begin{figure*}[htbp]
    \centering
    
    \begin{minipage}{0.26\textwidth}
        \centering
        \includegraphics[height=6.5cm]{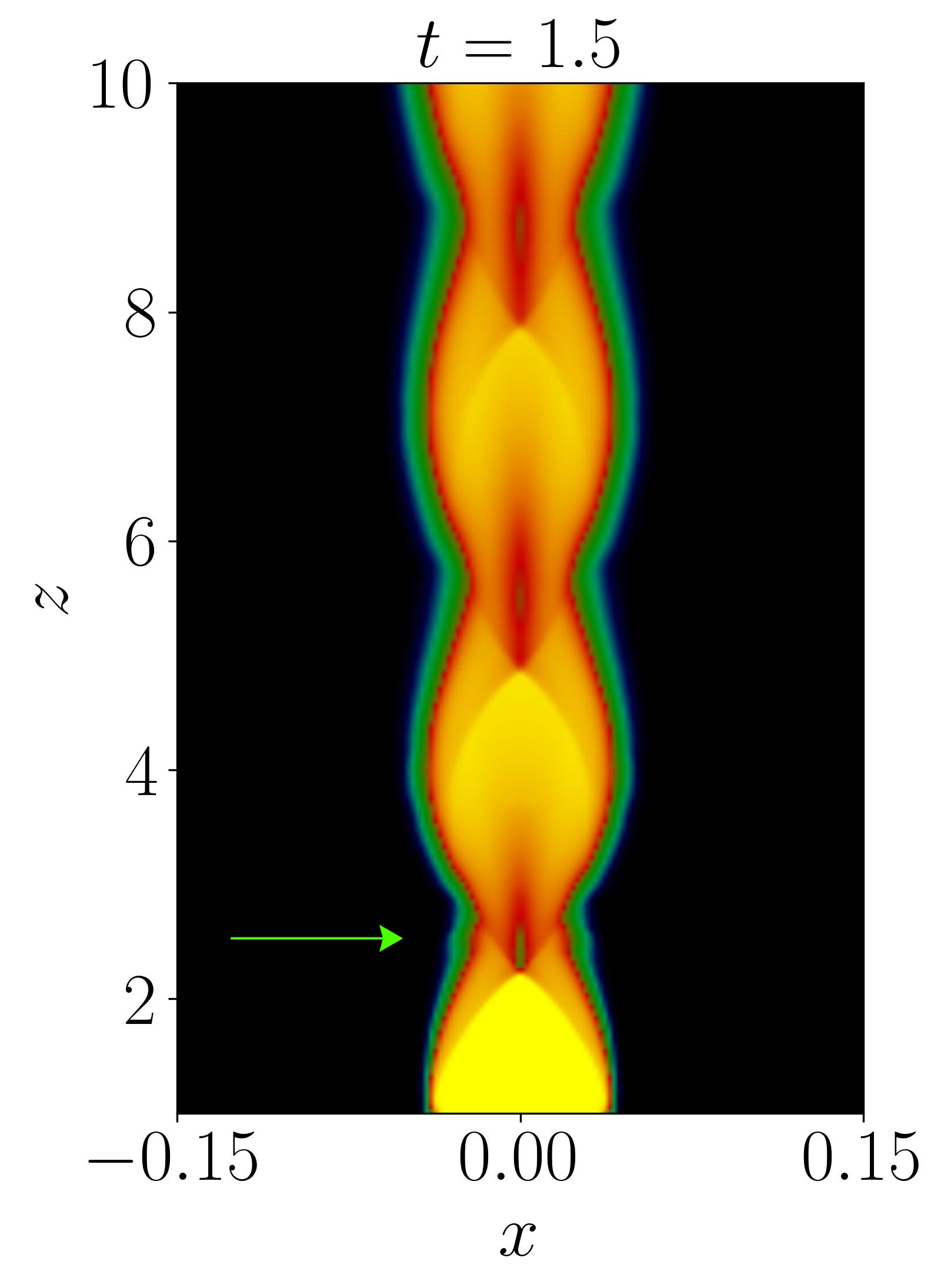}  
    \end{minipage}
    \begin{minipage}{0.23\textwidth}
        \centering
        \includegraphics[height=6.5cm]{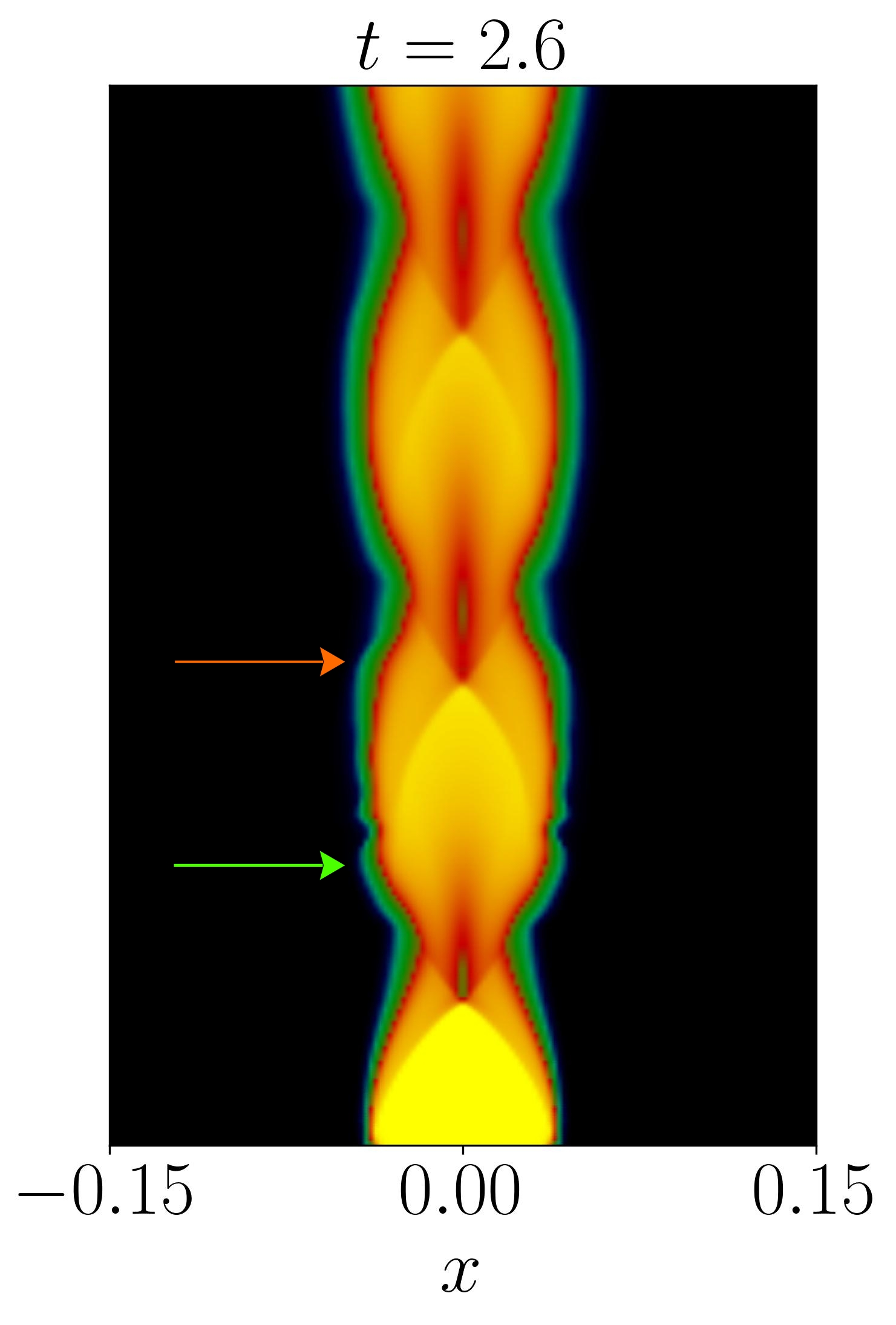}  
    \end{minipage}
    \begin{minipage}{0.23\textwidth}
        \centering
        \includegraphics[height=6.5cm]{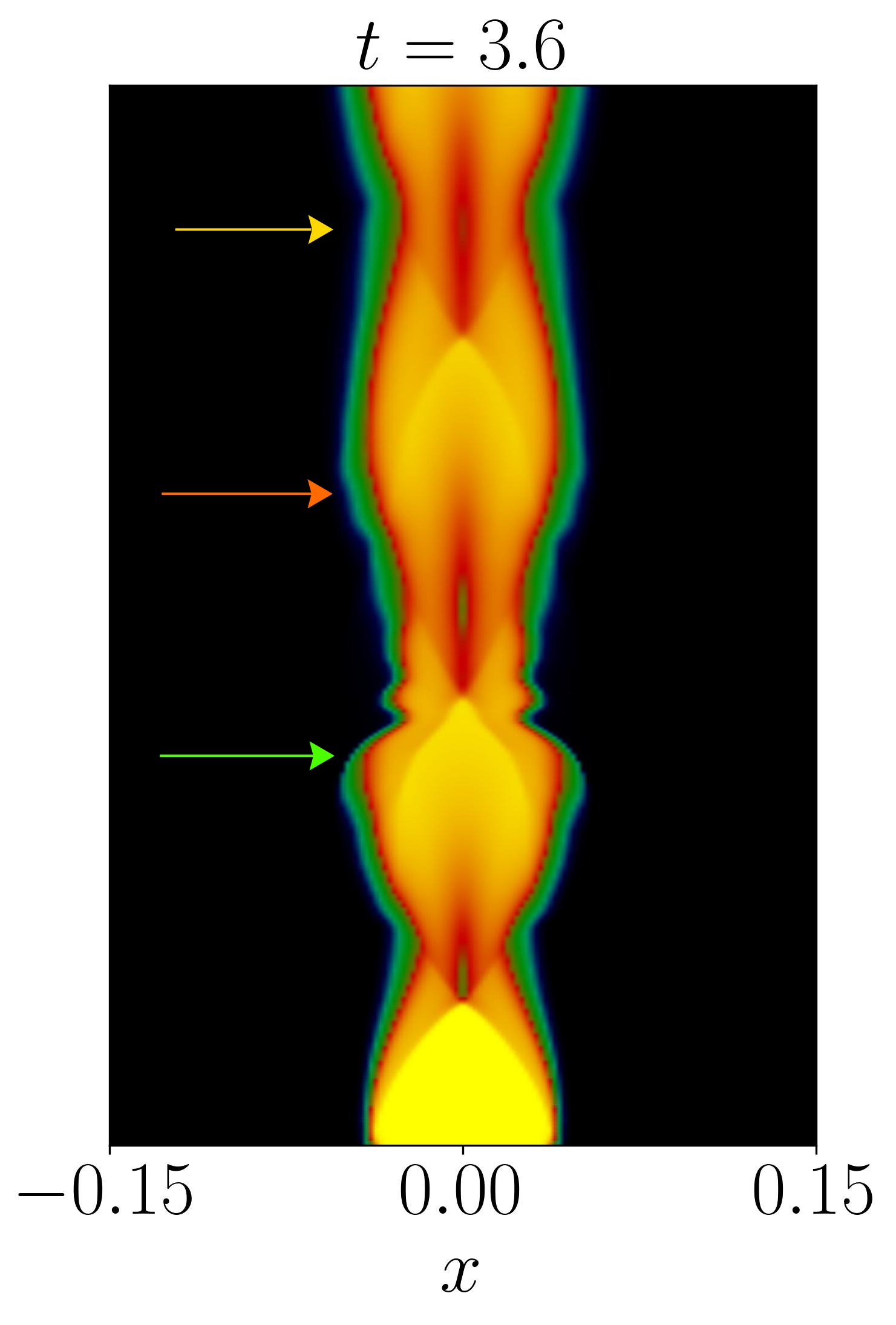}  
    \end{minipage}    
    \begin{minipage}{0.26\textwidth}
        \centering
        \includegraphics[height=6.5cm]{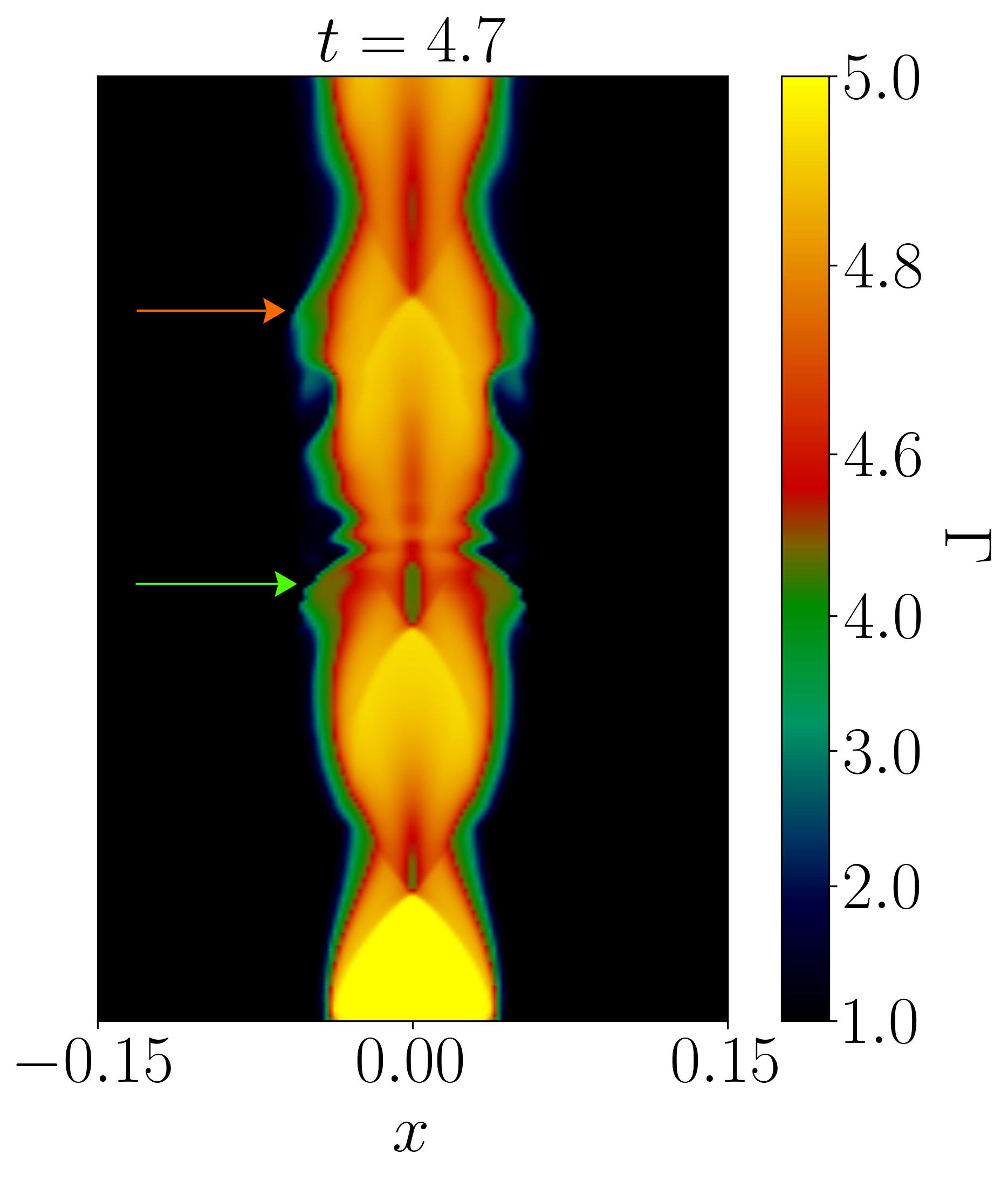}  
    \end{minipage}
    \caption{Sagittal maps of the Lorentz factor $\Gamma(x,y=0,z,t=t^*)$ for case SHC05.}
    \label{fig:SHC05_details}
\end{figure*}

All the previous results considered an axisymmetric jet, but in full 3D simulations the picture is fundamentally different: the jet starts to evolve because non-axisymmetric unstable perturbations may develop. In order to better understand the first phases of the instability evolution, we review in detail the jet  evolution for case  SHC05. 

Fig. \ref{fig:SHC05_details} shows maps of the Lorentz factor in the sagittal plane.
Videos of the initial phases of the evolution can be found at \href{https://drive.google.com/file/d/1Gxcvvi3a_d8w5VKYUEcOIetuEhFt1okH/view?usp=drive_link}{\textbf{SHC05.details.zoomed.mp4}}, showcasing the $4-$velocity component in the jet propagation direction, $u_z = \beta_z \Gamma$, and at \href{https://drive.google.com/file/d/1ciL7gGSmgToaot9w4O_O-3zrIkWGXvkJ/view?usp=drive_link}{\textbf{SHC05.details.mp4}}, that is essentially the same as the previous one, but in a larger grid that lets the reader better appreciate the dynamics at larger $z$. 

\begin{figure}[htbp]
    \centering
    
    \begin{minipage}{\columnwidth}
        \centering
        \includegraphics[width=\textwidth]{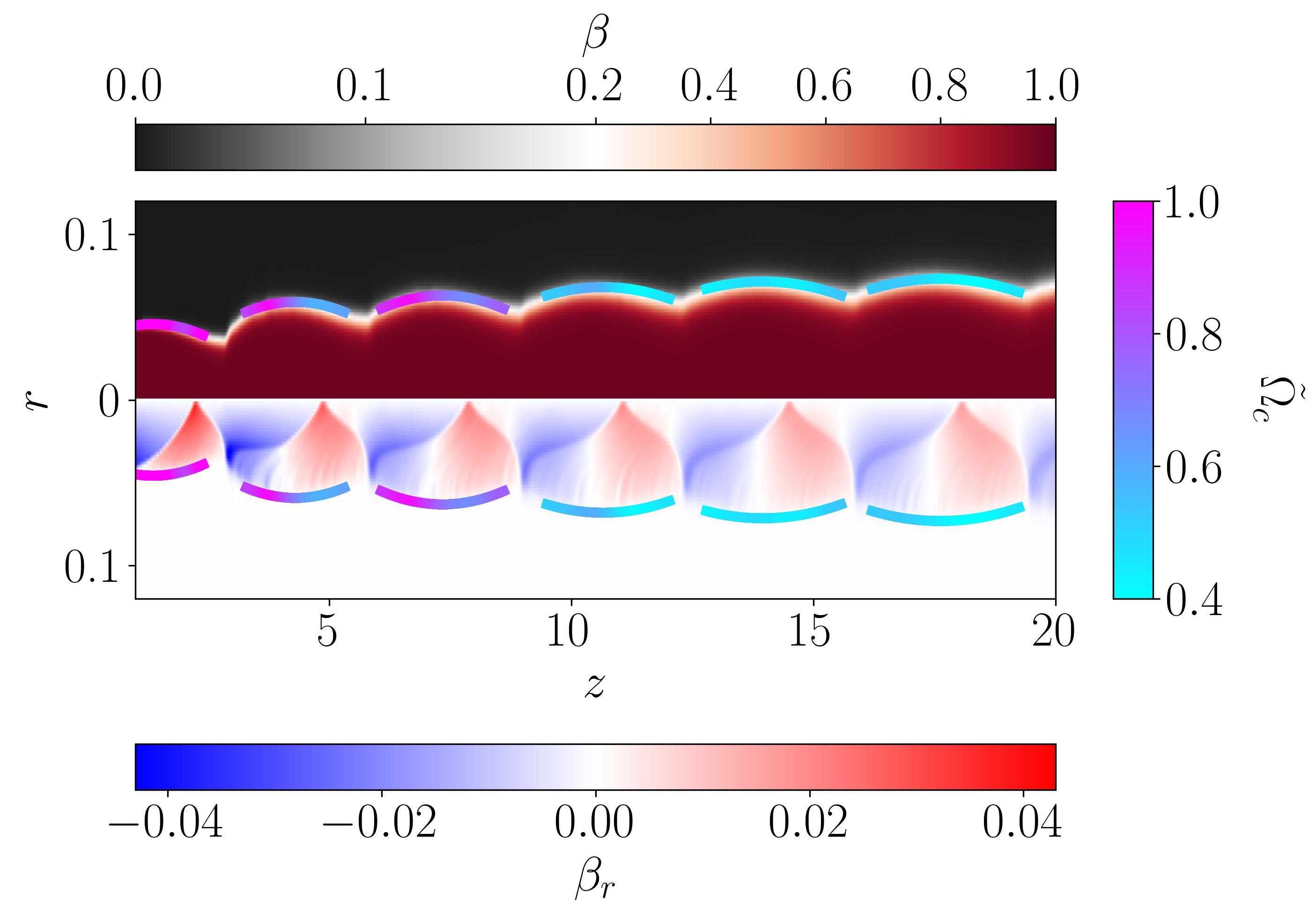}  
    \end{minipage}
     \caption{The spatial component of the velocity, $\beta= \sqrt{\beta_r^2+\beta_z^2}$ (top), and the radial component $\beta_r$ (bottom) at the end of the SHC05 2D simulation, and start of the 3D one. There are superimposed the values of angular velocity at the discontinuity, normalized to the maximum: $\tilde{\Omega}_{\text{c}}$.}
        \label{fig:SHC05_cfi1}
\end{figure}

The main instability that can develop in the steady configuration discussed in Section \ref{2D_results} is the CFI \citep{gourgkom18}, whose onset conditions are met  downstream of each recollimation shock, because of the curvature of the stream lines. Indeed, the perturbations indicated in Fig. \ref{fig:SHC05_details}, by green, orange and yellow arrows, develop downstream of the first, second and third recollimation shock respectively. 
The growth rate for the relativistic CFI has never been derived, however we can argue that the strength of the instability can be related to the magnitude of the centrifugal force and therefore to the angular velocity along the curved streamlines at the radial position  of the Lorentz factor discontinuity. Figure \ref{fig:SHC05_cfi1} shows the $r-z$ maps of the velocity magnitude $\beta$ (top) and of its radial component $\beta_r$ (bottom) in the axisymmetric steady state. Superimposed on the maps, we plotted partial circles (distorted in the pictures due to the aspect ratio) that fit the Lorentz factor discontinuity profile. We computed the radius of curvature $r_c$ of these circles and the flow velocity $v_c$ tangent to each circle, in order to estimate  $\Omega_c = v_c/r_c$, i.e. the angular velocity of the flow, whose magnitude is  displayed through the color of the curves, normalized to its maximum: $\tilde{\Omega}_c = \Omega_c/\max(\Omega_c)$. The angular velocity shows a clear decrease along the jet and, at the first recollimation shock,  is larger by more than a factor $2$ with respect to the last recollimation shock. Thus, we expect the instability to be stronger at the first shock and the perturbation indicated by the green arrow to grow faster than those indicated by the orange and yellow arrows.

\begin{figure*}[htbp]
\centering
    \includegraphics[width = \textwidth]{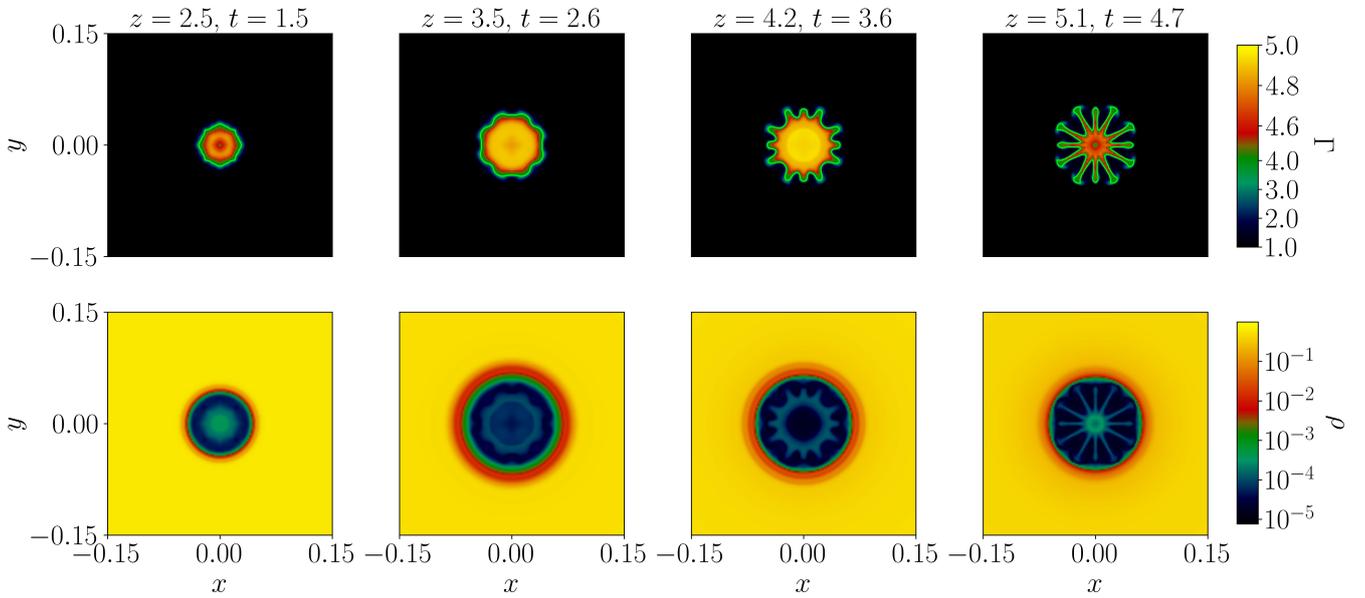}
    \caption{Transverse maps of the Lorentz factor, $\Gamma(x,y,z=z^*,t=t^*)$ (first row), and of the density $\rho(x,y,z=z^*,t=t^*)$ (second row), at different output times $t^*$, for case SHC05. The green contour on Lorentz factor maps is taken at $\Gamma(x,y,z=z^*,t=t^*)\simeq 3.5$.}
    \label{fig:SHC05_details_top}
\end{figure*}

We therefore follow in more detail the evolution of the first perturbation, as it is advected along the jet. Figure  \ref{fig:SHC05_details_top} displays transverse maps at the positions $z^*$ indicated by green arrows in  Fig. \ref{fig:SHC05_details}, at times $t=t^*$. The first row shows the  Lorentz factor, with a contour at $\Gamma=3.5$, while the second row represents the density. We can observe that CFI perturbations grow primarily in the $x-y$ plane, in contrast with the KHI modes that were formed at the contact discontinuity in 2D simulations. 
As the perturbation is advected, in addition to being amplified by the action of the CFI, it also experiences the action of the RMI as it crosses reflection shocks. This is shown by the two leftmost panels in Fig. \ref{fig:SHC05_details_top}. The panel at $z=2.4, t=1.5$ shows the perturbation before crossing the reflection shock, while the one at $z=3.5, t=2.6$ is after the crossing.  A comparison of the two panels shows that perturbations in the second are reflected with respect to the first. This is particularly visible for those along the $x$ and $y$ axes, and along $x=\pm y$ lines, that are inward oriented, while they were outward oriented before the reflection shock.


This reflection of perturbation  is due to the action of  the RMI \citeg{zhou_rmi}. In fact, the direction of the velocity of perturbations, $\partial \xi/\partial t$, depends on the sign of the Atwood number $\mathcal{A}$ with respect to the direction of the acceleration at the discontinuity, defined through $\Delta v$, more precisely
\begin{equation}
    \frac{\partial \xi}{\partial t} =\xi_{0} k \Delta v  \mathcal{A},
    \label{eq:rmi}
\end{equation}
where $\xi_{0}$ represents the amplitude of the displacement of the pre-existing perturbations and $k$ is the wave-number. Through the transverse maps of the density in the second row of Fig. \ref{fig:SHC05_details_top}, we can observe better the position of the growing perturbations that coincides with the Lorentz factor decrease, between the inner dense fast spine (denoted with $j,i)$ and the surrounding warm and slower region (denoted with $j,e$). For the jet, the standing reflection shock has the effect of decelerating the fluid and we expect $\Delta v<0 $. The classical Atwood number \citeg{chiuderivelli} is
\begin{equation}
    \mathcal{A} = \frac{\rho_{j,i}-\rho_{j,e}}{\rho_{j,i}+\rho_{j,e}}.
\end{equation}
There is no general consensus regarding the relativistic one; \cite{rti_jiang24} found
\begin{equation}
     \mathcal{A} = \frac{(\rho h)_{j,i} - (\rho h)_{j,e}}{(\rho h)_{j,i}+(\rho h)_{j,e}}.
\end{equation}
Both values of the Atwood number are positive in the SHC05 jet,  so 
the velocity of RMI perturbations should always be in the opposite direction relative to $\xi_{0}$, thus leading to a reflection of the perturbations. We further observe that Eq. \ref{eq:rmi} predicts that the RMI growth rate scales linearly with the wave-number. This is consistent with the fact that the dominant mode, $m_d$, progressively becomes larger: from $m_d=4-8$ at $t=1.5-2.5$, to $m_d=12$ at $t=3.6$.

The two rightmost panels in Fig. \ref{fig:SHC05_details_top} at $t=3.6$ and $t=4.7$ show the further evolution of the first mode, growing interpenetrating spike- and mushroom-like structures in the transverse plane. At the same times, we can observe in the sagittal plane, in Fig. \ref{fig:SHC05_details} and in video \href{https://drive.google.com/file/d/1ciL7gGSmgToaot9w4O_O-3zrIkWGXvkJ/view?usp=drive_link}{\textbf{SHC05.details.mp4}}, that the CFI-RMI perturbations deform as they get advected downstream, resulting in structures that are reminiscent of the KHI.

One by one, as can be best appreciated in video \href{https://drive.google.com/file/d/1ciL7gGSmgToaot9w4O_O-3zrIkWGXvkJ/view?usp=drive_link}{\textbf{SHC05.details.mp4}}, all the post-recollimation shock regions (orange and yellow arrows) become unstable and grow. In Fig. \ref{fig:SHC05_details}, we can follow in particular the growth of the perturbation indicated by the orange arrow, while that indicated by the yellow arrow exits the domain. 

\begin{figure*}[htbp]
    \centering
    
    \begin{minipage}{0.26\textwidth}
        \centering
        \includegraphics[height=6.5cm]{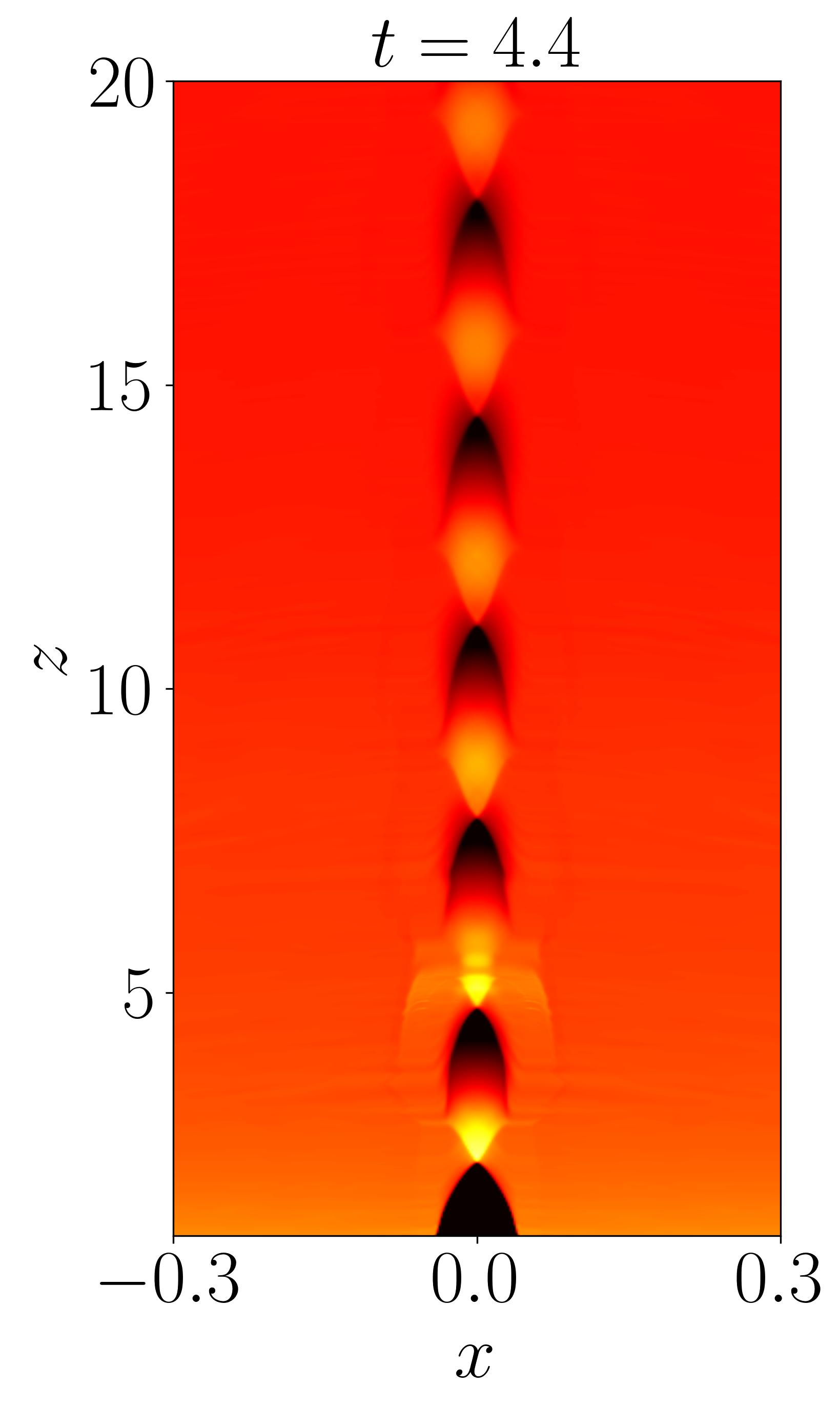}  
    \end{minipage}
    \begin{minipage}{0.22\textwidth}
        \centering
        \includegraphics[height=6.5cm]{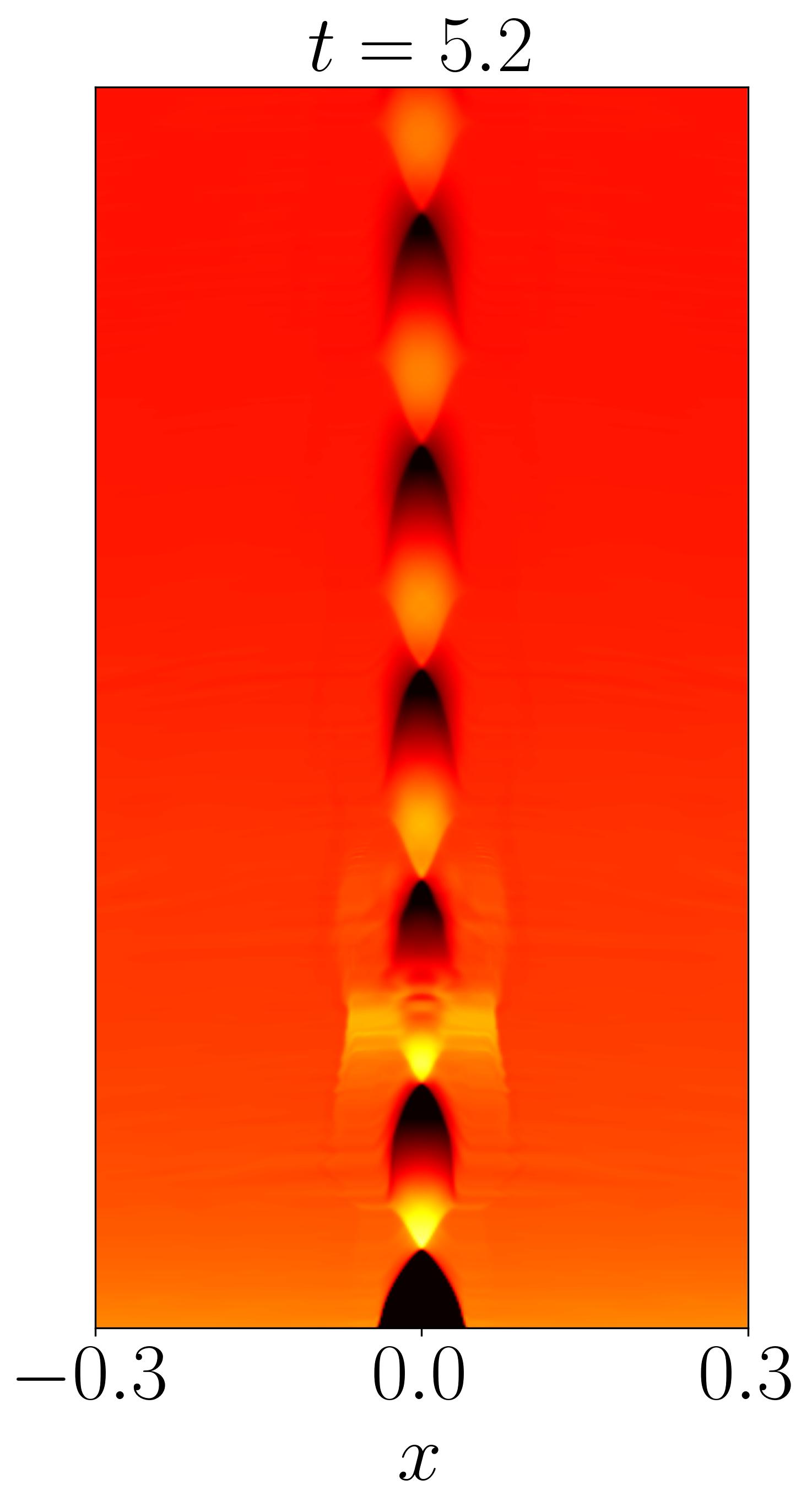}  
    \end{minipage}   
    \begin{minipage}{0.22\textwidth}
        \centering
        \includegraphics[height=6.5cm]{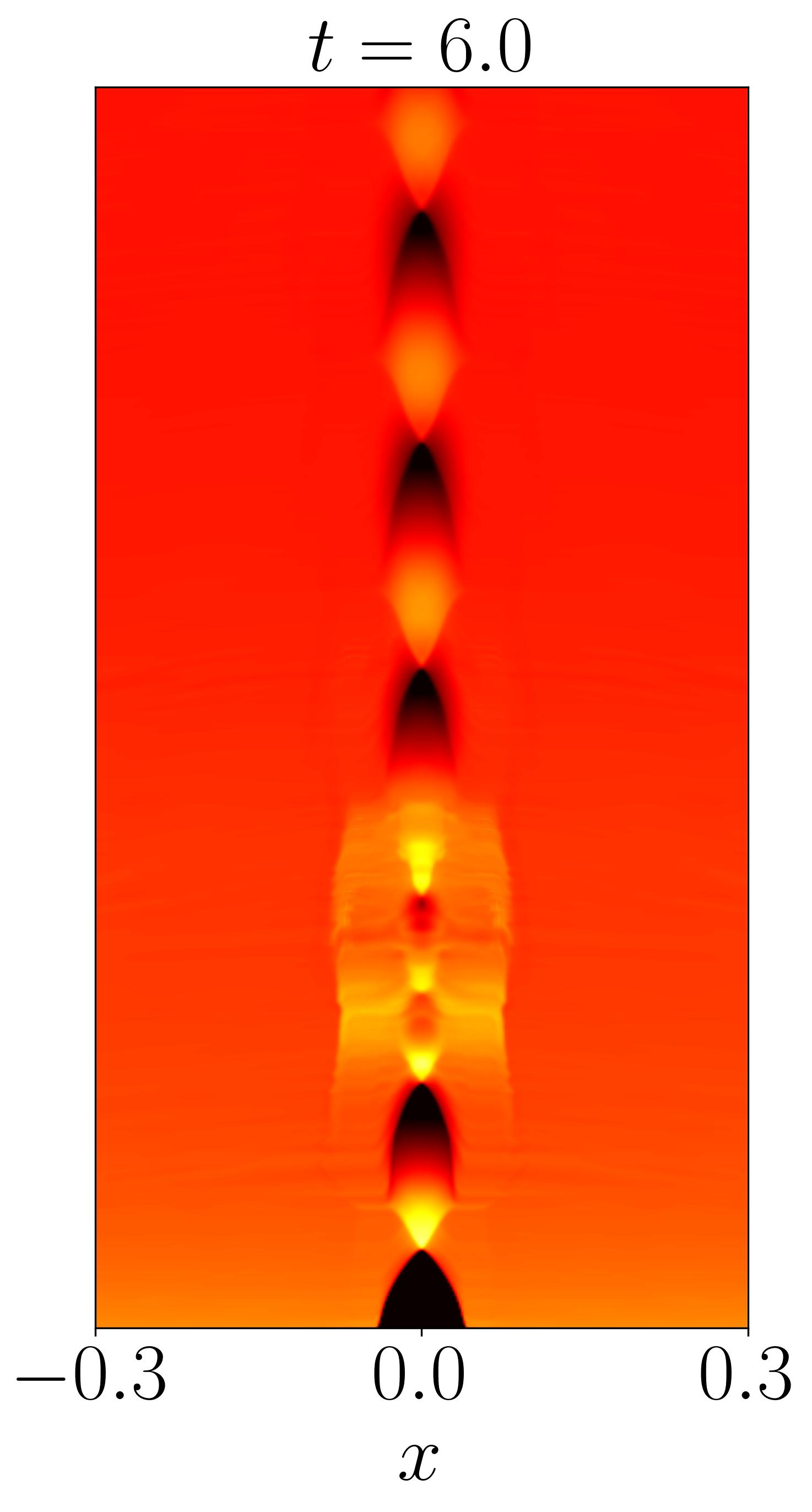}  
    \end{minipage}
    \begin{minipage}{0.28\textwidth}
        \centering
        \includegraphics[height=6.5cm]{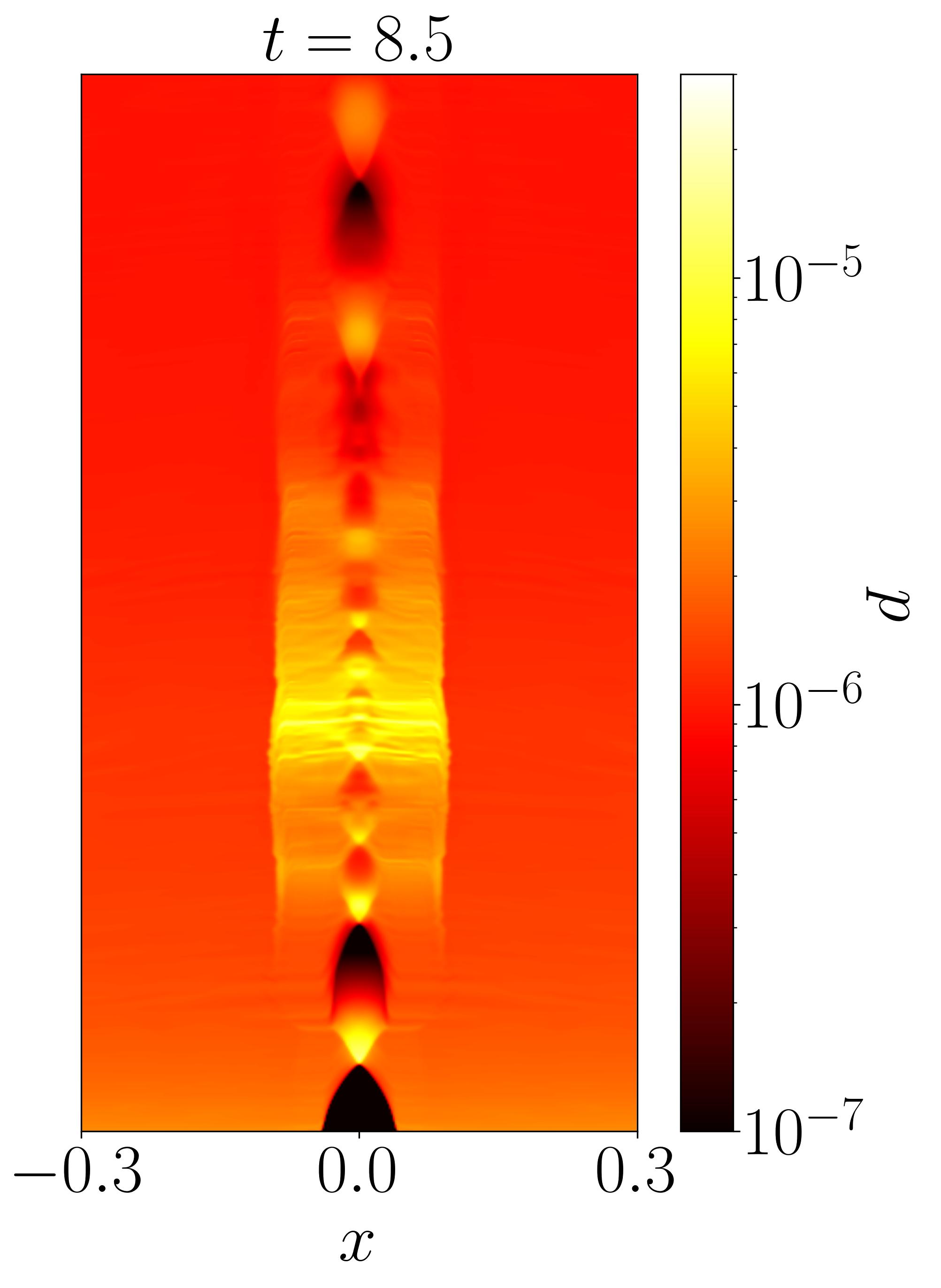}  
    \end{minipage}
    \caption{Sagittal pressure maps $p (x,y=0,z,t=t^*)$ for case SHC05.}
    \label{fig:SHC05_details_prs}
\end{figure*}

The growth of the first perturbation eventually leads to mixing and entrainment of the external material and therefore to a flow deceleration. The interaction of this slower region with the incoming faster jet causes the formation of shocks that heat the region surrounding the spine that is more subject to entrainment. This evolution, occurring at later times with respect to those shown in the previous figures, can be observed in the pressure maps shown in Fig. \ref{fig:SHC05_details_prs}, and in the video linked at \href{https://drive.google.com/file/d/1zCRQ15c_1x2B-vNdrBjrtN5VZ54r_8B2/view?usp=drive_link}{\textbf{SHC05.details.pressure.mp4}}. In these figures and video, we see the formation of an over-pressured region that surrounds and compresses the faster spine, where the recollimation shocks become much closer to one another. Eventually, as we will discuss in the next section, the recollimation shock chain is disrupted and we see the formation of a hot and turbulent region.

\section{Instability feedback on the jet \label{bubble}}
Eventually the instability evolution leads to the formation of a hot and turbulent region in which strong dissipation occurs and  to a  change in  the environment conditions around the jet. These effects may then modify the recollimation shock structure. Below we discuss in more detail the formation and growth of this region and of the effect that it has on the recollimation shocks.

\subsection{Case SHC05}
\begin{figure*}[htbp]
    \centering    
    \begin{minipage}{0.26\textwidth}
        \centering
        \includegraphics[height = 6.5cm]{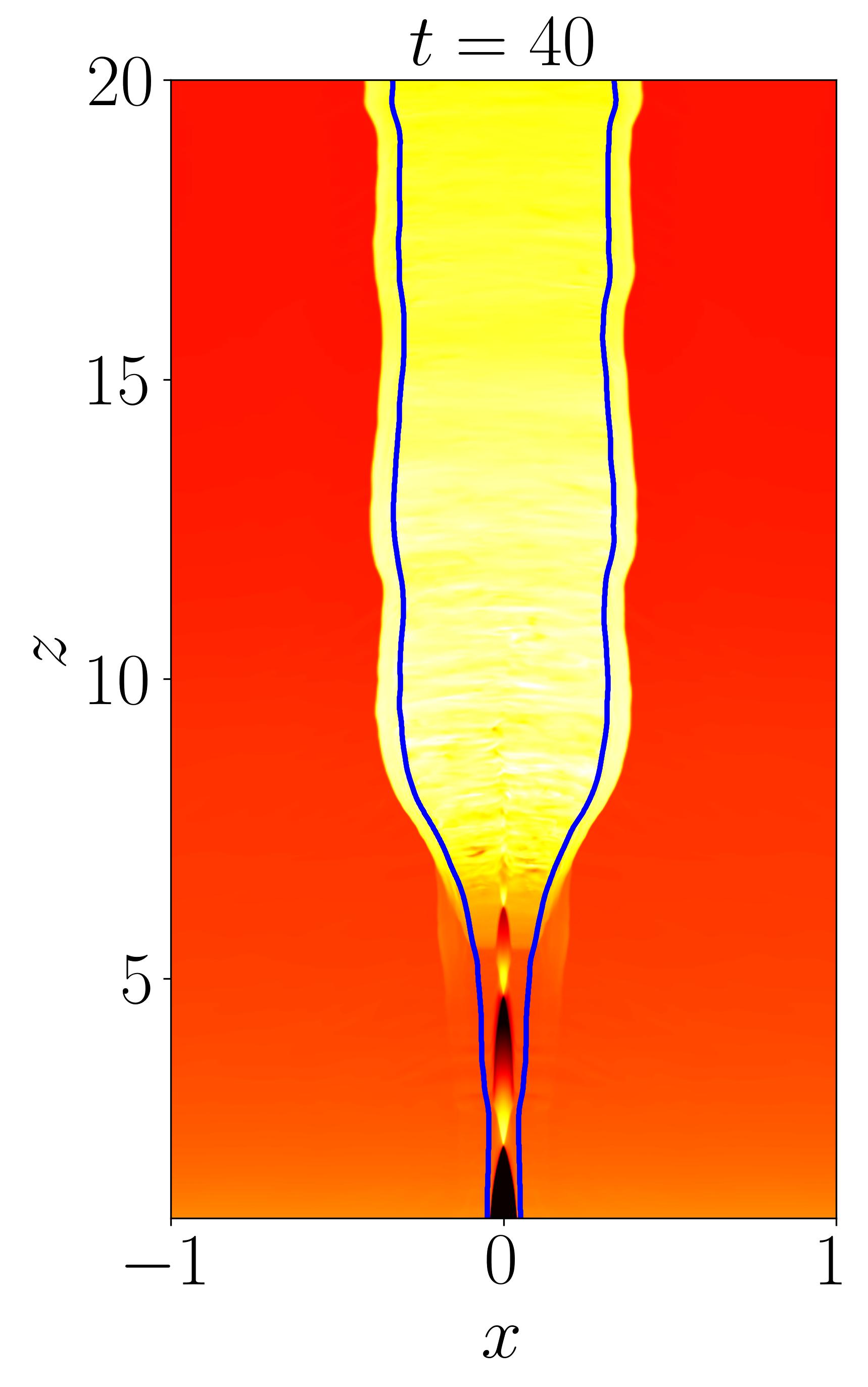}  
    \end{minipage}
    \begin{minipage}{0.22\textwidth}
        \centering
        \includegraphics[height = 6.5cm]{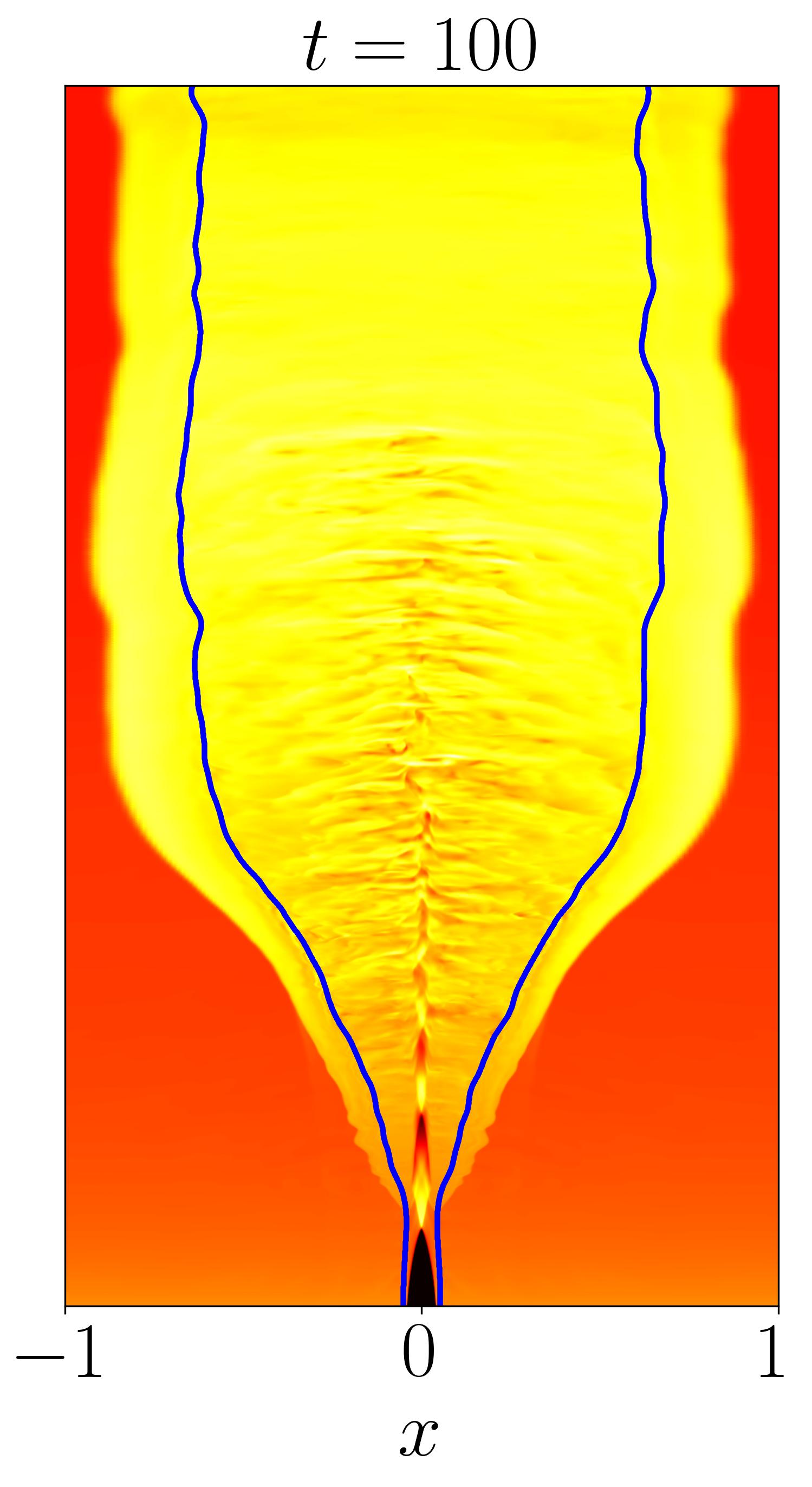}  
    \end{minipage}   
    \begin{minipage}{0.22\textwidth}
        \centering
        \includegraphics[height = 6.5cm]{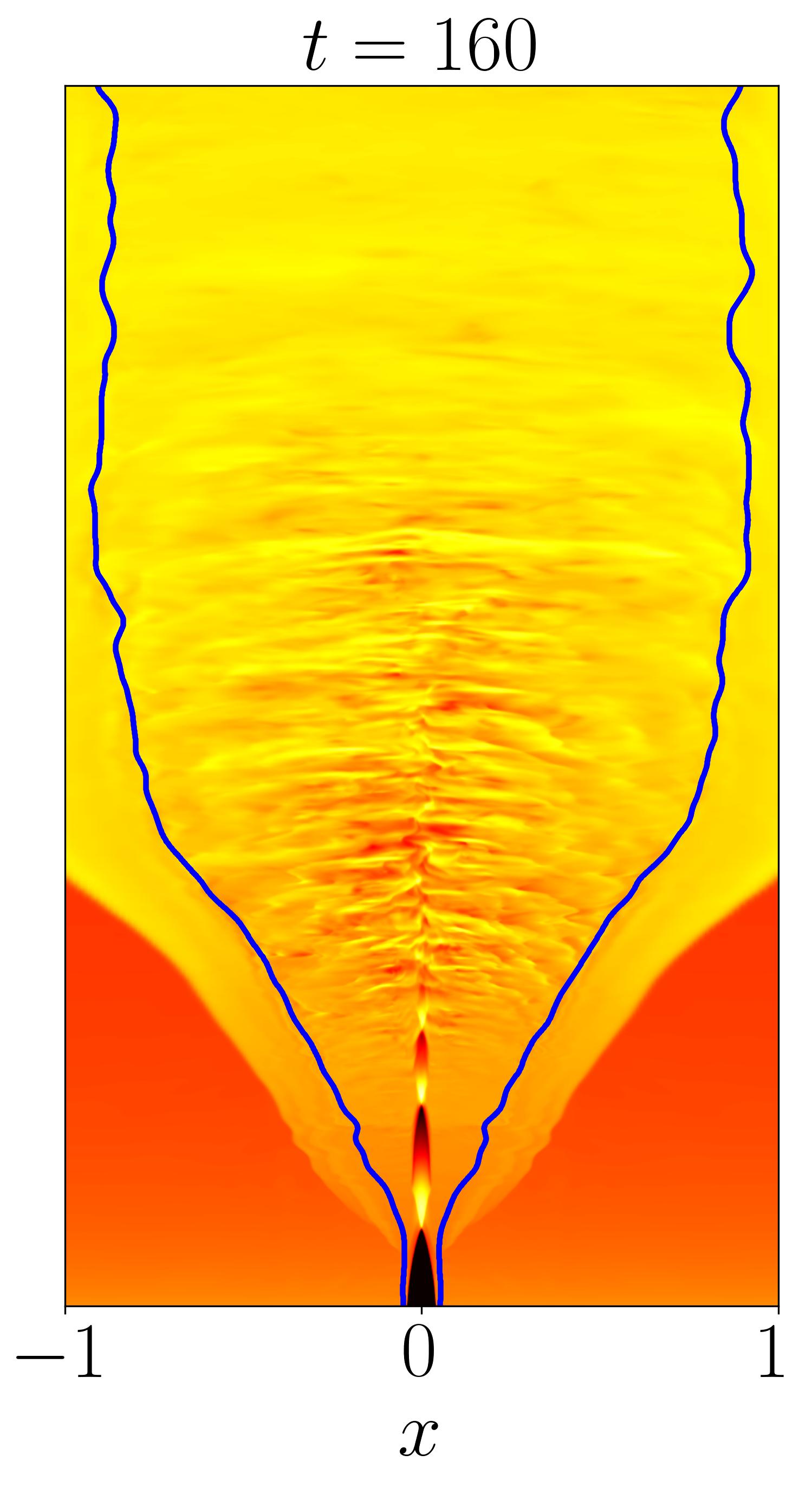}  
    \end{minipage}
    \begin{minipage}{0.28\textwidth}
        \centering
        \includegraphics[height = 6.5cm]{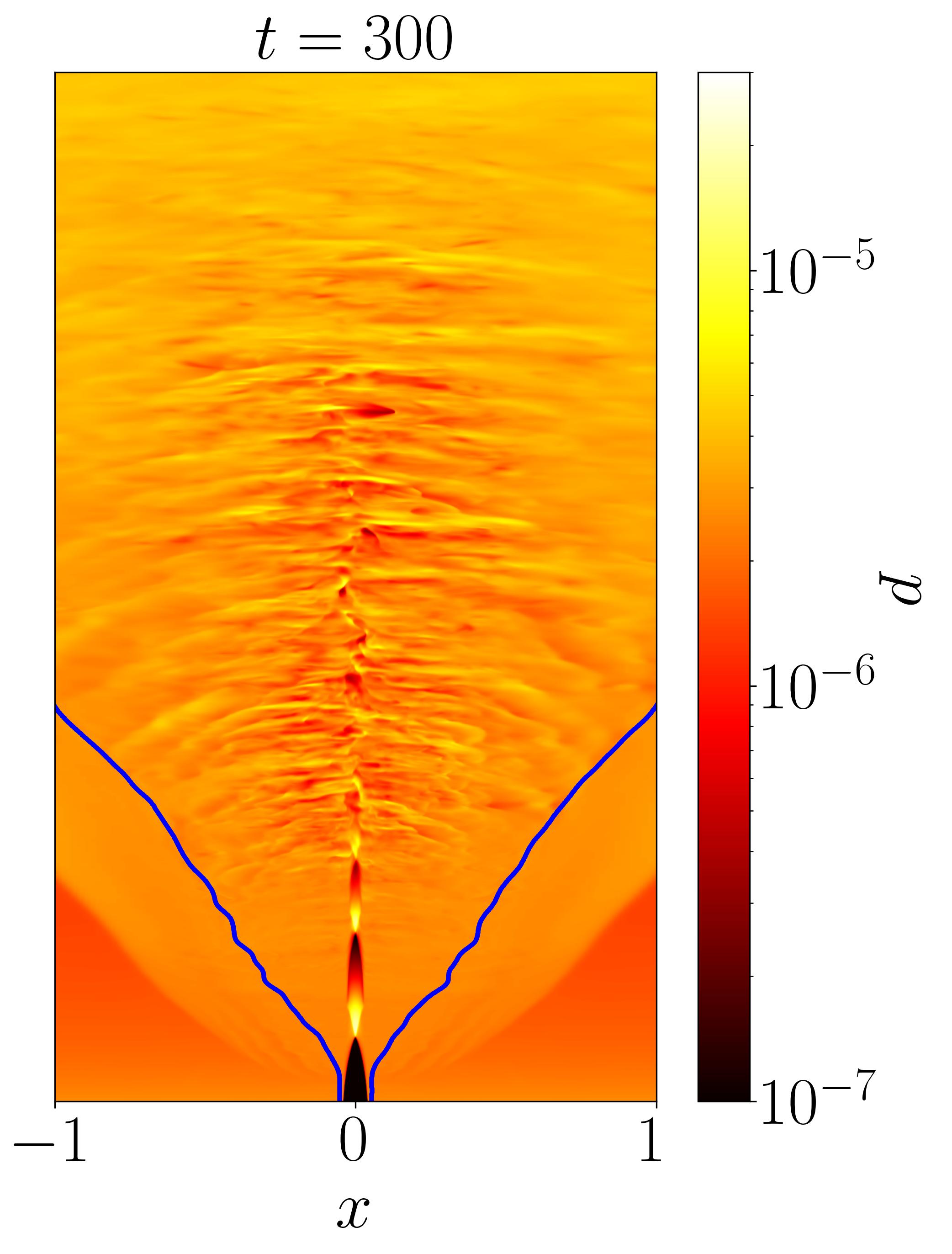}  
    \end{minipage}

    \caption{Sagittal pressure maps $p (x,y=0,z,t=t^*)$, at different output times $t^*$, for case SHC05. Blue curves are tracing the temperature transition at $\mathcal{T}(x,y=0,z,t=t^*) = 3\times 10^{-4}$.  Note that the extension of the $x$ axis is larger than in the previous figure. }
    \label{fig:SHC05_long_prs}
\end{figure*}

\begin{figure*}[htbp]
    \centering
    \begin{minipage}{0.26\textwidth}
        \centering
        \includegraphics[height = 6.5cm]{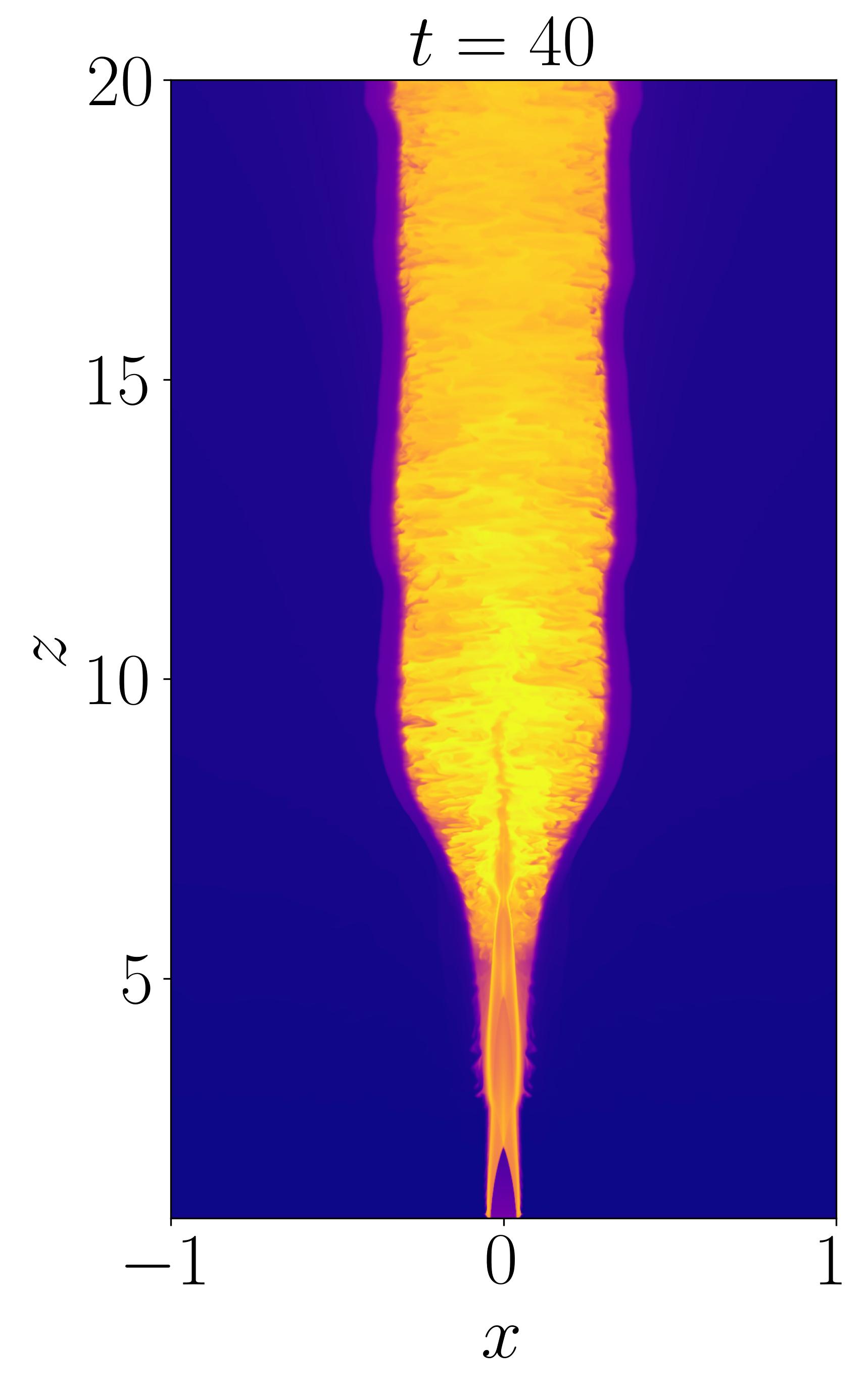}  
    \end{minipage}
    \begin{minipage}{0.22\textwidth}
        \centering
        \includegraphics[height = 6.5cm]{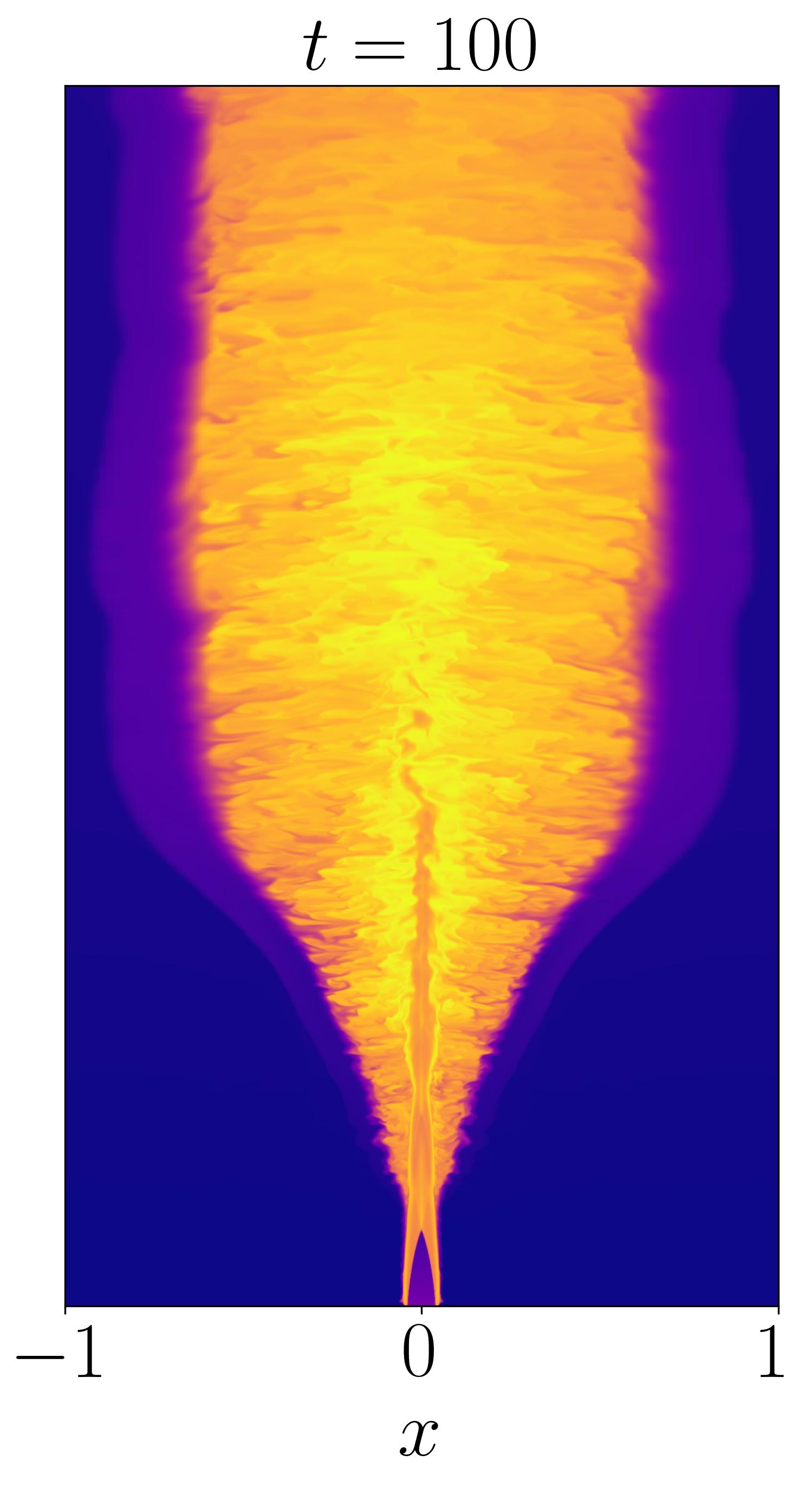}  
    \end{minipage}   
    \begin{minipage}{0.22\textwidth}
        \centering
        \includegraphics[height = 6.5cm]{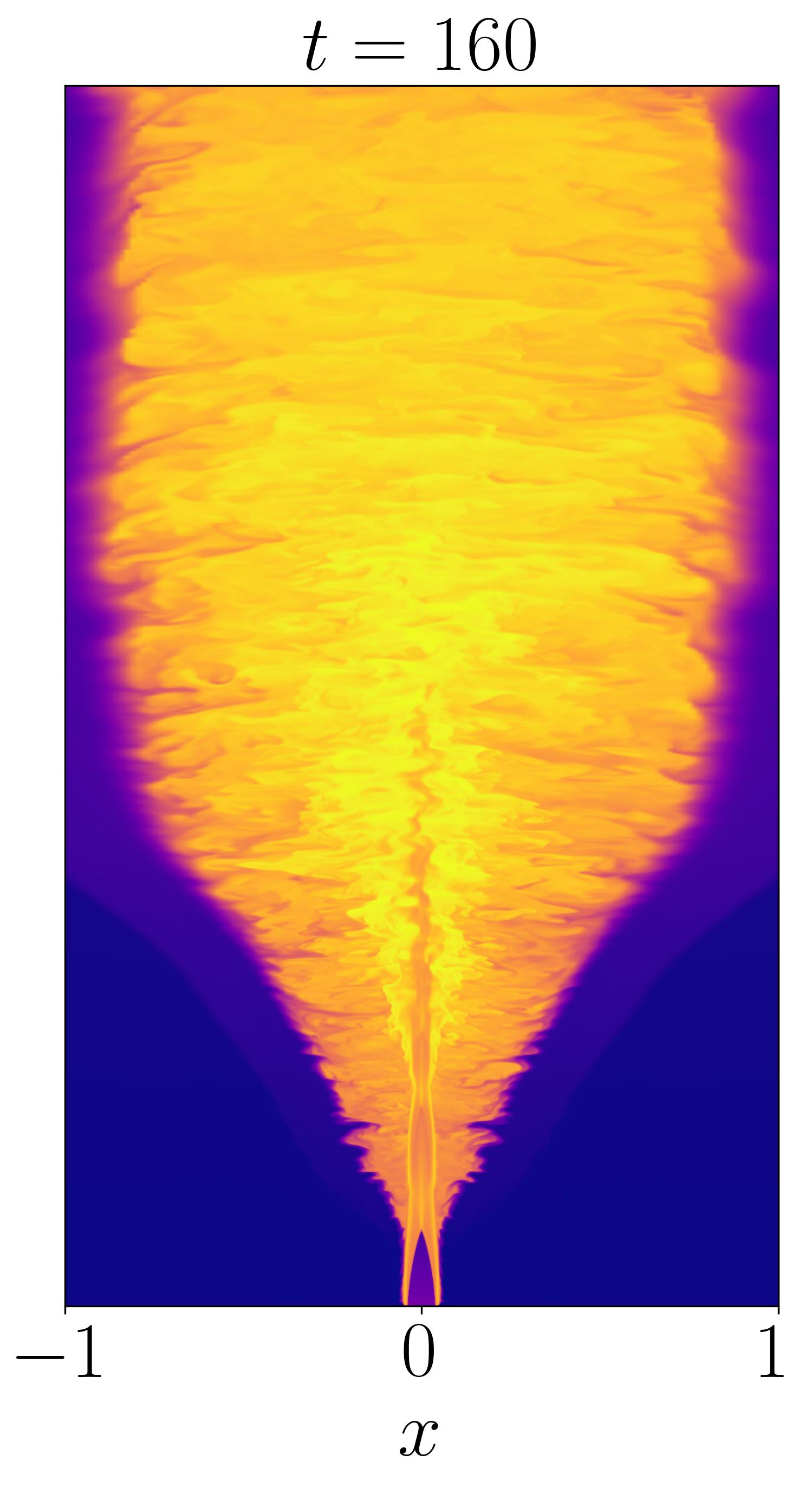}  
    \end{minipage}
    \begin{minipage}{0.28\textwidth}
        \centering
        \includegraphics[height = 6.5cm]{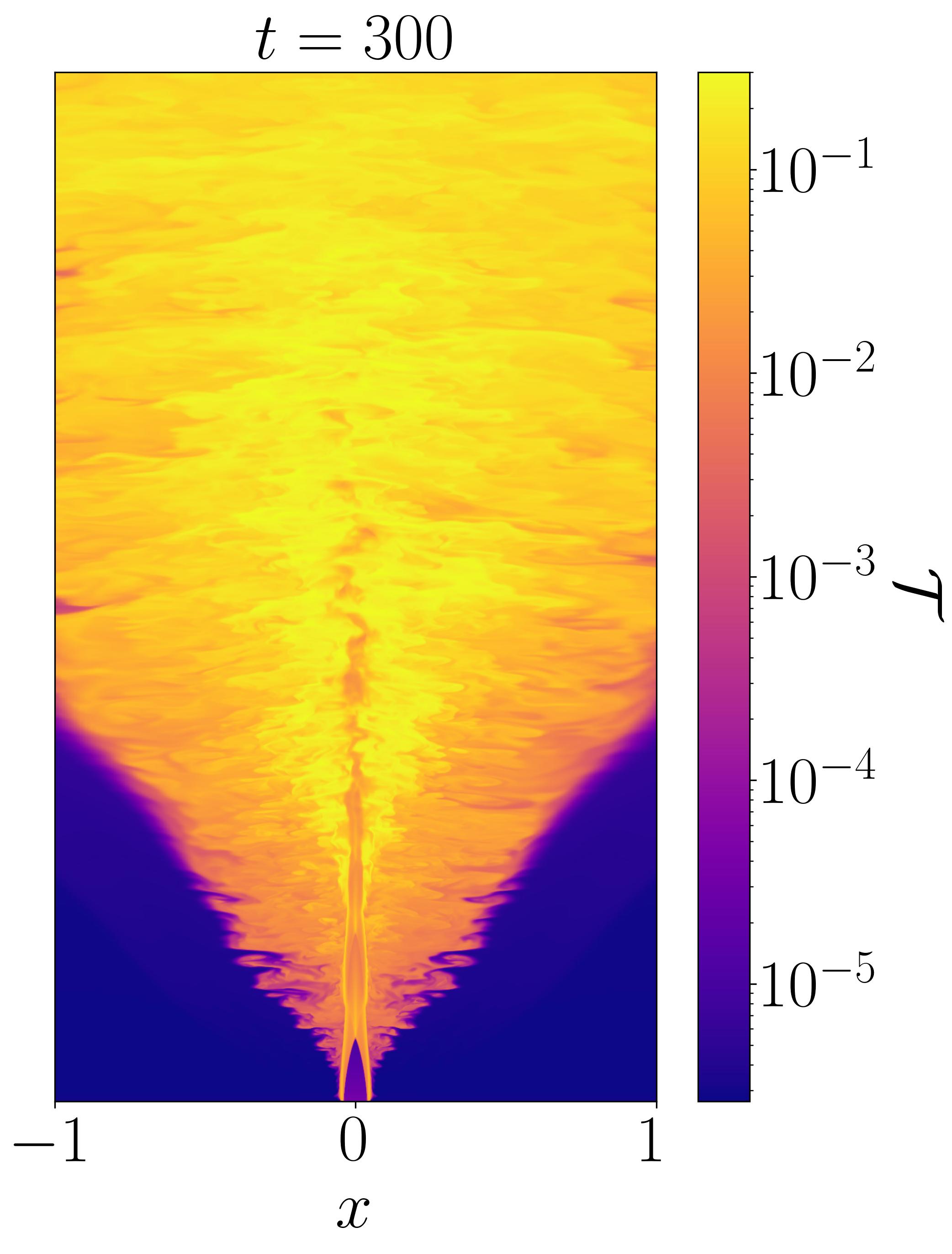}  
    \end{minipage}
    \caption{Sagittal maps $\mathcal{T} (x,y=0,z,t=t^*)$ of the temperature, at different output times $t^*$, for case SHC05.}
    \label{fig:SHC05_long_temp}
\end{figure*}

The long term evolution for  case SH05 is presented in Figs. \ref{fig:SHC05_long_prs} and \ref{fig:SHC05_long_temp}, where we show respectively pressure and temperature maps at three different times. Superimposed on the pressure maps, we plot a blue contour of the temperature at $\mathcal{T}=3\times 10^{-4}$. This value is chosen because it corresponds to a region of high temperature gradient. 
A video of the long term evolution of the pressure can be found at \href{https://drive.google.com/file/d/1BhjvFDv3wAEpy0Rvz3i79k2kUg4ZRU8u/view?usp=drive_link}{\textbf{SHC05.prs.long.mp4}}. As we discussed above, after the first stages of instability growth, the interaction between different perturbations leads to the formation of a hot turbulent sheath surrounding and compressing the inner faster jet, the spine. The hot sheath eventually destroys the multiple shock structure after a few recollimation shocks, and the whole jet becomes hot and turbulent, as it is possible to observe in the temperature maps. On the outside, this hot region drives a shock in the ambient medium, resulting in the formation of an over-pressured \textit{bubble} made of two distinct components: a hot turbulent interior (the sheath), and a more homogeneous outer part consisting of shocked external gas. The two components are separated by the blue contours drawn on the pressure maps. By comparing the three different times shown in the figures, we observe that the over-pressured bubble expands both sideways and backwards. At the same time, the position of the recollimation points shifts backwards as a consequence of the hotter medium through which the  spine propagates, see Eq. \ref{eq:recpoint_code}. The recollimation shock structure starts to be destroyed around $z \sim 7-8$. Above this region, the recollimating spine disappears, and the over-pressured bubble acquires a cylindrical shape. 

\begin{figure}[ht]
    \centering
    \includegraphics[width=0.9\columnwidth]{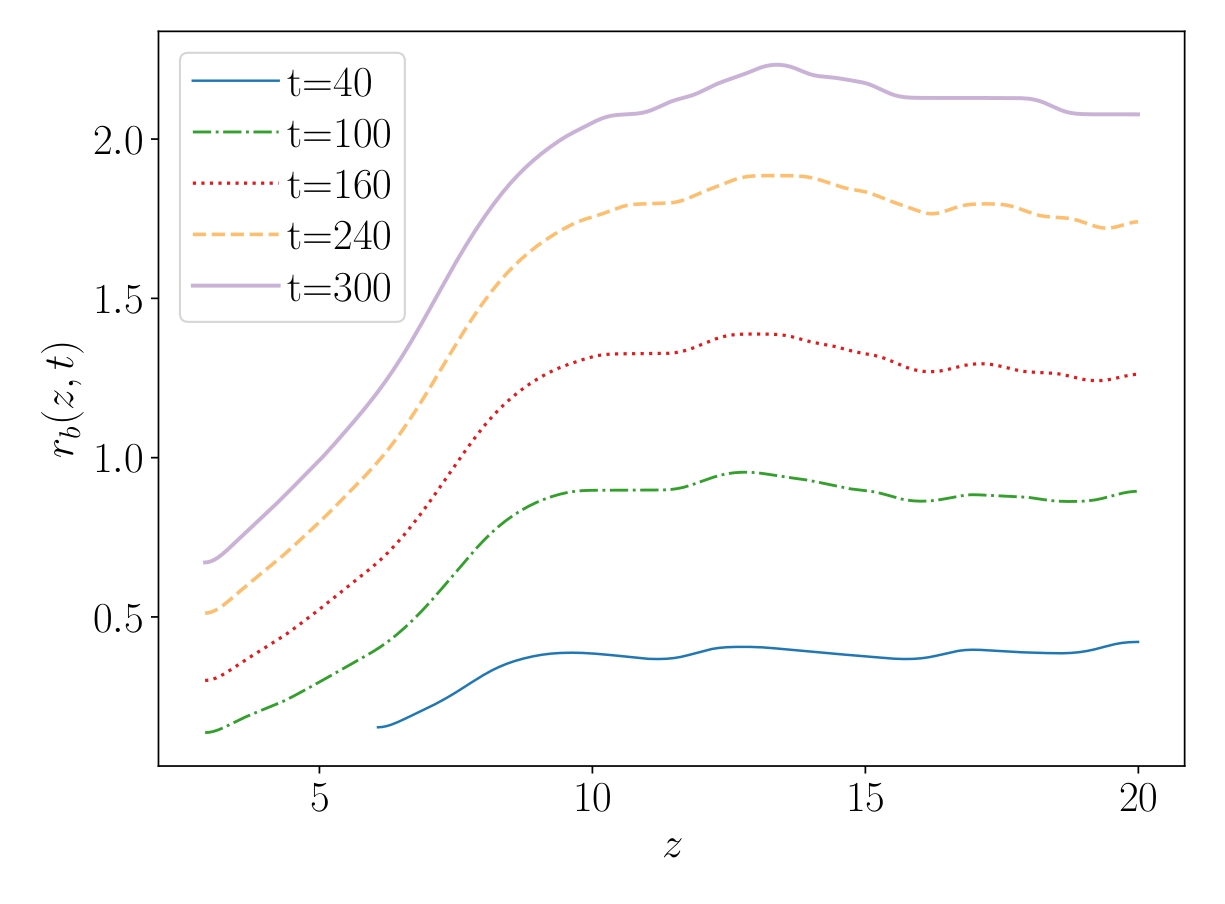}  
    \caption{The radial size of the bubble $r_b (y=0,z,t)$, for case SHC05.}
    \label{fig:SHC05_bubble}
\end{figure}

We can quantify more precisely the expansion of the over-pressured bubble by looking at Fig. \ref{fig:SHC05_bubble}, where we show its radial extension (i.e. the radius of the compression wave), $r_b$, as a function of the altitude $z$ and time $t$. The position $r_b(z,t)$ is defined by large values of the pressure gradient, as described in Appendix \ref{appendix_shock_det}. In the figure we can distinguish the two regimes already observed in Figs. \ref{fig:SHC05_long_prs} and \ref{fig:SHC05_long_temp}: for $z<7$ the radius grows linearly with the altitude $z$, while for $z\gtrsim 7-8$ it becomes independent from $z$. This might be connected to the disruption of the shock chain structure for $z \gtrsim 7$, and to the dissipation of the bulk kinetic energy carried by the spine above this point.   
As the bubble expands its mean pressure decays; as a consequence the expansion velocity driven by the internal pressure diminishes with time. Specifically, we observed that the lateral expansion speed of the shock, in the region at uniform expansion, at late times is smaller by approximately $50\%$  with respect to the values of the earlier phases.

\begin{figure*}[htbp]
    \centering
    \includegraphics[width=\textwidth]{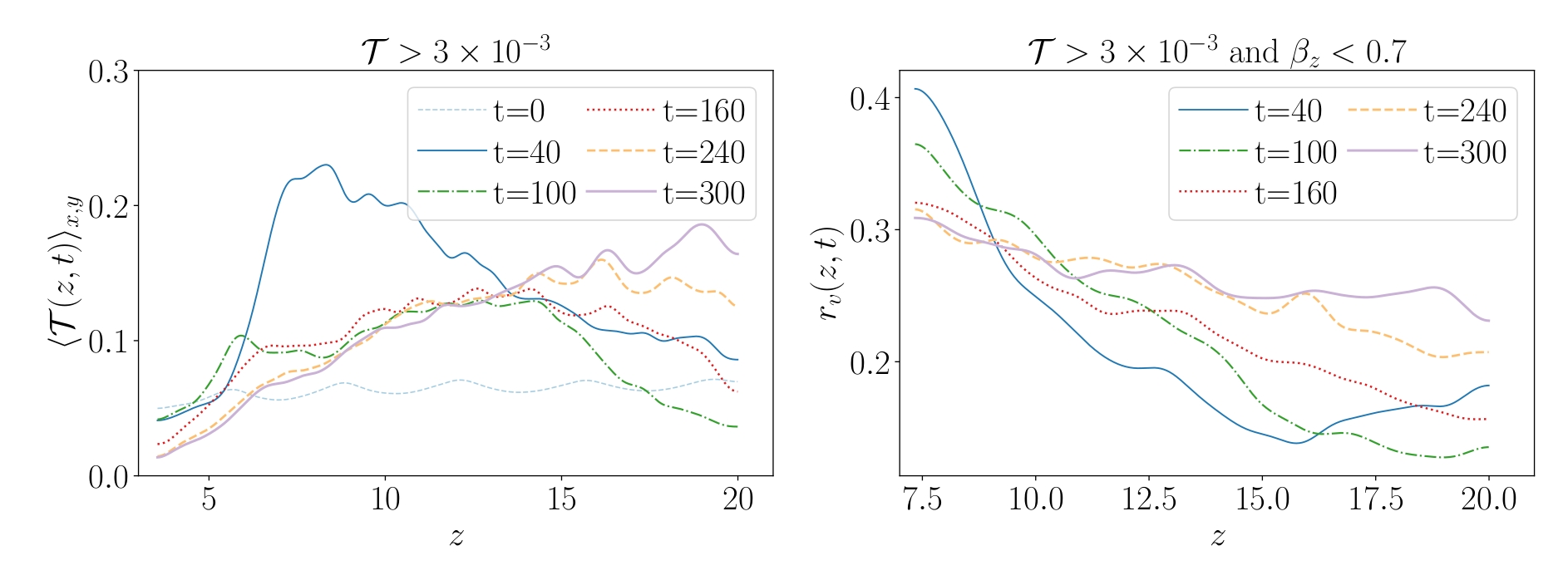}  
    \caption{Temperature and relative strength of velocity perturbations profiles averaged on transverse sections of the jet as functions of the altitude and distance, for case SHC05: $\langle \mathcal{T}(z,t)\rangle_{x,y}$ and $r_v(z,t)$.}
    \label{fig:SHC05_trends}
\end{figure*}

More details about the jet kinetic energy dissipation can be obtained by looking at Fig. \ref{fig:SHC05_trends}, where we plot, in the left panel, the average temperature of the sheath and, in the right panel, the r.m.s. value of the perturbation velocity normalized with the mean velocity, as a function of $z$ at different times. More precisely, in the left panel we plot  $ \langle \mathcal{T} (z,t)  \rangle_{x,y} $ which is computed by averaging $\mathcal{T}$ on the area $\tilde{A}(z,t)$ in the $x-y$ plane, at fixed $z$, where $\mathcal{T}>3\times 10^{-4}$, in order to select only the sheath. In the right panel we plot the quantity
\begin{equation}
    r_v(z,t) = \sqrt{\frac{\langle \delta v^2(z,t) \rangle_{x,y}} {\langle v^2(z,t)\rangle_{x,y}}},
    \label{eq:r_v}
\end{equation}
where 
\begin{equation}
  \delta v^2 = \sum_{i=x,y,z}\left(v_i- v_{0,i}\right)^2 \,,  
\end{equation}
and where the local mean $ v_{0,i}$ is obtained by smoothing each component of the velocity $v_i$ with a 3D Gaussian filter of standard deviation $\sigma$ equal to $1/6$ the initial jet radius (in this case $\sigma\simeq0.008$). Its averages are taken on the area $\tilde{A}_2(z,t)$ in the $x-y$ plane, at fixed $z$, where $\mathcal{T}>3\times10^{-4}$, but also where $\beta_z<0.7$, in order to select only the sheath and to exclude the inner high velocity spine.


From the temperature plot (left panel of Fig. \ref{fig:SHC05_trends}) we can see that the initial profile (dashed light-blue) is consistent with the 2D steady state of the jet in a multiple-shock configuration: the mean temperature increases and decreases due to shocks and adiabatic cooling respectively. The evolution in time shows, at first, at $t=40$ (solid blue), a strong heating of the jet where the instability evolution leads to the formation of shocks. At later times, when turbulence develops, the jet mean temperature  decreases, progressively reaching  (for $ t > 200$) a quasi-stationary configuration. The profiles at $t=240$ (dashed-dotted thick yellow line) and at $t = 300$ (solid thick grayish line) look very similar and show a weak dependence on $z$, with larger average temperatures at larger distances. 

In the right panel of Fig. \ref{fig:SHC05_trends}, we only show the profiles at $t>0$ and for $z\gtrsim7.5$, when and where the recollimation shock structure has been dissipated and $\langle \delta v^2 \rangle_{x,y}$ is mainly due to turbulence. At $t=40$ (solid blue) the value of $r_v$ peaks around $z \sim 7.5$ with a value $\sim 40 \%$; at later time the behavior of $r_v$ as a function of $z$ becomes flatter, with a value of about  $28 \%$ at $t=300$ (solid thick grayish curve).
This means that the energy dissipated from instabilities is not only converted into thermal energy, heating the jet, but a non negligible fraction is  also conveyed to disorderly kinetic fluctuations. 

In summary, in terms of distance $z$, we can distinguish  3 regions. The initial one at $z\lesssim3$, before the first recollimation point, remains stable and cold, and does not get significant feedback by the bubble. Indeed, the pressure of the bubble is comparable with that of the environment when it reaches the first recollimation shock, as shown in  Fig. \ref{fig:SHC05_long_prs} at $t=300$. A second region, $z\gtrsim 7$, is where dissipation mostly occurs. 
This results in a progressive increase of the mean temperature and of the r.m.s. of velocity fluctuations. In between, for $3\lesssim z\lesssim 7$, the jet is characterized at late times by a relativistic spine undergoing recollimation and reflection shocks, surrounded by the hot sheath that flows back from the upper regions. The presence of the hot sheath has an effect on the recollimation structure, pushing the recollimation point backward, this effect is minimal in this case, but is more pronounced for more powerful cases, as we will see next. 

\subsection{Case SLWw2}
\begin{figure*}[htbp]
    \centering
    \begin{minipage}{0.26\textwidth}
        \centering
        \includegraphics[height = 6.5cm]{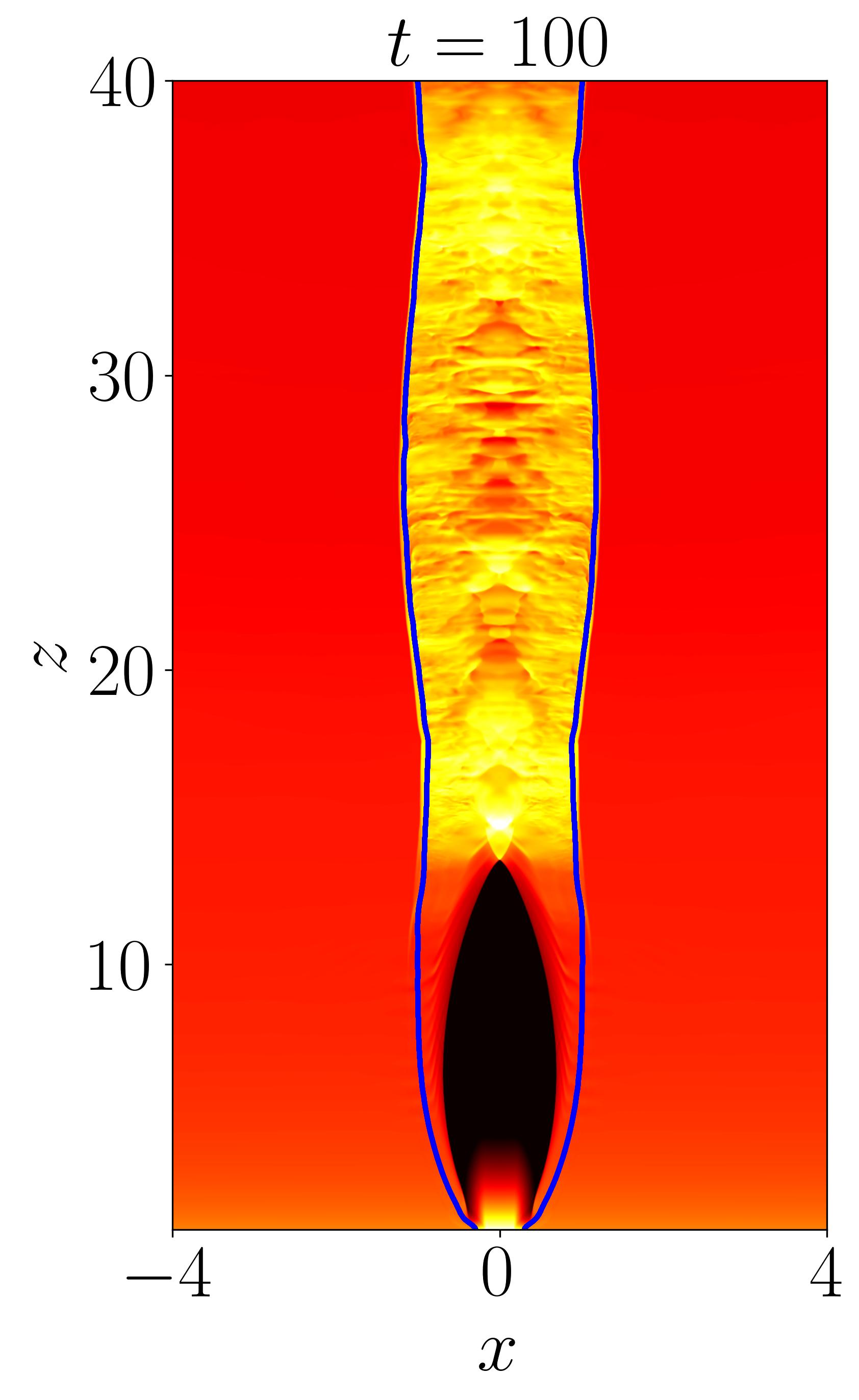}  
    \end{minipage}
    \begin{minipage}{0.22\textwidth}
        \centering
        \includegraphics[height = 6.5cm]{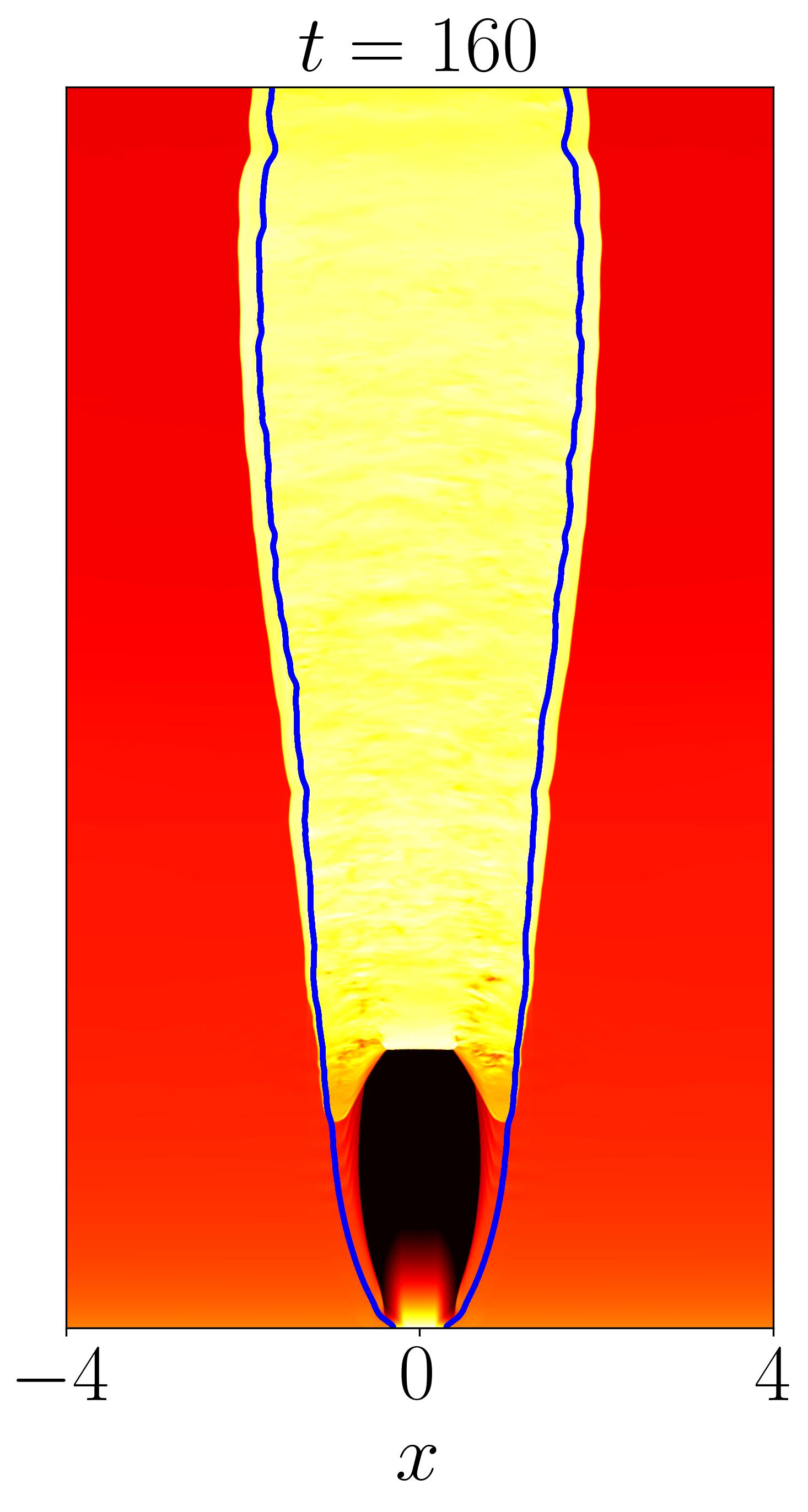} 
    \end{minipage}
    \begin{minipage}{0.22\textwidth}
        \centering
        \includegraphics[height = 6.5cm]{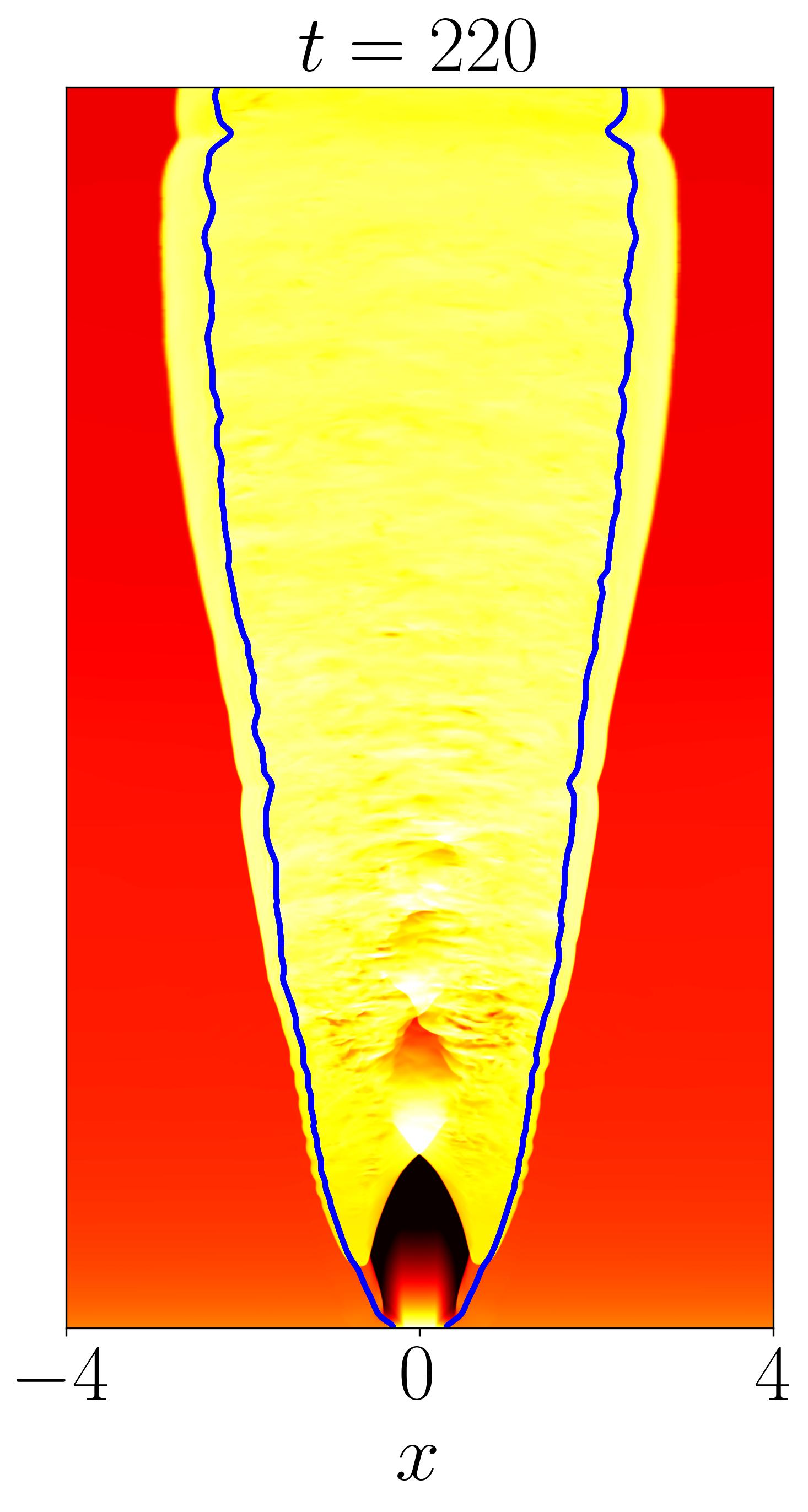}  
    \end{minipage}
    \begin{minipage}{0.28\textwidth}
        \centering
        \includegraphics[height = 6.5cm]{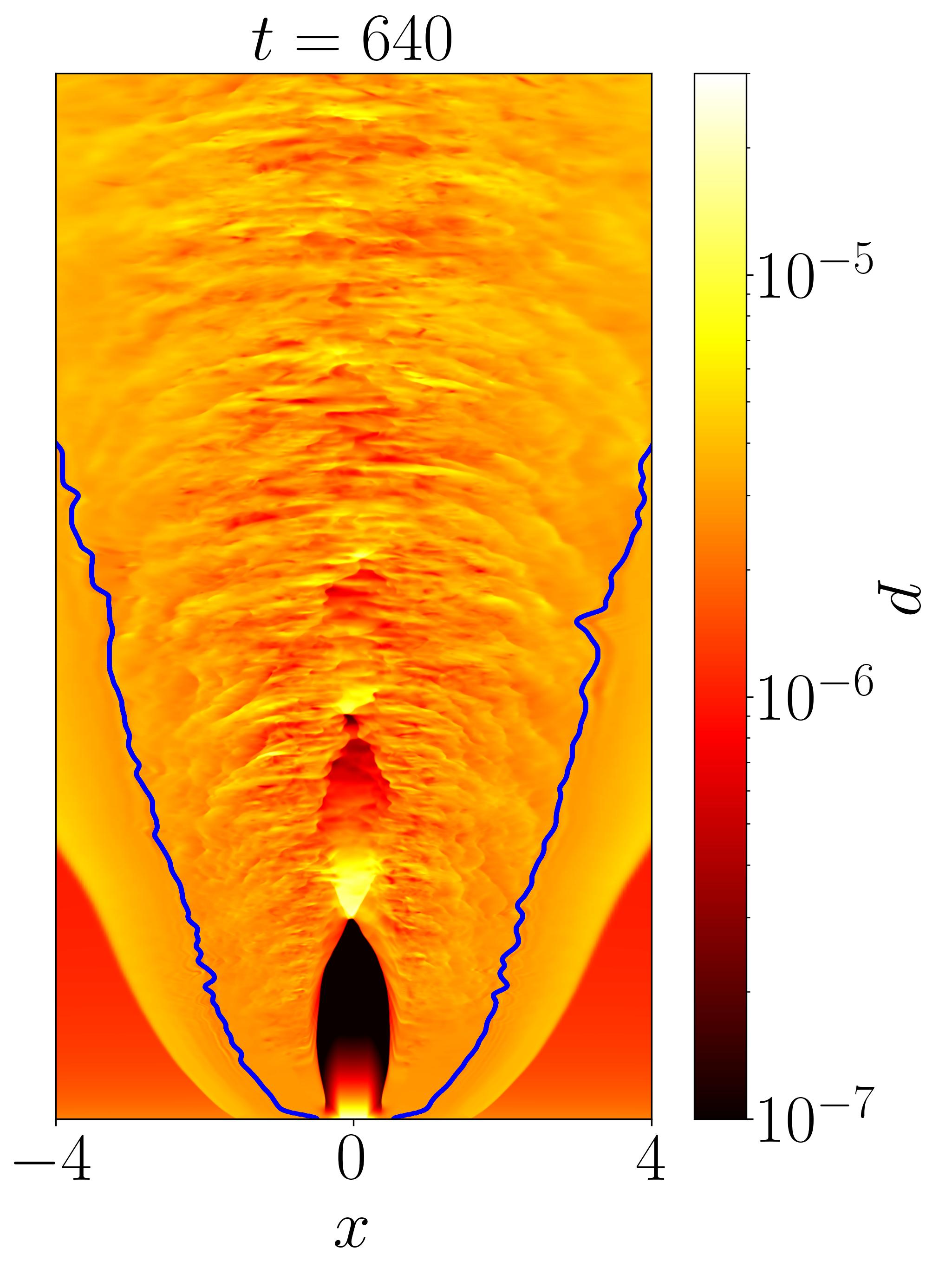}  
    \end{minipage}
    \caption{Sagittal maps of the pressure $p(x,y=0,z,t=t^*)$, at different output times, denoted with $t^*$ in case SLWw2. Blue lines correspond to $\mathcal{T}(x,y=0,z,t=t^*) = 10^{-3}.$}
    \label{fig:SLWw2_prs}
\end{figure*}

\begin{figure*}[htbp]
    \centering
    \begin{minipage}{0.26\textwidth}
        \centering
        \includegraphics[height = 6.5cm]{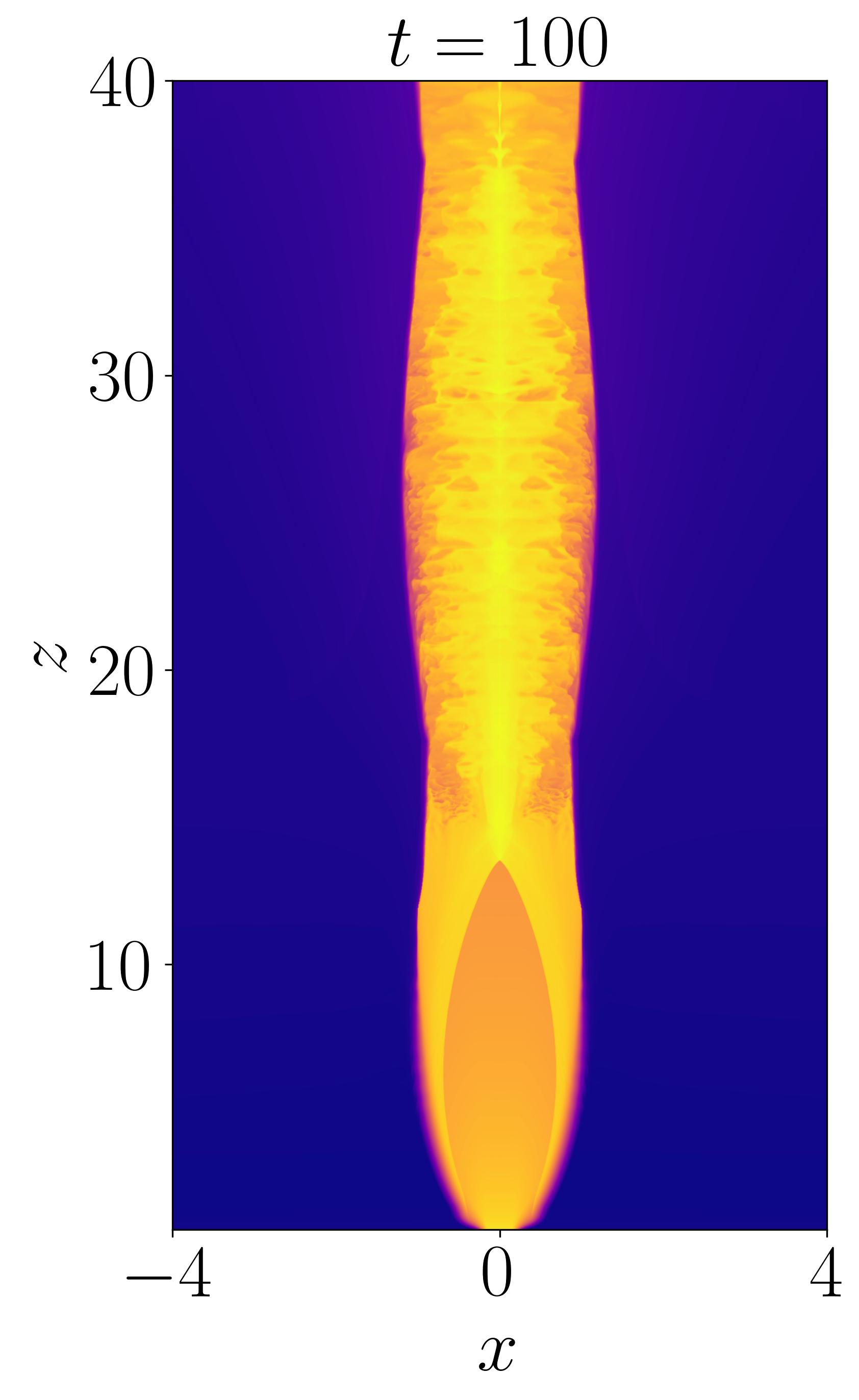}  
    \end{minipage}
    \begin{minipage}{0.22\textwidth}
        \centering
        \includegraphics[height = 6.5cm]{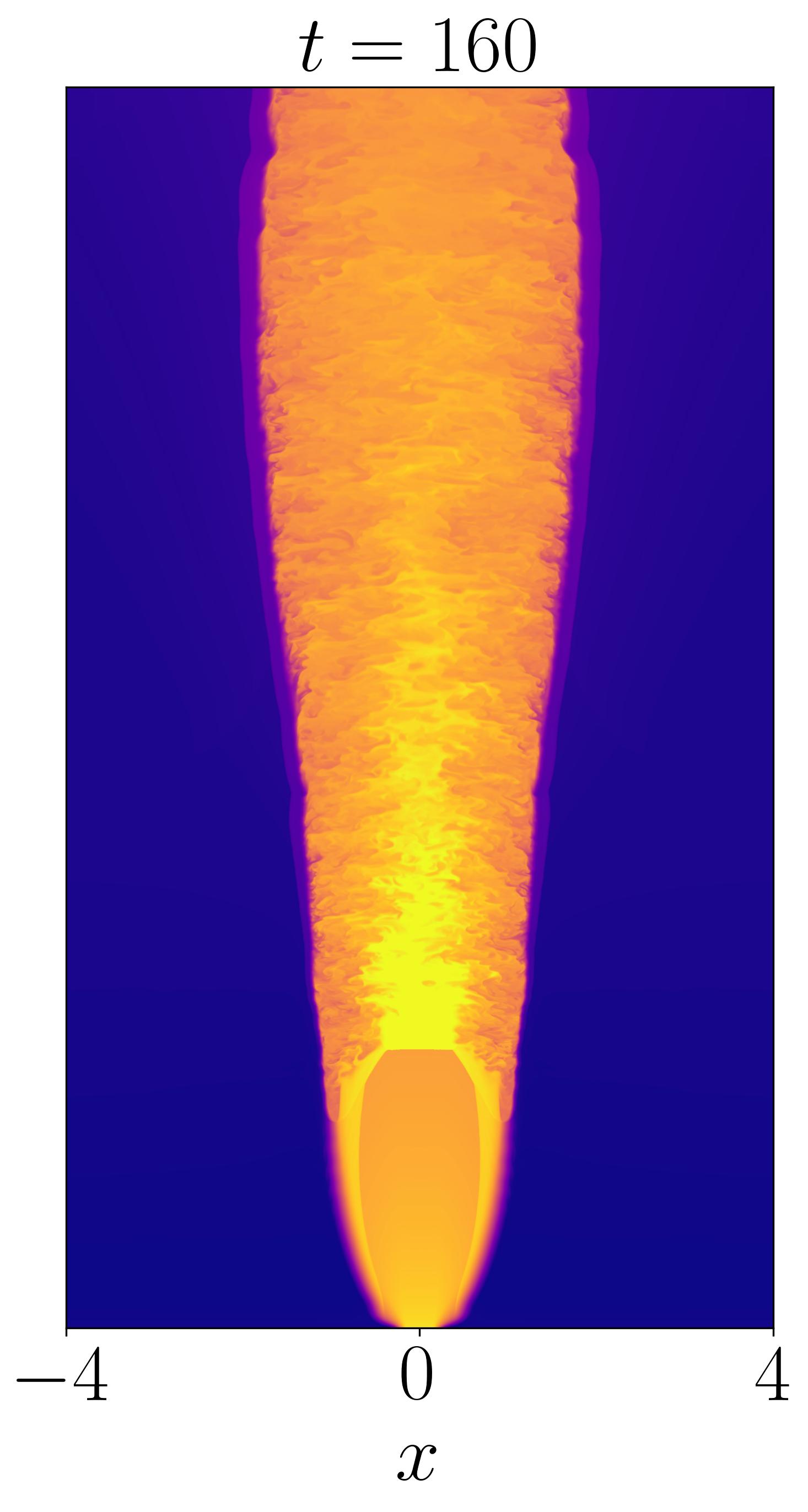} 
    \end{minipage}
    \begin{minipage}{0.22\textwidth}
        \centering
        \includegraphics[height = 6.5cm]{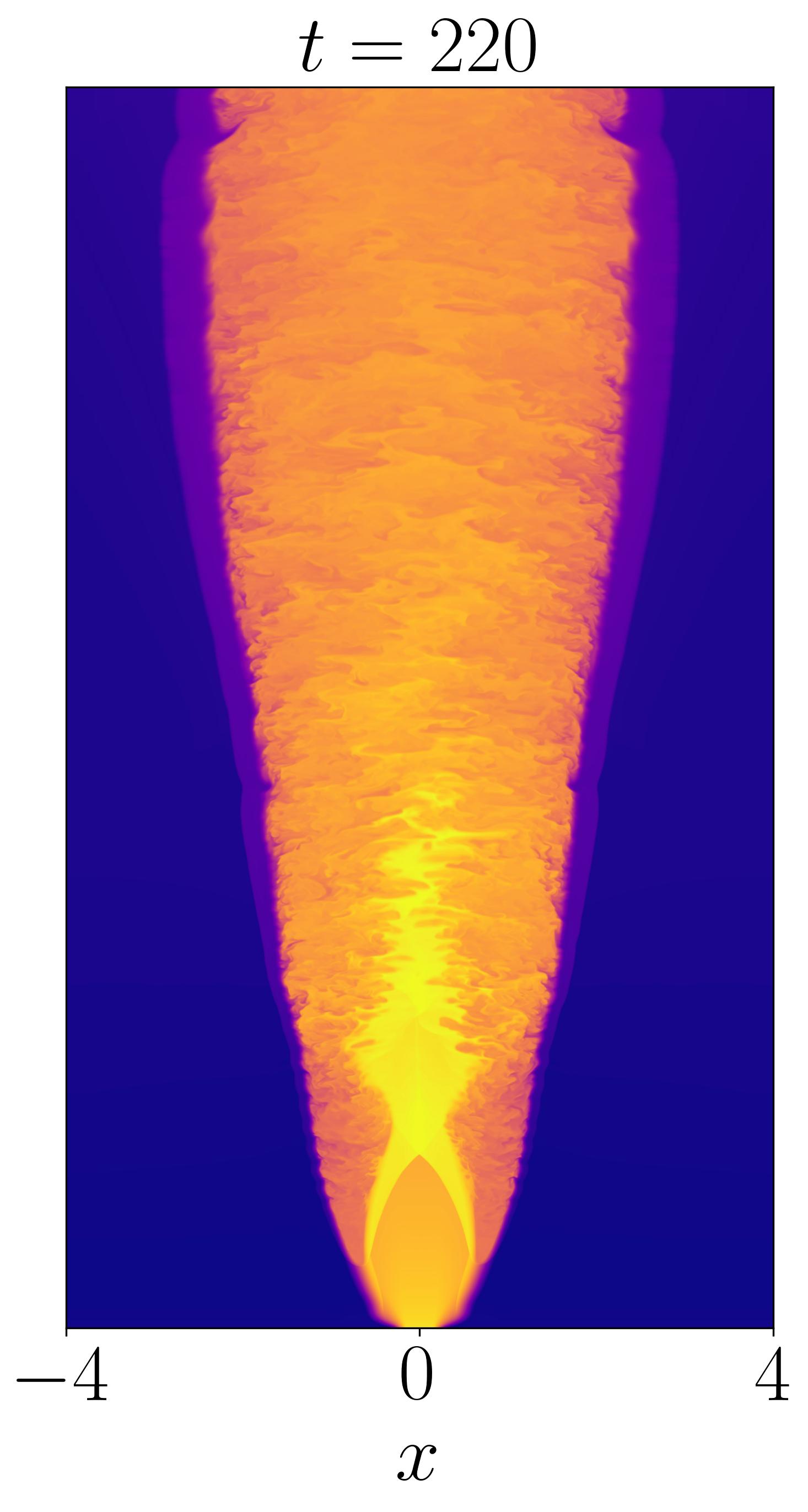}  
    \end{minipage}
    \begin{minipage}{0.28\textwidth}
        \centering
        \includegraphics[height = 6.5cm]{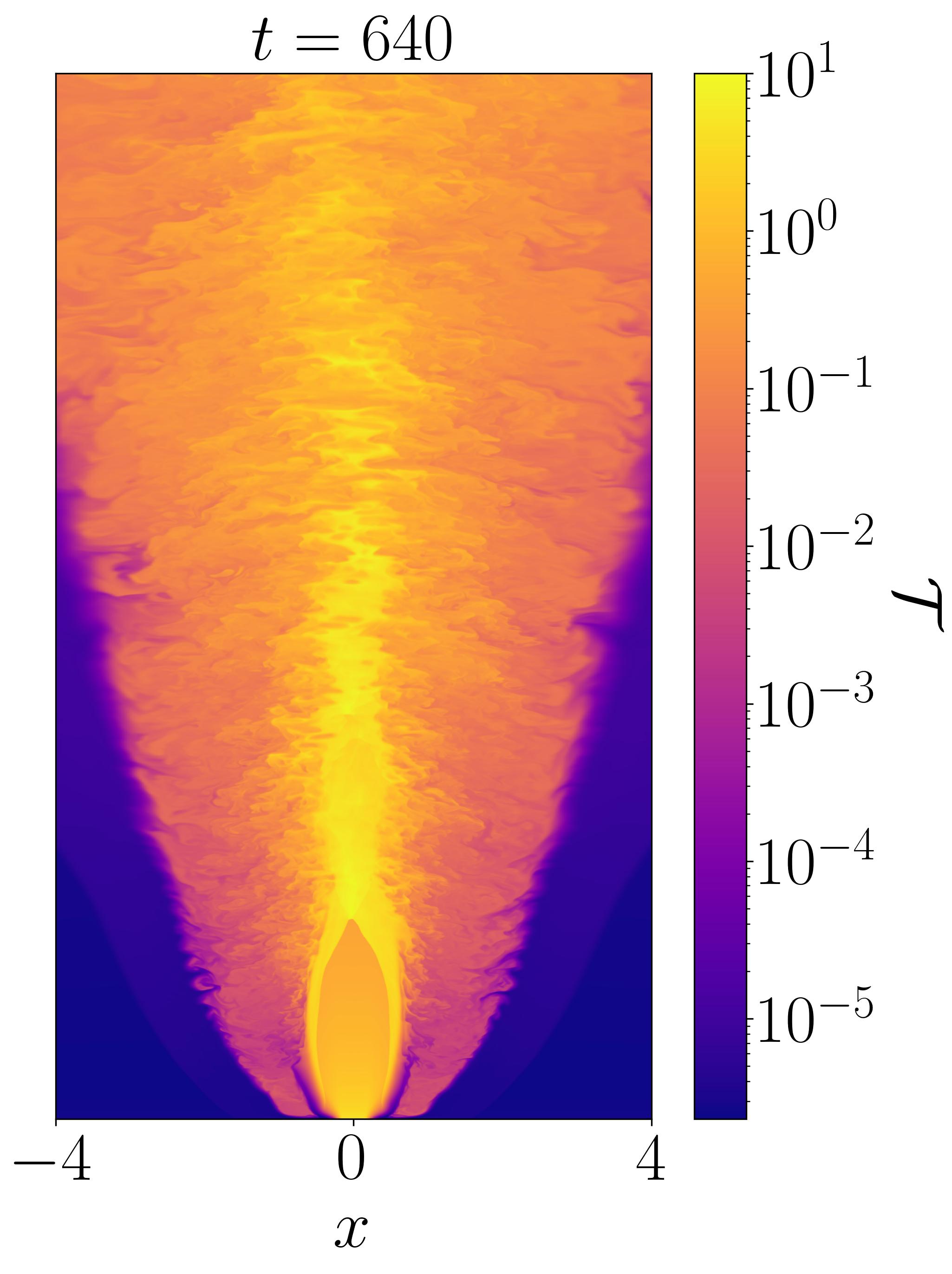}  
    \end{minipage}
    \caption{Sagittal maps of the temperature $\mathcal{T}(x,y=0,z,t=t^*)$, at different output times, denoted with $t^*$, in case SLWw2.}
    \label{fig:SLWw2_temp}
\end{figure*}

Case SLWw2, discussed in this subsection, is over-pressured with respect to the environment and it is enthalpy dominated, while the previous case was kinetically dominated. The energy flux in this case is $16$ times higher than in the previous case, with a large fraction related to the thermal energy. Figure \ref{fig:SLWw2_prs} and \ref{fig:SLWw2_temp} show sagittal maps of pressure and temperature for case SLWw2, analogous to Figs. \ref{fig:SHC05_long_prs} and \ref{fig:SHC05_long_temp}. On pressure maps, blue curves are plotted where $\mathcal{T} = 10^{-3}$ (corresponding again to the sharp temperature transition indicating the sheath-high pressure environment separation within the bubble) to indicate the boundary between the sheath and the outer regions of the over-pressured bubble.

There are some evident differences in the long-term evolution of the instability in jets SLWw2 and SHC05. The initial dissipation of energy leads to an extreme growth of  pressure and temperature; this drives backwards the recollimation region and leads to the formation of a Mach disk (see panels at $t=160$ of Figs. \ref{fig:SLWw2_prs} and \ref{fig:SLWw2_temp}) with the disappearance of the following shocks. At later times ($t=220$), the Mach disk disappears, the recollimation point is driven further backwards and the chain of recollimation shocks is partly reformed. As the pressure of the bubble decreases as a consequence of its expansion, the recollimation point shifts forwards and reaches a quasi-stationary position ($t=640$) at $z \sim 7-8$. We observe that this position is much closer to the injection point with respect to the initial steady axisymmetric solution, where the recollimation point was at $z \sim 14$ (Fig. \ref{fig:2D_maps_hot}). In fact, at this stage, the environment of jet propagation is not the external cold medium anymore, but the hot and turbulent sheath formed as a consequence of the instability evolution. The shock chain is completely destroyed after the second recollimation shocks ($z \sim 17$), that appears to be already strongly perturbed.  The over-pressured bubble, in this case, acquires a paraboloidal shape, contrary to the cylindrical shape observed in the previous case. This complex evolution requires a longer time and, for this reason, this simulation was run to later times. From the temperature maps we can also observe a strong contrast between the inner region of the bubble where most of the dissipation of the jet energy occurs, and the outer regions at considerably lower temperatures $\mathcal{T}\lesssim10^{-1}$. The contrast is much higher with respect to the SHC05 case for a kinetically dominated jet: the jet injection temperature is much higher in this case.

\begin{figure}[htbp]
    \centering
    \includegraphics[width=0.9\columnwidth]{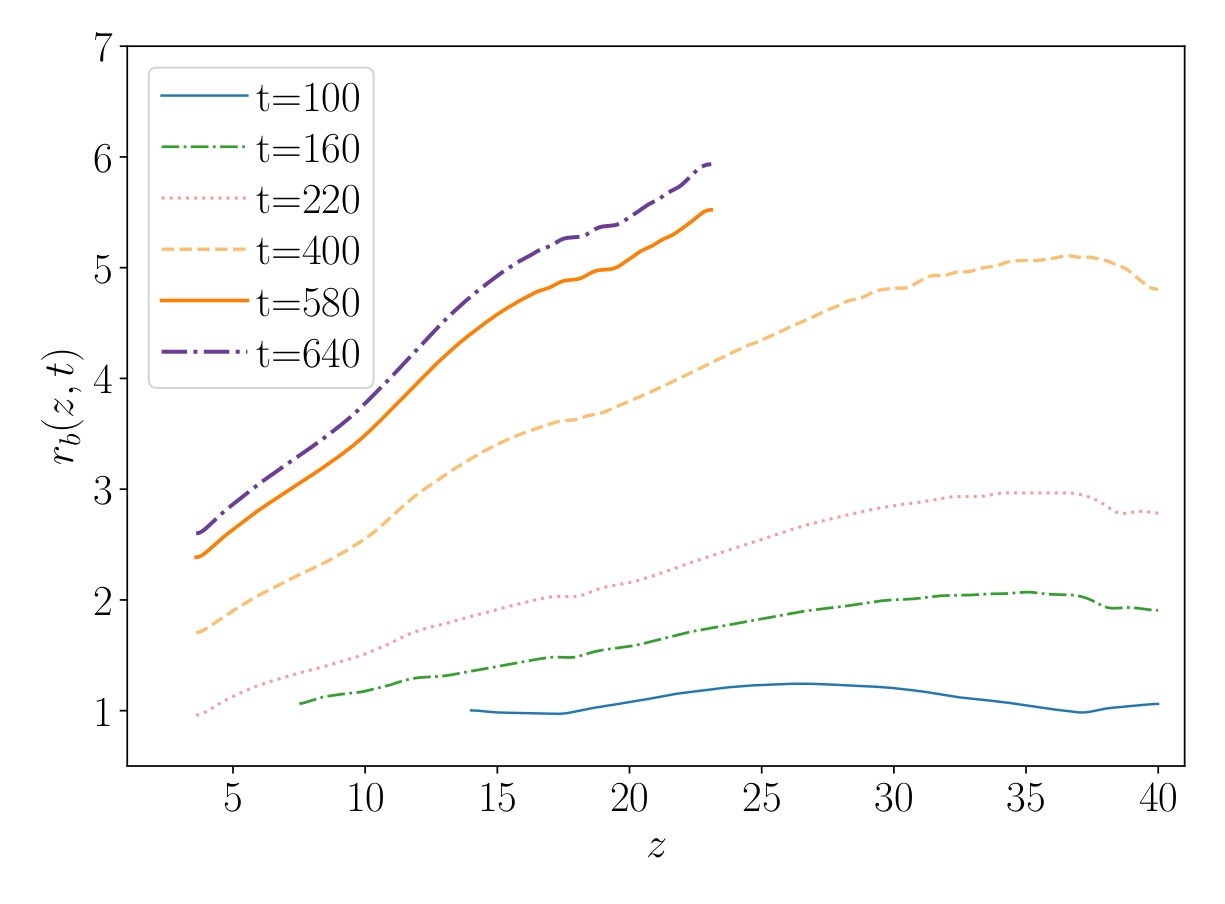}  
    \caption{The radial profile of the bubble $r_b(y=0,z,t)$, for case SLWw2. }
    \label{fig:SLWw2_bubble}
\end{figure}

\begin{figure*}[htbp]
    \centering
    \includegraphics[width=\textwidth]{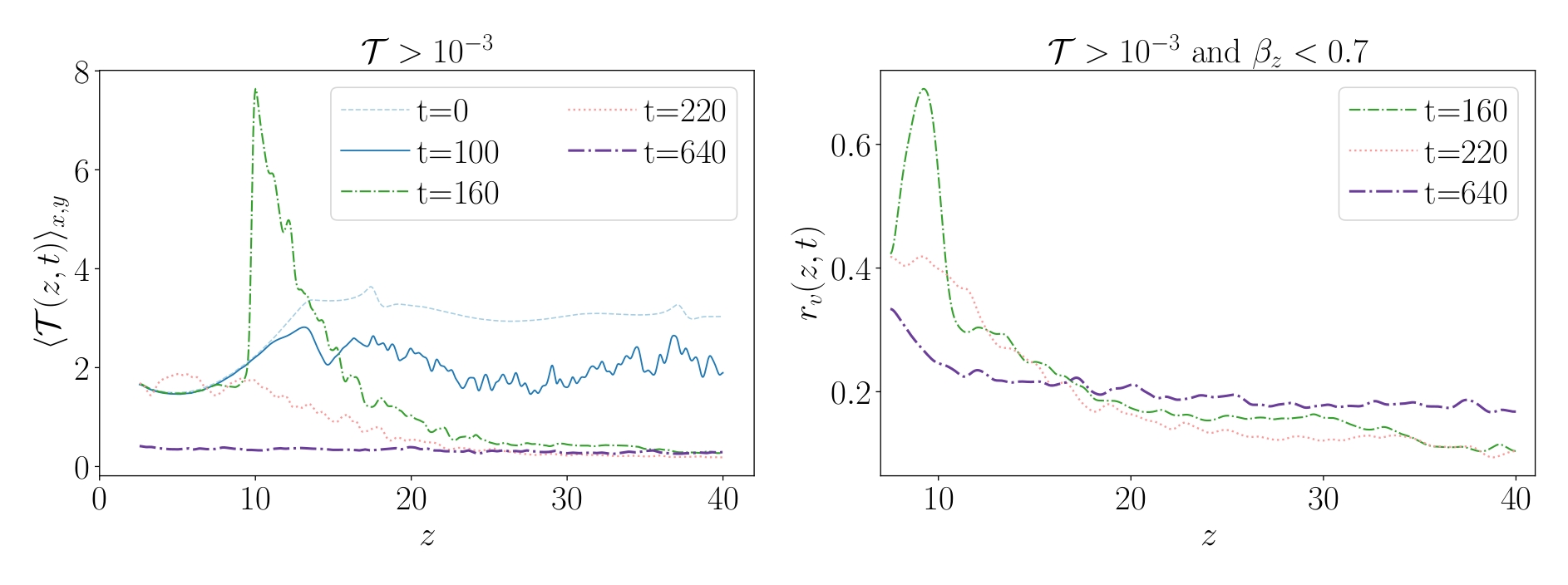}  
    \caption{Temperature and relative strength of velocity perturbations profiles averaged on transverse sections of the jet as functions of the altitude and distance, for case SLWw2: $\langle \mathcal{T}(z,t)\rangle_{x,y}$ and $r_v(z,t)$. }
    \label{fig:SLWw2_bubble_tem}
\end{figure*}

In Fig. \ref{fig:SLWw2_bubble} we plot the radial position $r_b (z,t)$ of the external shock as a function of $z$ at different times.
We remark that the curves at $t=580$ (solid thick orange) and $t=640$ (dashed-dotted thick purple) are cut at $z \sim 23$ because, at this time, the external shock has exited the side boundary. 
In contrast with case SHC05, where the radius of the bubble is more or less constant at large altitudes, this plot shows, as already noted, that the radius has a gradual growth over its length in the SLWw2 case, consistent with a more gradual energy dissipation, as we shall see in more detail in the next section. In this case too, as the pressure of the bubble decreases, its expansion decelerates over time by factors $\simeq 50\%$.

Similarly to what we did for the previous case, in Fig. \ref{fig:SLWw2_bubble_tem} we plot $\langle \mathcal{T} \rangle_{x,y}$ (left panel) and $r_v$ (right panel) as functions of $z$ at different times.  $\langle \mathcal{T} (z,t)\rangle_{x,y}$ and $r_v(z,t)$ are defined as previously, but in this case the selection condition on the temperature of the sheath is $\mathcal{T}(x,y,z^*,t^*)>10^{-3}$. The behavior of $\langle \mathcal{T} (z,t)\rangle_{x,y}$ is in stark contrast with the previous case. In fact, while for case SHC05 the average temperature at $t=0$ is lower than the average temperature at later times, in the present case we observe the opposite. This is again related to the fact that the SLWw2 jet is enthalpy dominated: it starts from a large temperature, and, at later time, the energy dissipation caused by the growth of instabilities is not sufficient to compensate the temperature decrease caused by the bubble expansion. Looking at the behavior of the average temperature along the jet, we see that at $t=160$ (dashed-dotted green) the curve peaks around $z \sim 10$, in correspondence to the downstream of the Mach disk, and then decreases steeply. At later times, the average temperature distribution tends to become more uniform along the jet, and its value at the end of the simulation ($t = 640$, dashed-dotted thick purple curve) is $\simeq 0.3-0.4$, higher than in the SHC05 case. A non negligible fraction of the dissipated jet kinetic energy is also conveyed to disordered fluctuations, as shown by the right panel of Fig. \ref{fig:SLWw2_bubble_tem}. At later times, the distribution of their energy along the jet tends to become more uniform, as it happens for the average temperature. The value of $r_v$ at the end of the simulation is around $20\%$, slightly less than in case SHC05.

As discussed for case SHC05, we can distinguish three regions along the SLWw2 jet as well: a first unperturbed region, $z\lesssim5$, a second region where perturbations grow $5\lesssim z\lesssim17$, and a third region, $z\gtrsim17$, where the recollimation chain and the relativistic spine are destabilized. There is however a difference in the extension of the over-pressured bubble that, in this case, extends down to the jet base. Nevertheless, in the first region, it does not perturb the relativistic jet spine. 

\subsection{Other simulations}

The considerations on the re-adjustment of the recollimation shock are not limited to setup SLWw2. In general, the stable part of the jet always reacts to its own dissipation, through the feedback caused by the hot sheath when it expands towards lower altitudes. However, in weaker unstable jets, like in setup SHC05, this effect is much less pronounced.

\begin{figure}[htbp]
    \centering
    \includegraphics[width=0.75\columnwidth]{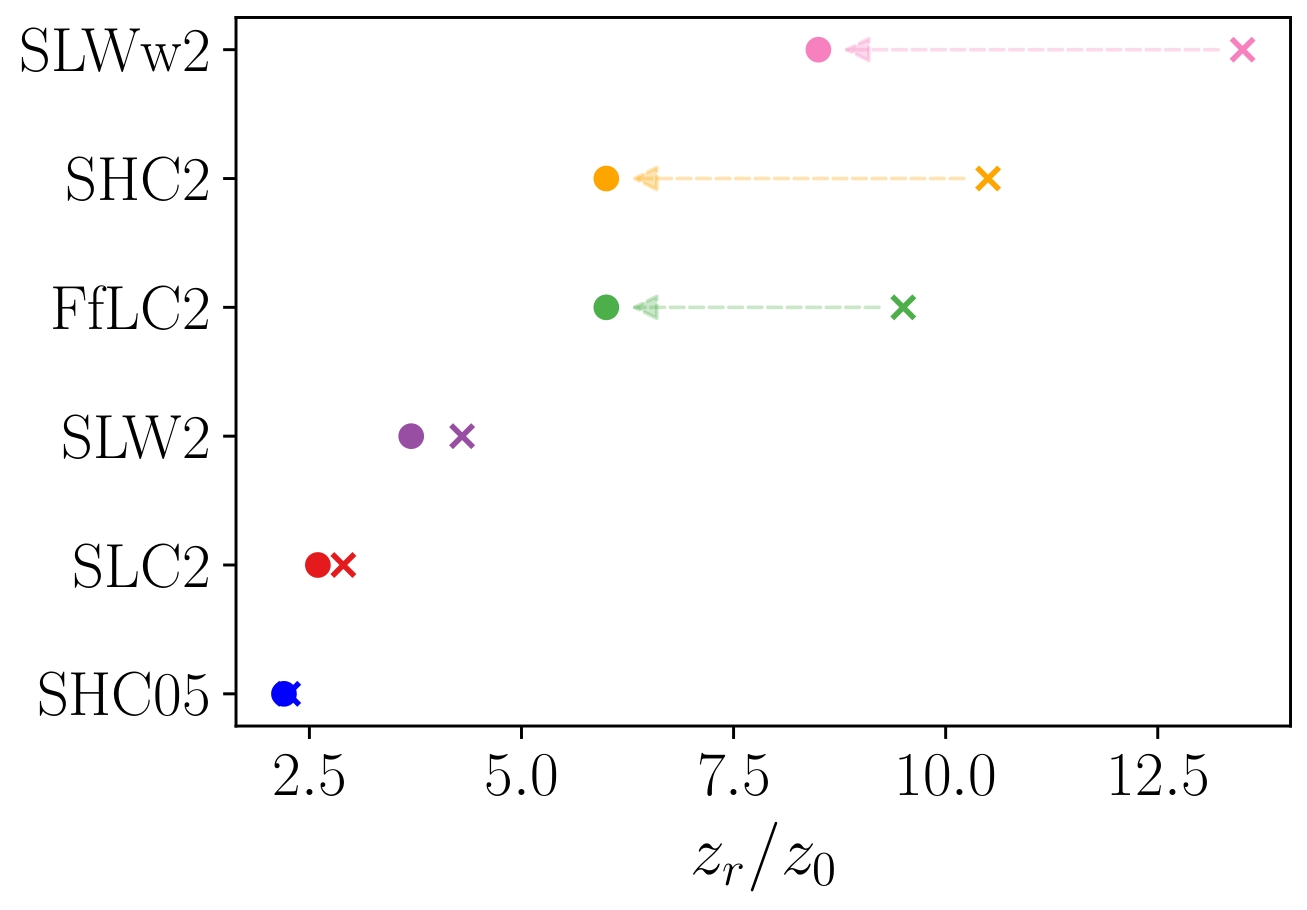}  
    \caption{Recollimation points in 2D ($\times$) and 3D (dot) simulations, for various setups at the end of the respective simulations. Plots are ordered along the y-axis. In case SHC05 the points almost coincide. The cases are ordered so that the jet power increases with the $y$ axis.}
    \label{fig:recpoint}
\end{figure}

In Fig. \ref{fig:recpoint} we compare the position of the recollimation point in the initial steady axisymmetric states (X) and in the 3D (dots) simulations, for all the setups. The larger differences correspond to more powerful jets (upper part of the plot), that recollimate later in the steady configurations, like FfLC2 (green), SLWw2 (pink), and SHC2 (yellow).
The result is that, in all 3D jets that we have simulated, the recollimation point always ends up being within $10 z_0$.

\section{Jet breaking, deceleration and entrainment \label{decentr}}

\begin{figure*}[htbp]
    \centering
 \begin{minipage}{\textwidth}
       \flushleft
       \includegraphics[width=\textwidth]{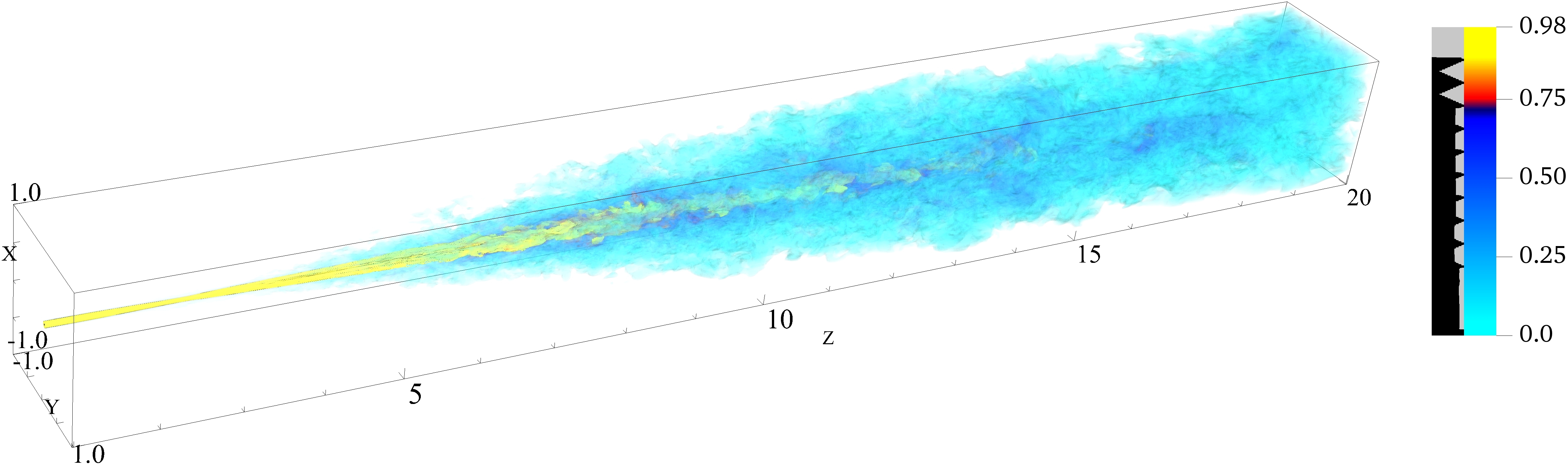}  
   \end{minipage}
    
   \begin{minipage}{\textwidth}
       \centering
       \includegraphics[width=\textwidth]{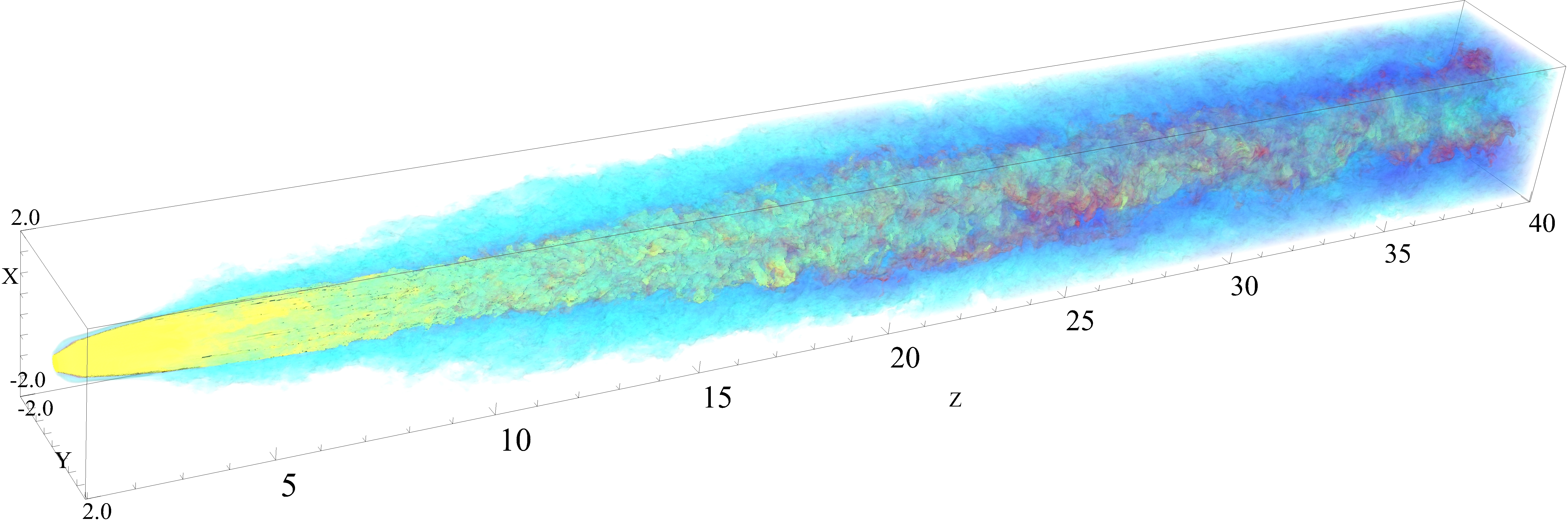}  
   \end{minipage}
    \caption{3D volume rendering of $\beta_z(x,y,z, t= t_{l,3D})$ at the end of the simulation for case SHC05 (top panel) and case SLWw2 (bottom panel). The black-gray bar on the left of the color-bar shows which values of $\beta_z$ are opaque (gray), and which are transparent (black). We note that the box is larger in the SLWw2 case.}
    \label{fig:SHC05_3D}
\end{figure*}

At the end of the 3D runs, the jets have reached a quasi-stationary configuration that is in general very different from the axisymmetric one. The over-pressured bubble discussed in the previous section still expands, but at a smaller and decreasing rate, and the profiles of averaged quantities along the jet show very little variations in time. The quasi-stationary configurations for cases SHC05 and SLWw2 are shown in Fig. \ref{fig:SHC05_3D}, where  we display a 3D rendering of the component of the velocity in the jet propagation direction, at the end of the run. The figure highlights the relativistic component of the jet, in yellow, and the surrounding slower jet regions in blue. 

In case SHC05 (top panel) one can observe a few expansion-confinement stages associated with the multiple-shock structure up to  $z \sim 7$ , where the jet starts to break, then the relativistic components completely disappear for $z \gtrsim 15$. In case SLWw2 (bottom panel), both the relativistic spine and the slower regions are wider, as it can be expected since the opening angle is larger and, in addition, the higher injection pressure leads to further expansion. The first recollimation region is clearly visible up to $z \simeq 10$.  The second recollimation region between $z \simeq 10$ and $z \lesssim 17$ is still visible, but is perturbed. For $z \gtrsim 17$ the relativistic region is strongly perturbed and signs of recollimation completely disappear. At the end of the domain, $z=40$, the relativistic spine is starting to disappear, but some relativistic material is still present. 

As we discussed above, in the final quasi-stationary configuration, we can  divide the jet length into three regions: in the first, the jet propagates almost undisturbed, with its sequence of recollimation and reflection shocks, in the second, perturbations grow leading to distortions of the relativistic jet spine, associated with the beginning of the entrainment process, and, in the third, the outcome of the development of instability evolution is a turbulent entrainment process, with the consequent dissipation of the  bulk kinetic energy. We then observe a deceleration of the relativistic flow possibly to sub-relativistic velocities accompanied by the loss of coherence of the high velocity spine. In the next subsections we will discuss in more detail the spatial development of the perturbations, that lead to the jet loss of coherence for the two representative cases, SHC05 and SLWw2, and the process of entrainment and deceleration.



\subsection{Jet structure}

\begin{figure*}[htbp]
    \centering
    \includegraphics[width = \textwidth]{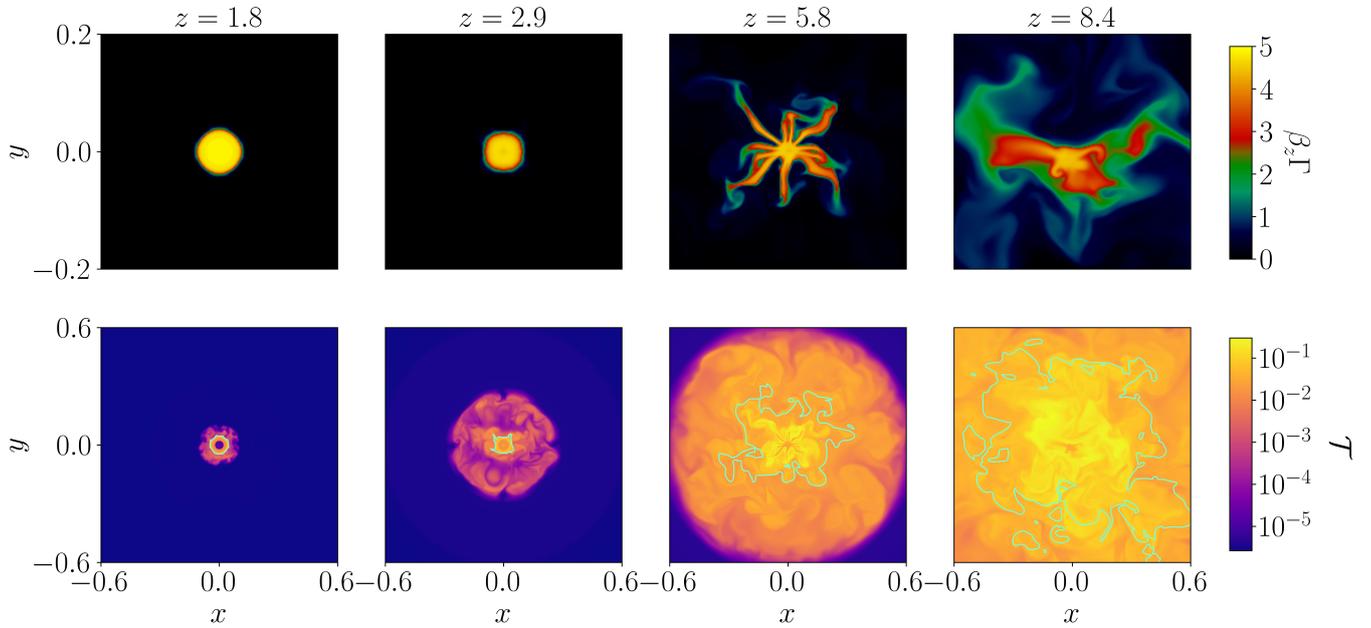}  

    \caption{$(\beta_z\Gamma)(x,y,z=z^*,t=t_{l,3D})$ (upper row) and $\mathcal{T}(x,y,z=z^*,t=t_{l,3D})$ (lower row) maps for case SHC05. The displayed domain is larger for temperature figures. Light blue curves are contours of $\beta_z(x,y,z=z^*,t=t_{l,3D}) = 0.1.$}
    \label{fig:SHC05_transverse}
\end{figure*}

\begin{figure*}[htbp]
    \centering
    \includegraphics[width = \textwidth]{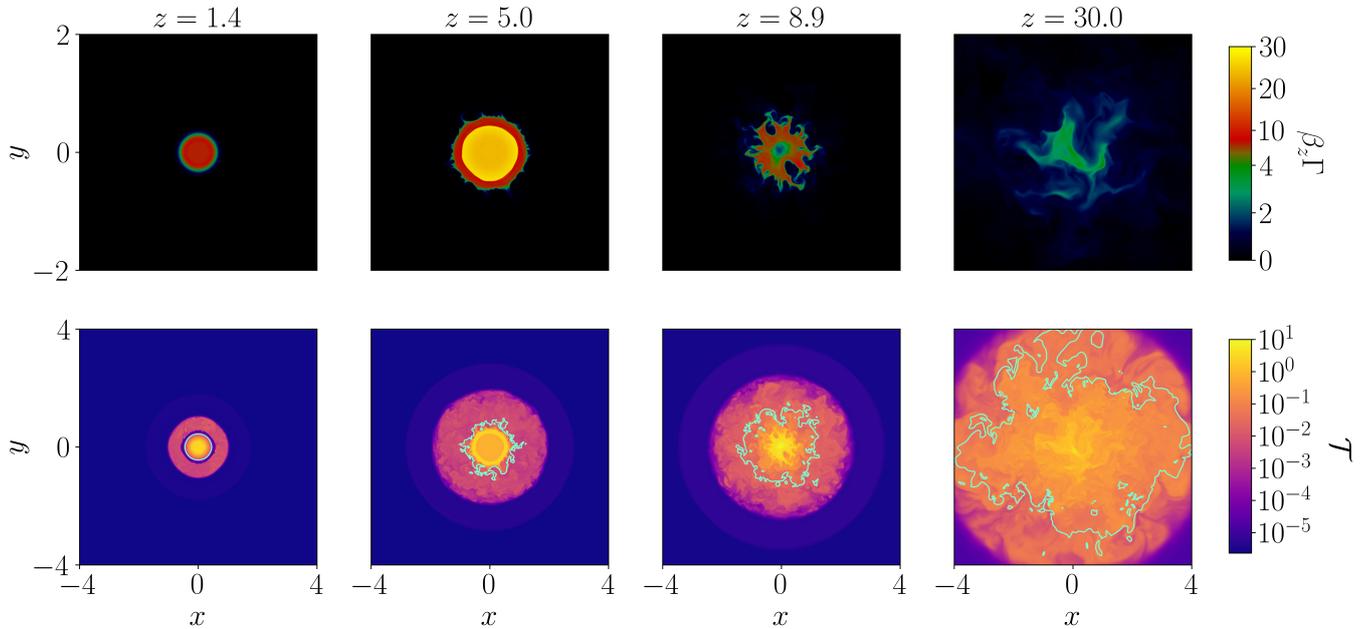}  

    \caption{Maps of $(\beta_z\Gamma)(x,y,z=z^*,t=t_{l,3D})$ (upper row), and $\mathcal{T}(x,y,z=z^*,t=t_{l,3D})$ (lower row), with the light blue contours of $\beta_z (x,y,z=z^*,t=t_{l,3D})= 0.1$, for case SLWw2. The size of the figure is larger with respect to the respective Fig. (\ref{fig:SHC05_transverse}) of case SHC05, and temperature maps are displayed in a larger box than Lorentz factor maps.}
    \label{fig:SLWw2_transverse}
\end{figure*}

In Figs. \ref{fig:SHC05_transverse} and \ref{fig:SLWw2_transverse} we display transverse maps, at different heights, of the component of the four velocity parallel to the jet axis, $\beta_z\Gamma$, (upper panels) accompanied by temperature maps, $\mathcal{T}$, (lower panels, displayed in a larger domain) for cases SHC05 and SLWw2, respectively. The upper panels show the evolution of the relativistic spine, whereas the lower panels highlight the structure of the hot and turbulent sheath. In order to capture the sheath extension, the temperature maps are shown in a larger domain compared to the $4$-velocity maps. On the temperature maps, we also display a light blue contour where the longitudinal component of velocity is $\beta_z = 0.1$. This contour indicates the region where substantial entrainment occurs. However, we notice that the entrainment occurs also in a larger region, with lower velocities. From the figures, we can examine in detail how the jet structure changes as we move along its length, at the final time of the simulations. We point out that the values of the height at which we display cuts are lower for case SHC05 than for case SLWw2, since, in the second case, the position of the recollimation point is shifted upward and the jet entrainment process starts later. 

In case SHC05 (Fig. \ref{fig:SHC05_transverse}), the first two top panels on the left ($z=1.8$ and $z=2.9$) show the relativistic spine undisturbed, but modulated by the effects of the recollimation and reflection shocks. The two lower left panels, instead, show the hot sheath that surrounds the spine as a consequence of a backflow from the upper regions. However, from the velocity contour we see that the entrainment process has hardly started.  At $z = 5.8$, the spine starts to be highly distorted, with the Lorentz factor distribution (top row) becoming very corrugated and characterized by thin relativistic fingers penetrating the slower sheath.  The velocity contour, in the bottom panel, shows the development of entrainment, accompanied by a strong energy dissipation. In fact, we see that the sheath temperature increases from values $\sim 10^{-3}$ at $z = 2.9$ to values $\gtrsim 10^{-2}$ at $z = 5.8$. The rightmost panels of Fig. \ref{fig:SHC05_transverse}, at $z = 8.4$, display how the process further proceeds, at larger distances, with a strong distortion of the spine (upper panel), a widening of the velocity contour, and of the high temperature region (bottom panel).

In case SLWw2 (Fig. \ref{fig:SLWw2_transverse}), at $z=1.4$  and $z=5.0$ we see, as in the previous case, that the relativistic spine is almost undisturbed and, especially at $z= 5.0$, surrounded by the sheath that is flowing back from the upper regions. The spine is accelerating as a result of the hot jet expansion. At $z=8.9$, the spine is corrugated, and the temperature of the sheath is strongly increased as a consequence of dissipation due to entrainment. These effects are more pronounced at $z=30.0$.

\subsection{Deceleration and entrainment across the different simulations}
\begin{figure}[htbp]
    \centering
    \includegraphics[width=\columnwidth]{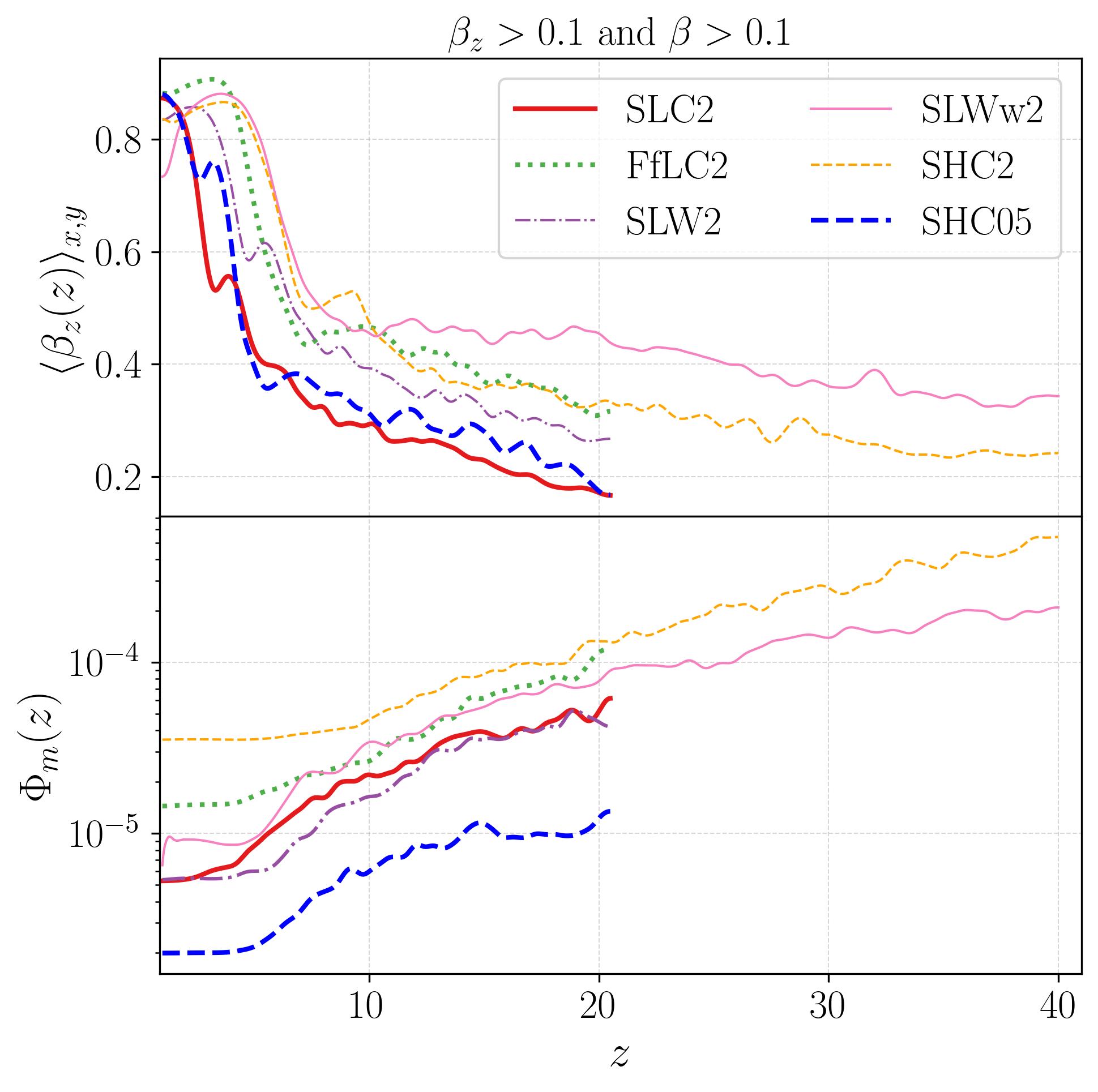}  
    \caption{Deceleration and entrainment of the jet across the different setups, shown through plots of $\langle \beta (z,t_{l,3D}) \rangle_{(x,y)}$ (upper panel), and $\Phi_m(z,t)$ (lower panel), defined in the text. }
    \label{fig:scan_decel}
\end{figure}

The entrainment process leads to a redistribution of the jet momentum to the external heavier material, and therefore to a jet deceleration. In order to discuss the deceleration properties across the parameter space, we plotted, in the upper panel of Fig. \ref{fig:scan_decel}, the average of the velocity component along the jet propagation direction, $\langle \beta_z \rangle_{x,y}$, at the end of each simulation, $t= t_{l,3D}$. In particular, we calculated the mean on transverse sections of the jet where $\beta_z>0.1$ (to exclude the environment and the external parts of the sheath at very low velocities). Since deceleration is connected to the entrainment of external gas, we quantified this process by plotting in the lower panel of Fig. \ref{fig:scan_decel} the mass flux in the propagation direction at the end of the simulations, $t=t_{l,3D}$. The flux is calculated on the same jet area $\tilde{A}$ as for the velocity average, with the condition $\beta_z>0.1$:
\begin{equation}
    \Phi_m(z,t_{l,3D}) =\int_{\tilde{A}} (\rho \Gamma v_z)(x,y,z,t_{l,3D}) \,dxdy .
\end{equation}
This is done to completely exclude the environment from the calculation: because of the significant density contrasts ($\nu<10^{-4}$ for all cases), small velocities in the environment could significantly affect the result. 

The  variations of the average velocity close to the jet base (see top panel of Fig. \ref{fig:scan_decel}), up to $z \sim 3-5$, are different between lower power jets (cases SLC2 and SHC05), in which the recollimation shock leads to the initial decrease of the average velocity, and higher power jets, where we observe an initial increase, due to the jet expansion. Starting from $z \sim 3-5$ up to $z \sim 7-10$, the strong decrease of the average velocity is connected to the starting of the entrainment process, as can be seen by looking at Fig. \ref{fig:scan_decel}. For $z \gtrsim 7-10$, both the decrease of the average velocity and the entrainment proceed, but at a lower rate. These three different phases loosely correspond to the three different jet regions defined previously. As one would expect, the jets that decelerate the fastest are the least powerful ones, in particular SLC2 (solid thick red) and SHC05 (dashed thick blue). 

\begin{figure}[htbp]
    \centering
    \includegraphics[height=7cm]{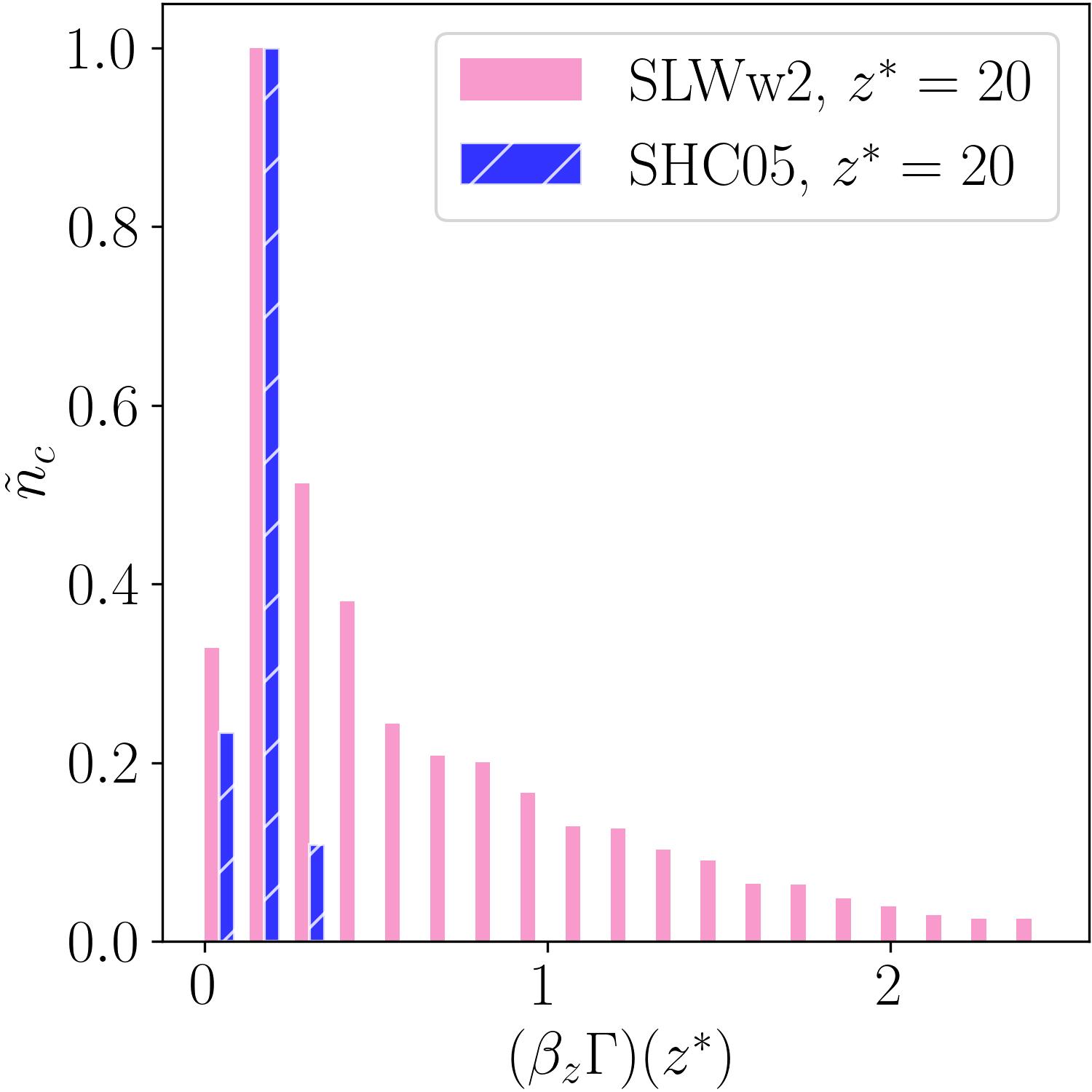}  
    \caption{The distribution of $(\beta_z\Gamma)(x,y,z=z^*,t=t_{l,3D})$ at $z^*=20$ for cases SHC05 and SLWw2. The maximum is normalized to 1 in each case.}
    \label{fig:scan_histo}
\end{figure}

\begin{figure}[htbp]
    \centering
    \includegraphics[width=0.7\columnwidth]{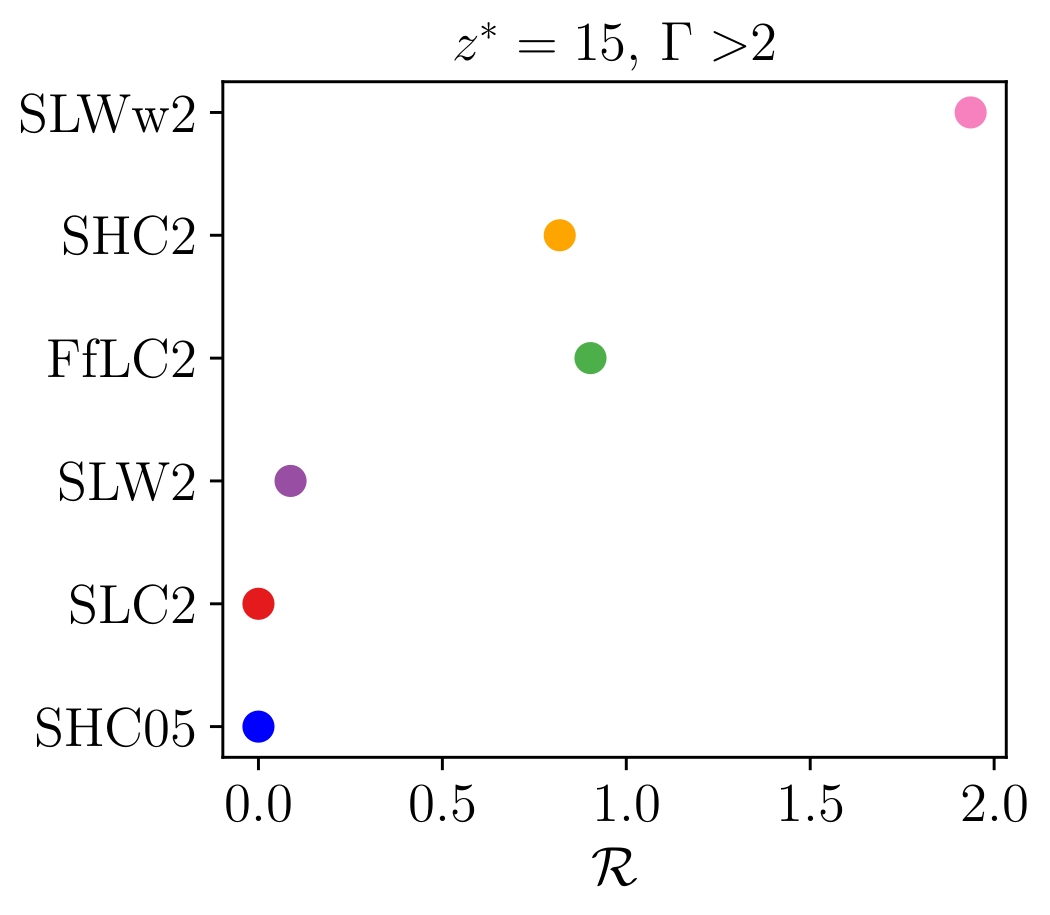}  
    \caption{Plot of  $ \mathcal{R} = \tilde{\Phi}_m(z=z^*,t=t_{l,3D})/\tilde{\Phi}_m(z=1,t=t_{l,3D}):$ the mass flux of the fast jet, selected by $\Gamma(z=z^*,t=t_{l,3D})>2$, normalized to the value at injection, for the various cases at the end of the simulations and at $z^*=15$.}
    \label{fig:scan_mflux}
\end{figure}

The average velocity does not give us a full information about the jet velocity structure, in fact from the average velocity we cannot know whether relativistic material is still present and to which extent. This is more clearly displayed in Fig. \ref{fig:scan_histo}, where we show the distribution of $\beta_z\Gamma(x,y)$, for the two cases at the end of the simulations, at the same position $z = 20$.  The histograms show a striking difference between the two cases, in case SHC05 (blue), there is no material with a velocity larger than $\beta_z = 0.3$, while in case SLWw2 (pink) we still observe material with $\Gamma \beta_z \gtrsim 2$.
An overview on the presence of relativistic material across the different cases can be obtained by looking at Fig. \ref{fig:scan_mflux}, where we plot the ratio $\mathcal{R}$ between the mass flux of the material moving with $\Gamma > 2$, $\tilde{\Phi}_m$, at $z=15$, normalized to the jet mass flux at the base. The figure displays a strong difference between the three lower power cases, in which the relativistic material has completely disappeared, the intermediate power cases and the highest power jet (SLWw2) in which the ratio is about 2. At larger distances, $z \gtrsim 20$, the highest power case stands even more apart,  with the ratio $\mathcal{R}$ remaining 2 for this case, while it falls to zero for all the other cases.

\section{Cylindrical case \label{remarks}}

All hydrodynamic jets we have simulated develop recollimation instabilities, which exert a feedback on the  recollimating structure and induce entrainment and deceleration. The specific behavior generally falls in between what we have observed for setups SHC05 and SLWw2, which stand at the opposite ends on many aspects of the jet parameter space we explored: SHC05 and SLWw2. SHC05 is cold, narrow, kinetically dominated and the least powerful one, while SLWw2 is wide, warm, enthalpy dominated, and it is the most powerful jet we simulated.

\begin{figure}[htbp]
    \centering    
     \includegraphics[width = \columnwidth]{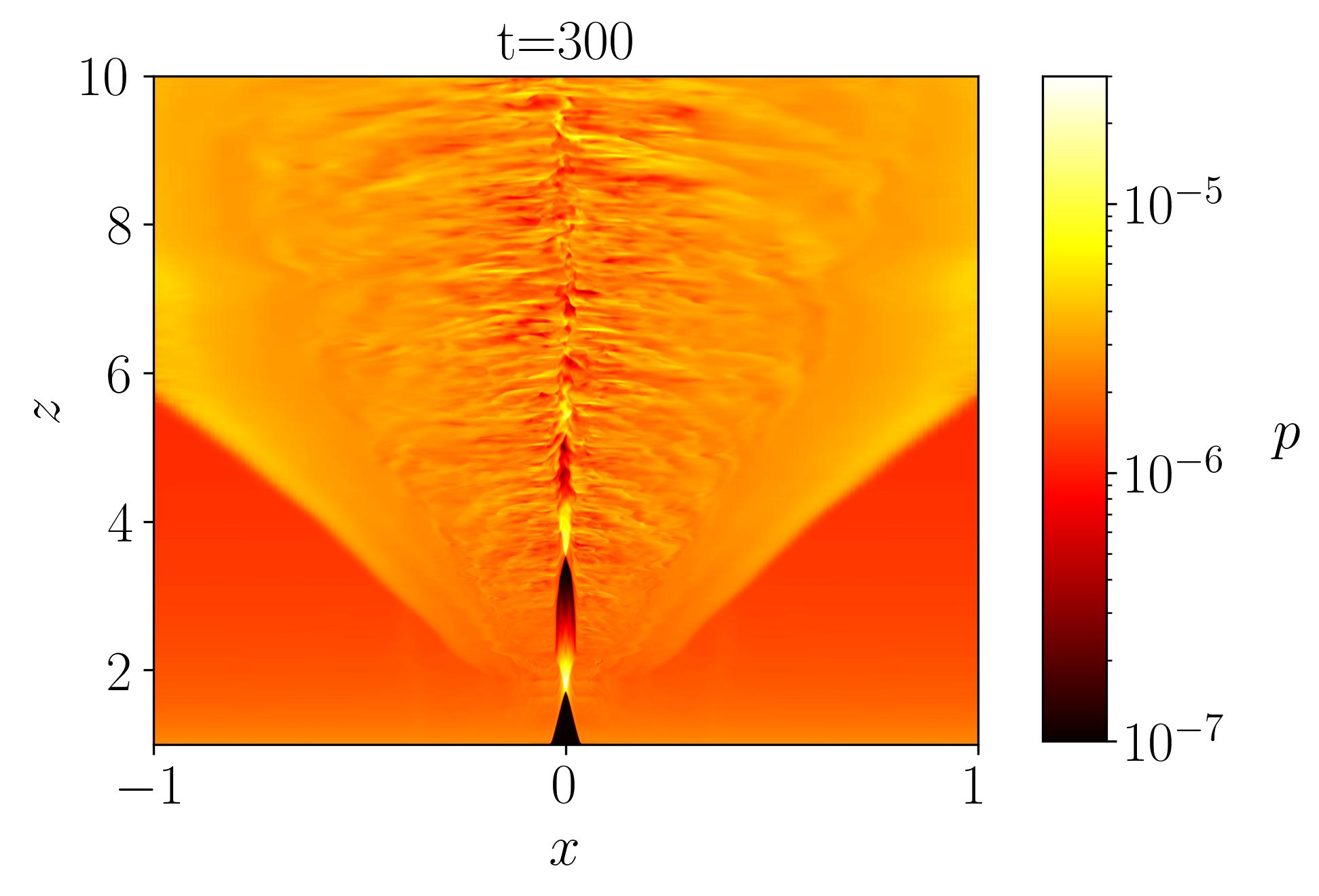}  
     \caption{$p(x,y=0,z,t=t_{l,3D})$ map in case SHCcyl, displayed in the same color-scale of case SHC05.}
        \label{fig:SHCcyl_maps_prs}
\end{figure}

\begin{figure}[htbp]
    \centering    
     \includegraphics[height=8cm]{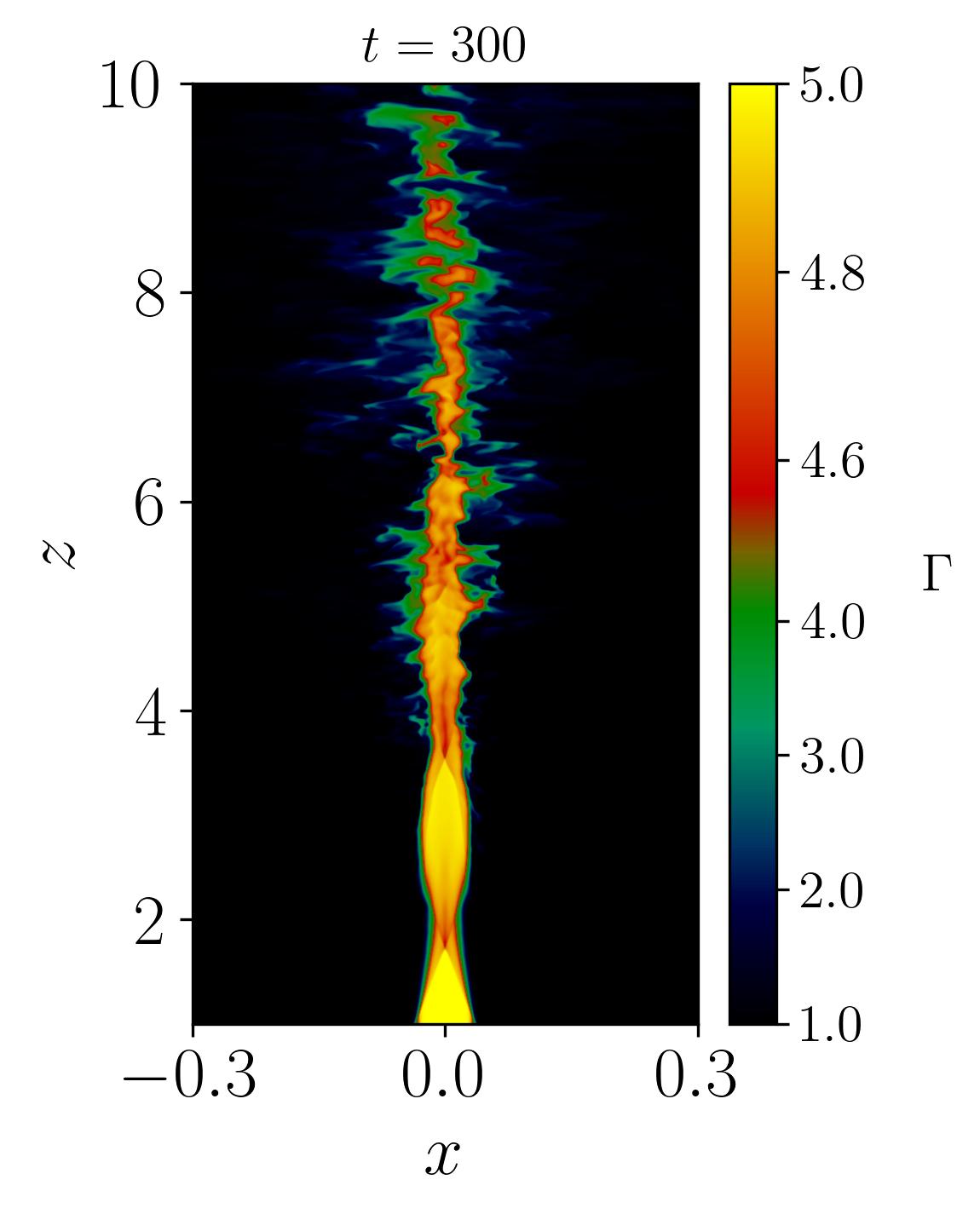}  
     \caption{$\Gamma(x,y=0,z,t=t_{l,3D})$ map in case SHCcyl.}
        \label{fig:SHCcyl_maps_lf}
\end{figure}

All the simulations we  presented  assume a conical jet, but we have also performed one simulation of a cylindrical jet. Sagittal maps of the pressure and of the Lorentz factor at $y=0$, at the end of the simulation $t=t_{l,3D}=300$, are shown in Figs. \ref{fig:SHCcyl_maps_prs} and \ref{fig:SHCcyl_maps_lf}. 

The SHCcyl jet is peculiar: it does not expand in the beginning, since it is cylindrical and under-pressured, in contrast with conical simulations, which are all characterized by a significant ram-pressure due to a relativistic radial velocity component. The first recollimation shock is not curved, and it does not trigger the CFI. Nevertheless, the jet develops the chain of shocks, as in the conical case, and the subsequent recollimation shocks are curved, leading to the onset of the CFI. As in the conical case, the growth of instabilities leads to the formation of an over-pressured bubble, whose interior part is turbulent (the sheath). Turbulence in the sheath induces entrainment and jet deceleration. Even if the first two recollimation shocks are almost undisturbed (up to $z \sim 3.5)$, the third one becomes perturbed, and for $z \gtrsim 5$ the recollimation chain disappears. 

We can compare this simulation with SHC05, that differs from this case only for the jet opening angle $\theta_j$ (and for the pressure ratio $p_r$, but in a non-significant way). In the SHC05 case, the  spine loses its coherency at greater distances, for $z\gtrsim7$. Looking at Figs. \ref{fig:SHC05_long_prs} and \ref{fig:SHCcyl_maps_prs} at $t=300$, we observe that this happens because the SHCcyl jet first recollimates over shorter distances, with the first recollimation point being located at $z_{c,1}\simeq1.7$, slightly less than the respective point $z_{c,\text{SHC05},1}=2.2$ for case SHC05.

This simulation allows us to compare our work with results in literature that considered cylindrical under-pressured jets that recollimate, as in jet-cocoon system \citeg{rossi20,ayan24}. In particular, we observe good agreement with the results of \cite{ayan24}, where there are differences due to slightly different parameters, but the qualitative results are consistent, with the deceleration and disruption of the spine happening after a few recollimation shocks. We also see agreement with \cite{rossi20} where, even if later, the instability develops in a very similar way, with analogous features. This proves that cylindrical jets are not fundamentally different from conical ones in the way they respond to recollimation instabilities. In addition, they can recollimate faster, becoming unstable over shorter scales with respect to conical jets.

\section{Discussion \label{discussion}}
In this work we investigated the development and outcome of hydrodynamic instabilities of recollimation shocks in relativistic jets, by performing three-dimensional numerical simulations with the PLUTO code.  We performed a parameter study and, although specific details differ in the various setups, we can robustly conclude that low-power hydrodynamic jets are all unstable at recollimation due to a combination of CFI, RMI and KHI. Most of our jets are injected with an expanding conical profile, but we also performed a simulation of a cylindrical jet, to allow a comparison with other works that assume such configuration.

We presented a detailed view of the onset of the recollimation-induced instabilities. Our results support the CFI and RMI nature of the instabilities. In particular, the growth times of initial perturbations downstream of recollimation shocks grow monotonically with decreasing angular velocity of curved streamlines, as one would expect for the CFI; moreover, we found evidence of the RMI by comparing the evolution of corrugations at the reflection shock crossing with the theory of the RMI \citep{richt1960,meshkov,meyerblewett72}.

We focused on the non-linear phase of the instability growth and studied the dynamical evolution of the jet in reaction to the effects of the instabilities. The dissipation of the bulk kinetic energy of the flow leads to the formation of an expanding high-pressure bubble, that consists of an internal hot and turbulent sheath composed by jet material moving at low Lorentz factor and of an external region composed by the cold heavy ambient material. This bubble provides a new confining condition for the inner relativistic jet, the spine, i.e. the inner faster region where recollimation occurs. The recollimation shock chain is disrupted after just a few shocks, and those shocks which do survive are now located closer to the central to the central engine. In general, this re-adjustment is greater in more powerful jets (that according to the axisymmetric picture should instead recollimate later), that provide more thermal energy to the sheath. In our simulations we observe that, after the readjustment, the position of the first recollimation point is always below $10\,z_0$, where $z_0$ is the distance at which reconfinement starts.

We further studied the properties of the decelerated and heated jets. In general, in terms of the distance from the central engine, we distinguished three regions: a first during which the jet propagates and reconfines almost undisturbed by instabilities, a second during which the jet spine is still undergoing shocks, while being compressed by the hot and turbulent sheath, finally the overall jet slows down in the third region, with velocities averaged across the jet becoming sub-relativistic. We note that, although the averaged velocity becomes sub-relativistic, a faster spine, that may be still relativistic in the more powerful jet cases, can persist for significant distances into the third region. 

Examining the influence of the various parameters, we studied the effects of the  density contrast, jet temperature, Lorentz factor and opening angle, by choosing at least two values for each parameter. We fixed the environment conditions. We found that the factors that influence  most the jet stability are the jet opening angle, and the jet power (that depends both on the jet density contrast and Lorentz factor). Indeed, we showed that  jets decelerate to sub-relativistic velocities after a few recollimation and reflection shocks, with the exception of our most powerful case, SLWw2, that maintains a largely undisturbed relativistic spine over its whole computational domain (up to $40\,z_0$). We note that our setup and parameters have been chosen to describe low-power jetted AGNs, like FRIs and especially FR0s, that could recollimate at blazar scales. We expect that the combination of different stabilizing properties results in a progressive stabilization of the relativistic ejecta, eventually allowing hot, fast and dense, and therefore more powerful, jets to reach kpc distances. This is consistent with the typical view of the FRI-FRII dichotomy, that FRIIs are characterized by more powerful jets \citeg{gourgkom18}, with higher Mach numbers \citeg{deyoung93,laing02}, and/or magnetization \citeg{matsumoto21}, ensuring the stability required to survive the recollimation instabilities and reach the kpc scale. We observe that different choices for the environment would influence the jet dynamics too. Nevertheless, we did not perform dedicated simulations to study the effect of the environment, as with steeper pressure profiles (larger $\eta$) or lower external temperatures, $\mathcal{T}_\text{ext}$, since the jet would recollimate later, or not at all if $\eta\gtrsim2$, or if the environment changed at greater distances. Studying in detail recollimation in jets in such an environment would then require larger domains and increase the computational costs of the simulations; at the same time it is out of our interest, since we wanted to focus on blazar-scale recollimating jets.

Comparing our results with the previous literature, we found general agreement with findings from several earlier 3D RHD simulations of unstable recollimation shocks, where it was observed that jets become unstable to CFI \citep{gourgkom18}, that narrow, fast, and heavy/powerful recollimated jets tend to be less subject to recollimation-induced instabilities \citep{gourgkom18,matsumasa19,GottliebHD,ayan24} and that light jets are unstable \citep{gourgkom18,ayan24} as well, even if the conditions for the onset of the RTI are not satisfied.
It is instead difficult to compare our work with \citealt{abolbrom23,hu25}, as they focused on heavier jets and interpreted the primary instability as RTI, while we focused on lighter jets. Comparing our simulations of steady-state naked jets with simulations of propagating jets, it is possible to interpret the nature of some of the instabilities observed in these cases \citeg{rossi20} as recollimation-induced. Furthermore, there are some differences in considerations about the impact of a high jet temperature \citep{matsumasa19}, that in our simulations does not appear to have a destabilizing effect. Discrepancies may arise because of differences in modeling and partly because these simulations typically adopt lower resolutions due to employing larger domains.

In this paper we ignored the effects of magnetic fields. This assumption makes the results meaningful for contexts where the magnetic fields are unimportant for the jet dynamics,  i.e. for cases in which the magnetization is low and the jet is kinetically dominated with respect to Poynting dominated, like in lower power blazars \citep{sciacca24}. Nevertheless, the importance of including magnetic fields in the treatment has been stressed in a few works on recollimation instabilities, where they showed that strong toroidal fields can stabilize recollimation instabilities due to magnetic tension \citep{matsumoto21,GottliebMHD}, and suggested the possible development of kink instabilities that could destabilize the flow in an alternative way \citep{hu25}. Therefore, it will be important to include magnetic fields in the treatment in future works (Boula et al. in prep.). 

Relativistic magneto-hydrodynamic simulations will also be instrumental in characterizing consistently particle acceleration, emission and polarization of these sources \citeg{sciacca25}. In fact, our simulations show complex dynamic features at the recollimation point that are interesting for phenomenological modeling. An interesting first aspect is that instability naturally leads to the formation of a velocity structured flow, with a fast core (or spine) surrounded by a slower sheath. This supports early scenarios for the emission of low power BL Lac and radiogalaxies based on the radiative interplay between the two structures \cite{ghisellini05}. Moreover, the turbulent  sheath and  sheath-spine interface represent the ideal places where efficient acceleration can occur via stochastic \citeg{tavecostasciacca}) or shear \citeg{rieger19}) mechanisms, possibly leading to the hard electron distributions required to reproduce extreme BL Lacs \citeg{biteau20}. Further numerical studies, including also the treatment of relativistic particle evolution and emission \citeg{vaidya18}, can provide important clues on these relevant issues.

\begin{acknowledgements}
This work has been supported by the Spoke-1 "FutureHPC \& BigData” of the ICSC – Centro Nazionale di Ricerca in High Performance Computing, Big Data and Quantum Computing – and hosting entity, funded by European Union – NextGenerationEU. We acknowledge financial support by Universita' degli Studi dell'Insubria, by INAF Theory Grant 2022 (PI F. Tavecchio) and the PRIN 2022 (2022C9TNNX) project. We acknowledge support by CINECA, through ISCRA and Accordo Quadro INAF-CINECA, and by PLEIADI, INAF – USC VIII, for the availability of HPC resources. Plots have been made with VisIt and Python. We have used the following Python libraries: Numpy \citep{numpy}, Matplotlib \citep{matplotlib}, Scipy \citep{scipy}, and PyPluto \cite{pypluto25}. We thank Konstantinos N. Gourgouliatos and Sergei Komissarov for useful discussions on simulations and on the CFI.

\end{acknowledgements}

\bibliographystyle{aa}
\bibliography{Bibliography}

\appendix

\section{Setup details}
\subsection{Details of the 2D setup \label{2D_details}}

In order to reduce numerical noise at the jet-environment boundary, we smooth the initial profiles of Lorentz factor, density and pressure with functions of the type \citep{mukh20}
\begin{equation}
    q(r,z) = q_{\text{ext}}(r,z) + \left(q_j(r,z)-q_{\text{ext}}(r,z)\right)\sech\left[\left(\frac{r}{z\theta_q}\right)^{\alpha_q}\right],
    \label{eq:smooth}
\end{equation}
for the conical jet, and
\begin{equation}
    q_{\text{cyl}}(r,z) = q_{\text{ext}}(r,z) + \left(q_j(r,z)-q_{\text{ext}}(r,z)\right)\sech\left[\left(\frac{r}{r_q}\right)^{\alpha_q}\right]   
    \label{eq:smooth_cyl}
\end{equation}
for the cylindrical one. 

The parameters $\alpha_q$, $\theta_q$ and $r_q$ define the width and angular or radial position of the transition for each quantity $q$. The Lorentz factor one is sharper and happens at lower jet angle/radius with respect to the density and pressure transitions. This is chosen to avoid the formation of unnatural maxima and minima in the profiles of conserved quantities \citeg{abolbrom23}. Their values for each case can be read in Tab. \ref{tab:grid}, that describes a set of details of the 2D and 3D setups for all the simulated cases.

The 2D simulation domain is $[0,r_{\max,2D}]\times[z_{\min,2D}, z_{\max,2D}]$, where the lengths are in units of $z_0$, and the specific values of $r_{\max,2D}$ and $z_{\max,2D}$ depend on the requirements of each simulation, with typical values of $20$ and $30$ respectively. The grid is uniform only in the inner regions of the jet, in $[0,r_{u,2D}]\times[z_{\min,2D},z_{u,2D}]$, while it is stretched in the outer regions, $[r_{u,2D},r_{\max,2D}]\times[z_{u,2D},z_{\max,2D}]$, with typical values of $n_{r,2D} =1400 = 1000\text{u}+ 400\text{s}$ and $n_{z,2D} = 5000\text{u}+350\text{s}$. The notation indicates the number of points in the uniform and geometrically spaced sections of the grid with `u' and `s' respectively. The details of the grid are also listed in table \ref{tab:grid}. 

We observe that $z_\text{min,2D}<1$ in all cold jet cases. Indeed, the computational domain was augmented in order to fix the injection conditions of the jet in the initial phases of the propagation: the fluxes of the Riemann solver got set to zero in the region $[0,z \theta_j ]\times [0.5,1]$ in the case of conical simulations, and $[0,r_0]\times [0.5,1]$ for the cylindrical case. Outside, in $[z\theta_j,r_{\max,2D}]\times[0.5,1]$, the environment is free to evolve. This has been done to separate the evolution of the contact discontinuity from the lower boundary where there may be numerical effects due to reflection of waves. In the case of warm jets, this was not necessary.

\subsection{Details of the 3D setup \label{3D_details}}
The 3D domain in Cartesian coordinates covers the region $[-L_{\max,3D},L_{\max,3D}]\times[-L_{\max,3D},L_{\max,3D}]\times[1,z_{\max,3D}]$, with a uniform grid only in $[-L_{u,3D},L_{u,3D}]\times[-L_{u,3D},L_{u,3D}]\times[1,z_{u,3D}]$. The typical number of points is $n_{x,3D} = n_{y,3D} = 350u+200s$, where $100s$  points are for $[-L_{\max,3D},-L_{u,3D}]$ and $100s$ for $Lx_{u,3D},L_{\max,3D}]$, and $n_{z,3D} = 1500u+350s$, see Tab. \ref{tab:grid}.

The computational domain of 3D simulations was reduced with respect to 2D simulations in the lateral $x$ and $y$ directions, since the jet is already collimated at the start of the 3D simulation, and a great lateral extension was not necessary any more. We also had to reduce the overall resolution of the simulations, since a 3D one is extremely more expensive from a computational point of view. In the lateral direction, where the CFI and RMI modes are expected to grow the most, we adopted a higher resolution than in the longitudinal direction, with typically $\Delta x_{3D} = \Delta y_{3D} = 0.007$, while $\Delta z_{3D} = 0.01$. The number of grid points per jet radius is typically $n_{j,3D} \simeq 30$.

\begin{table*}
    \begin{tabular}{|c|c|c|c|c|c|c|c|c|}
        \hline
        \diagbox[width=2.5cm,height=1.1cm]{Param.}{Case} & 
        \textbf{SLC2} & 
        \textbf{FfLC2} & 
        \textbf{SLW2} & 
        \textbf{SLWw2} & 
        \textbf{SHC2} & 
        \textbf{SHC05} & 
        \textbf{SHCcyl} \\
        \hline
        \hline
        \rowcolor{cleangray} $\alpha_\Gamma$ & 8 & 8 & 8 & 8 & 8 & 13 & 9 \\
        \hline
        $\theta_\Gamma$ / $r_\Gamma$ & 0.16 & 0.16 & 0.16 & 0.16 & 0.16 & 0.04 & 0.07 \\
        \hline
        \rowcolor{cleangray} $\alpha_{\rho} = \alpha_p $ & 10 & 10 & 10 & 10 & 10 & 14 & 14 \\
        \hline
        $\theta_{\rho} = \theta_p $ / $r_\rho= r_p$ & 0.29 & 0.29 & 0.29 & 0.29 & 0.29 & 0.07 & 0.16 \\
        \hline
        \hline
        \rowcolor{cleangray} $r_{\max,2D}$ & 20 & 20 & 20 & 20 & 20 & 3 & 0.5 \\
        \hline
        $z_{\max,2D}$ & 30 & 30 & 30 & 80 & 80 & 30 & 15 \\
        \hline
        \rowcolor{cleangray} $z_{\min,2D}$ & 0.5 & 0.5 & 1 & 1 & 0.5 & 0.5 & 0.5 \\
        \hline
        $r_{u,2D}$ & 1.5 & 1.5 & 1.5 & 10 & 10 & 1.5 & 0.3 \\
        \hline
        \rowcolor{cleangray} $z_{u,2D}$ & 20 & 20 & 20 & 40 & 40 & 20 & 10 \\
        \hline
        $n_{r,2D}$ & 1000u+400s & 1000u+400s & 1000u+400s & 2000u+400s & 2000u+400s & 1000u+400s & 1000u+400s \\
        \hline
        \rowcolor{cleangray} $n_{z,2D}$ & 3000u+700s & 3000u+700s & 3000u+700s & 3000u+700s & 3000u+700s & 3000u+700s & 3000u+700s \\
        \hline
        $t_{l,2D}$ & 3000 & 3000 & 3000 & 6000 & 6000 & 3000 & 3000 \\
        \hline
        \hline
        \rowcolor{cleangray} $L_{\max,3D}$ & 5 & 5 & 5 & 6 & 6 & 2.5 & 5 \\
        \hline
        $z_{\max,3D}$ & 30 & 30 & 30 & 60 & 60 & 30 & 30 \\
        \hline
        \rowcolor{cleangray} $L_{u,3D}$ & 1 & 1 & 1 & 2 & 2 & 0.3 & 0.3 \\
        \hline
        $z_{u,3D}$ & 20 & 20 & 20 & 40 & 40 & 20 & 20 \\
        \hline
        \rowcolor{cleangray} $n_{x,3D}$ & 350u+200s & 350u+200s & 350u+200s & 600u+200s & 600u+200s & 350u+250s & 300u+250s \\
        \hline
        $n_{z,3D}$ & 1500u+350s & 1500u+350s & 1500u+350s & 3000u+700s & 3000u+700s & 1500u+350s & 1500u+350s \\
        \hline
        \rowcolor{cleangray} $t_{l,3D}$ & 368 & 300 & 300 & 640 & 300 & 300 & 300 \\
        \hline
    \end{tabular}
    \captionsetup{width=\textwidth} 
    \captionof{table}{Setup and run details in the different simulations.}
    \label{tab:grid}
\end{table*}

\section{Calculation of averages \label{appendix_average}}

In the paper we have plotted averages of a few jet quantities, calculated on transverse sections of portions of the jet in the $x-y$ plane. Depending on the specific case, the area $\mathcal{A}$ was selected based on a certain condition $\mathcal{K}(x,y,z,t)$. In general, the average of the quantity $q(x,y,z,t)$ is calculated as
\begin{align}
    \langle q(z,t)\rangle_{x,y} = & \frac{1}{\mathcal{A}} \int_{\mathcal{A}} q(x,y,z,t)\,dx\,dy =\\
    =&\frac{\int_{x,y} q(x,y,z,t) g(x,y,z,t)\,dx\,dy}{\int_{x,y} g(x,y,z,t)\,dx\,dy},
\end{align} 
where we defined the step function 
\begin{equation}
    g(x,y,z,t) = \begin{cases}
        0\quad\mbox{if }\mathcal{K}(x,y,z,t)=\text{False}\\
        1\quad\mbox{if }\mathcal{K}(x,y,z,t)=\text{True}
    \end{cases}.
    \label{eq:stepfunc}
\end{equation}

\section{Shock/wave detection \label{appendix_shock_det}}

Shocks and wave fronts discussed in the paper are localized through a detection algorithm \citep{mignoneAMR} based on a condition on the pressure gradient:
\begin{equation}
    \frac{\boldsymbol{\nabla} p}{p}\cdot d\mathbf{x} > k(z,t).
    \label{eq:compratio}
\end{equation}

The exact value of the adimensional parameter $k(z,t)$ has been chosen based on specific conditions. For waves discussed in Section \ref{bubble}, it has adaptively been set to values between $0.1-1$, since the waves strength varied with time and altitude. 

\section{Maps of non-displayed simulations}
\begin{figure}[htbp]
    \centering
    
    \begin{minipage}{0.49\columnwidth}
        \centering
        \includegraphics[width=\textwidth]{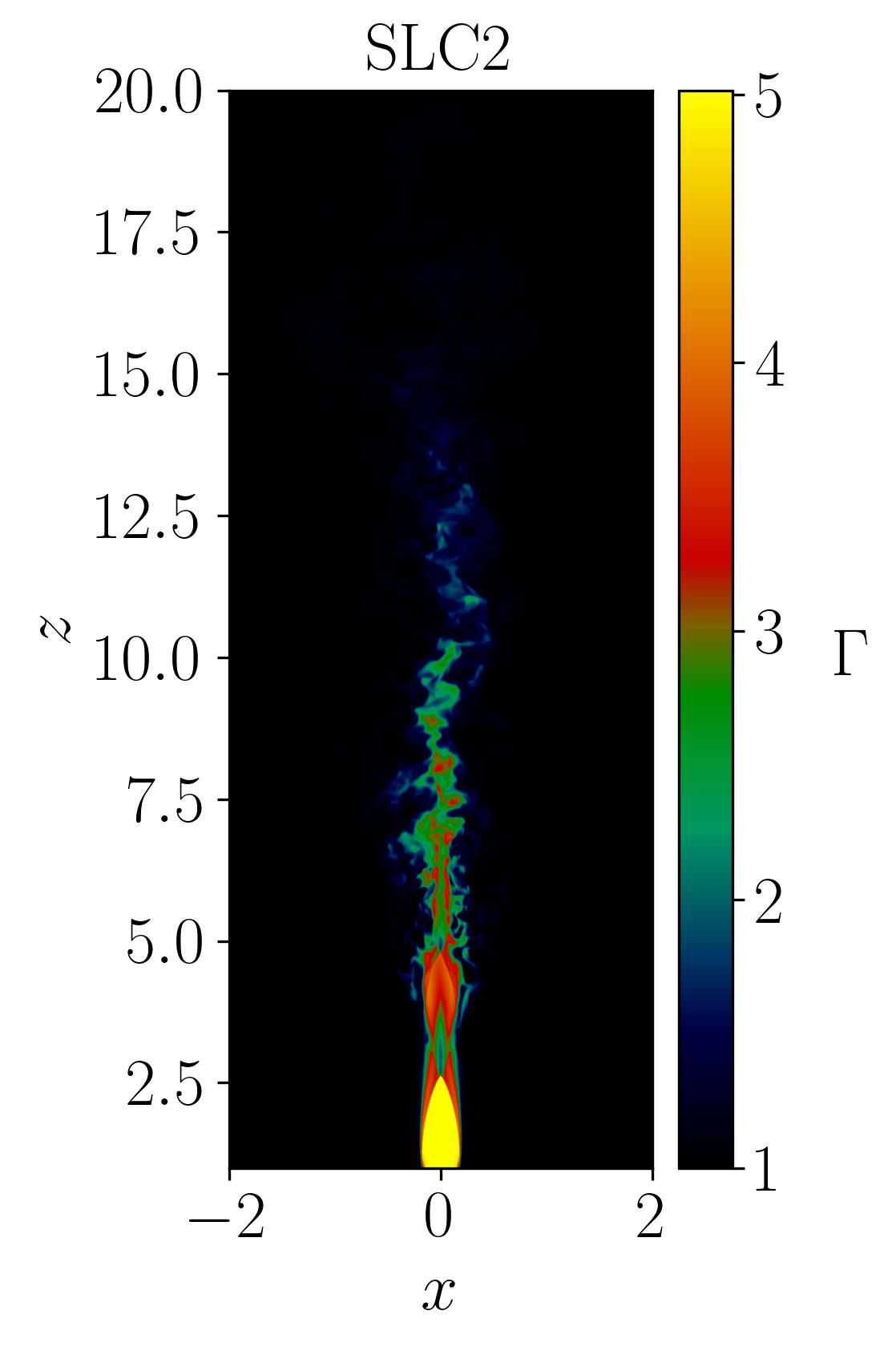}  
    \end{minipage}
    \begin{minipage}{0.49\columnwidth}
        \centering
        \includegraphics[width=\textwidth]{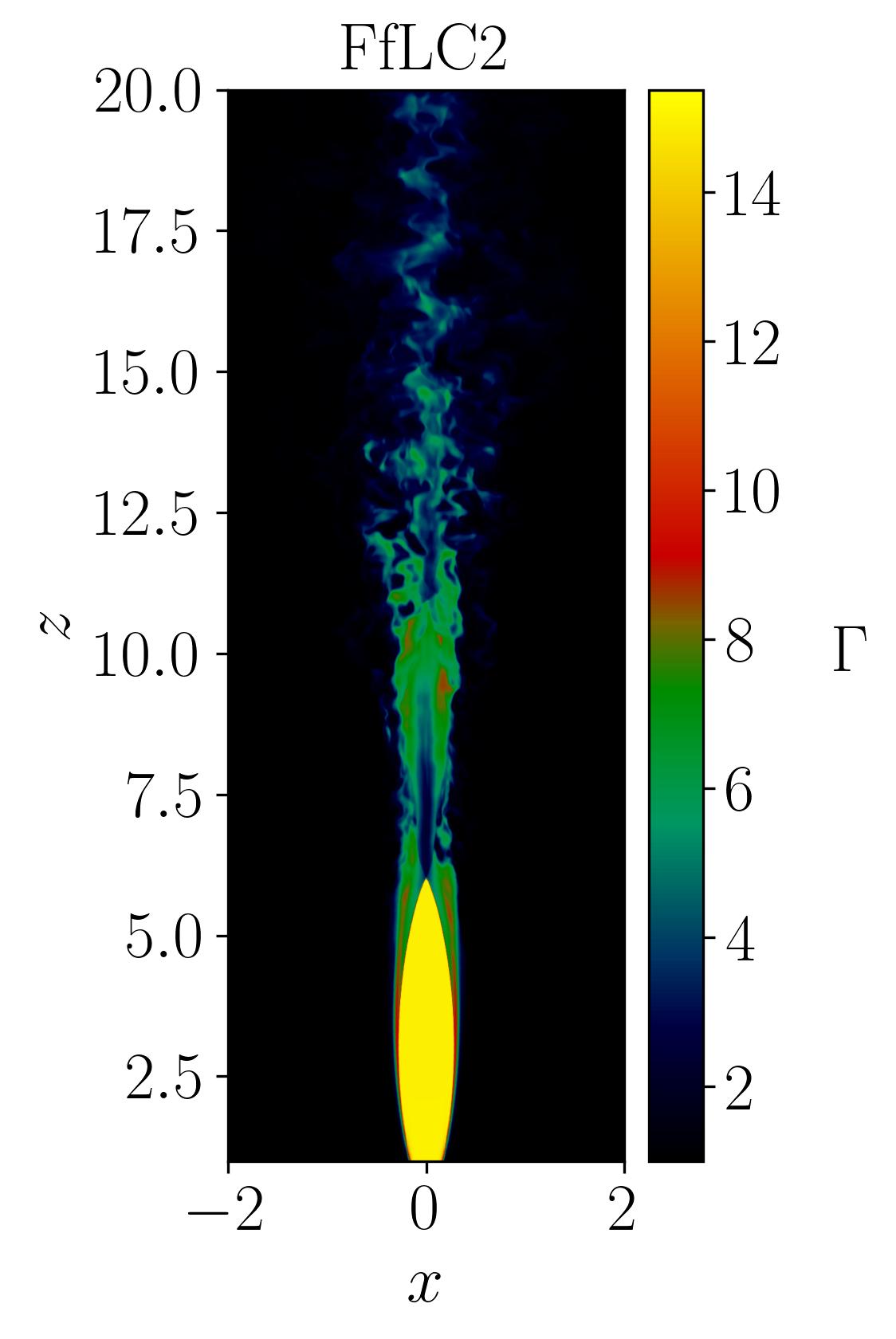}  
    \end{minipage}
    
    \begin{minipage}{0.49\columnwidth}
        \centering
        \includegraphics[width=\textwidth]{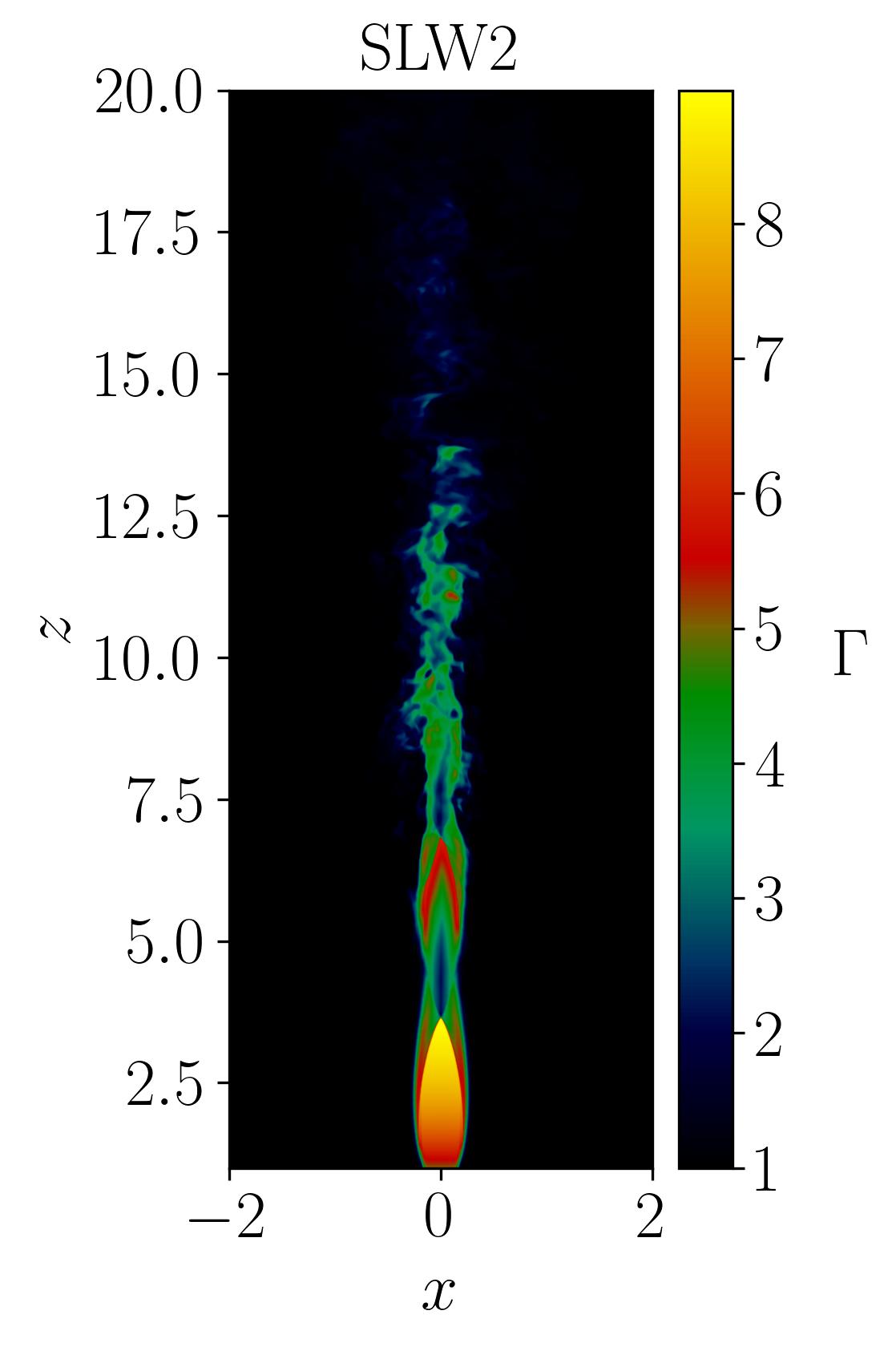}  
    \end{minipage}
    \begin{minipage}{0.49\columnwidth}
        \centering
        \includegraphics[width=\textwidth]{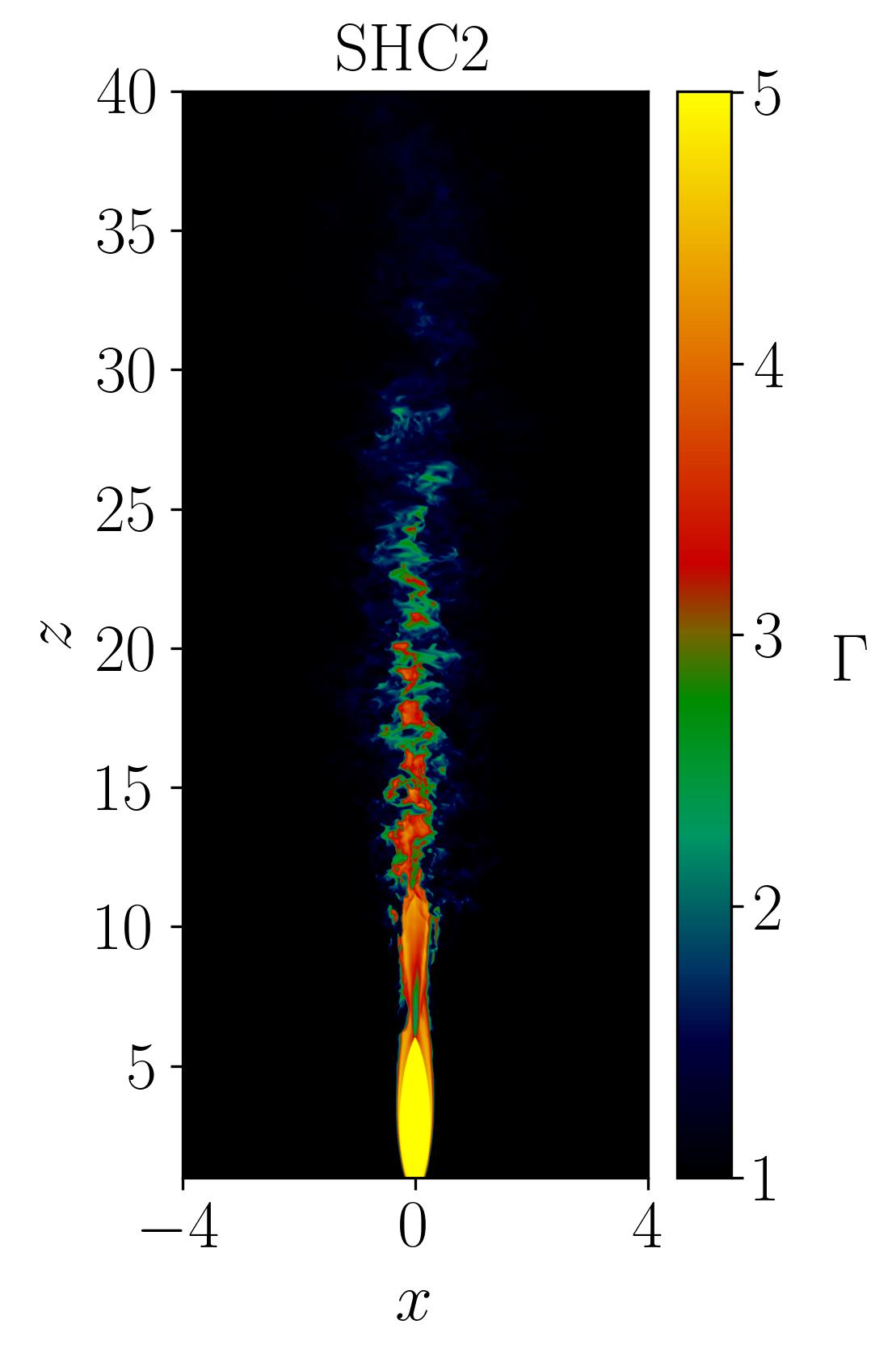}  
    \end{minipage}
    \caption{$\Gamma(x,y=0,z,t=t_{l,3D})$ for cases SLC2, FfL2, SLW2, SHC2.}
    \label{fig:scan_maps}
\end{figure}

For reference, in  \ref{fig:scan_maps} we report complementary maps of the Lorentz factor for the 4 conical setups that we did not describe individually: SLC2, FfLC2, SLW2, and SHC2. Case SHC2 is displayed on a longer box, up to $z=40$. The maps are made at the end of the respective runs, at $y=0$ and show that more powerful jets, by means of Lorentz factor (FfLC2), temperature (SLW2), or density (SHC2), can propagate relativistically to larger distances with respect to the weakest wide one: SLC2.

\end{document}